\documentclass[a4paper,11pt]{article}
\usepackage{jheppub} 
\usepackage{lineno}
\usepackage{float}
\usepackage{slashed}
\usepackage{xfrac,bigints}
\usepackage{physics}
\usepackage{subfigure}
\usepackage{booktabs} 
\usepackage{mathtools}
\usepackage{enumitem}
\usepackage[dvipsnames]{xcolor}
\usepackage{tikz-cd}
\usepackage{fontspec}
\usepackage{comment}

\usepackage{titlesec}

\titleformat{\paragraph}
{\normalfont\normalsize\bfseries} %
{\theparagraph}                   %
{1em}                             %
{}                                %

\titlespacing*{\paragraph}
{0pt}                             %
{3.25ex plus 1ex minus .2ex}      %
{1.5ex plus .2ex}                 %

\newfontfamily\junirunic{Junicode}[Path=./, Extension=.ttf, UprightFont=*-Regular, BoldFont=*-Bold, ItalicFont=*-Italic, BoldItalicFont=*-BoldItalic]

\newcommand{\jrunes}[1]{{\junirunic #1}}
\newcommand{\pertho}[0]{{\text{\jrunes{ᛈ}}}}
\newcommand{\ehwaz}[0]{{\text{\jrunes{ᛖ}}}}
\newcommand{\algiz}[0]{{\text{\jrunes{ᛉ}}}}

\newcommand{\fehu}[0]{{\text{\jrunes{ᚠ}}}}
\newcommand{\berkanan}[0]{{\text{\jrunes{ᛒ}}}}
\newcommand{\thurisaz}[0]{{\text{\jrunes{ᚦ}}}}

\newcommand{\shiftedfillupsphere}{\mathord{\raisebox{-0.15\height}{\includegraphics[height=5ex]{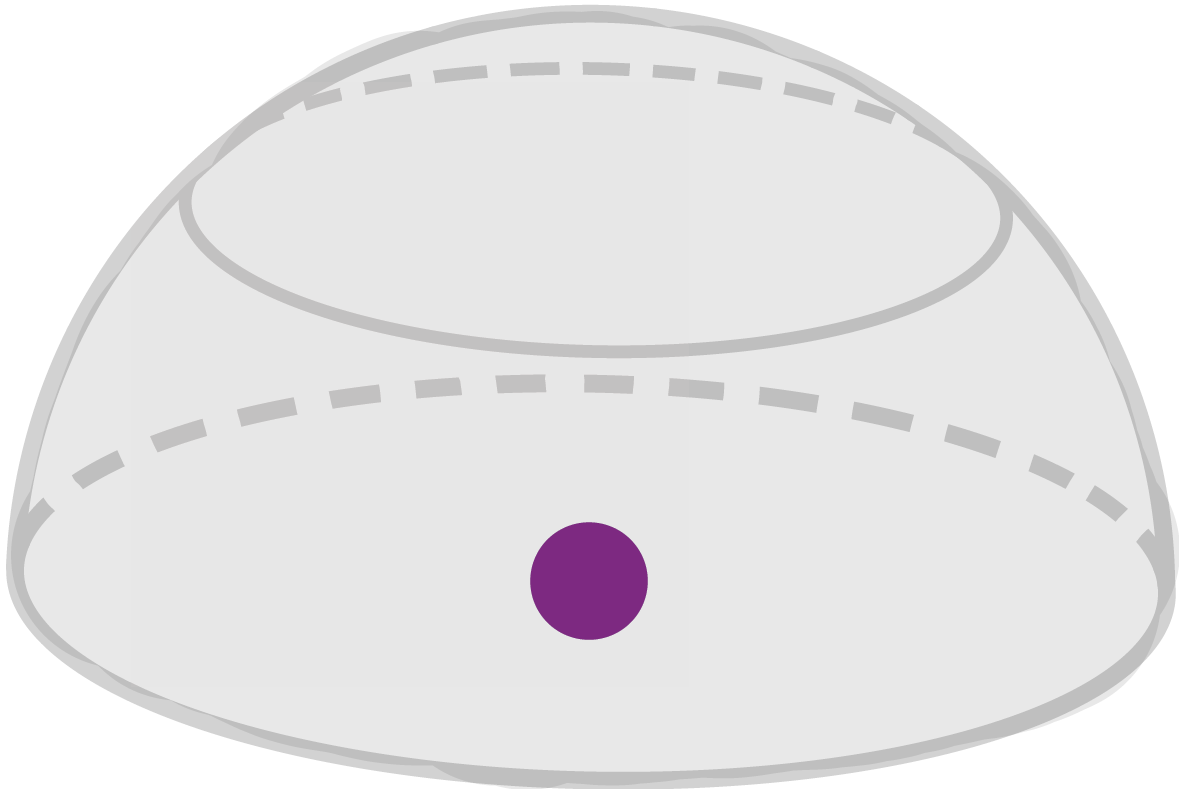}}}}
\newcommand{\shiftedfilldownsphere}{\mathord{\raisebox{-0.65\height}{\includegraphics[height=5ex]{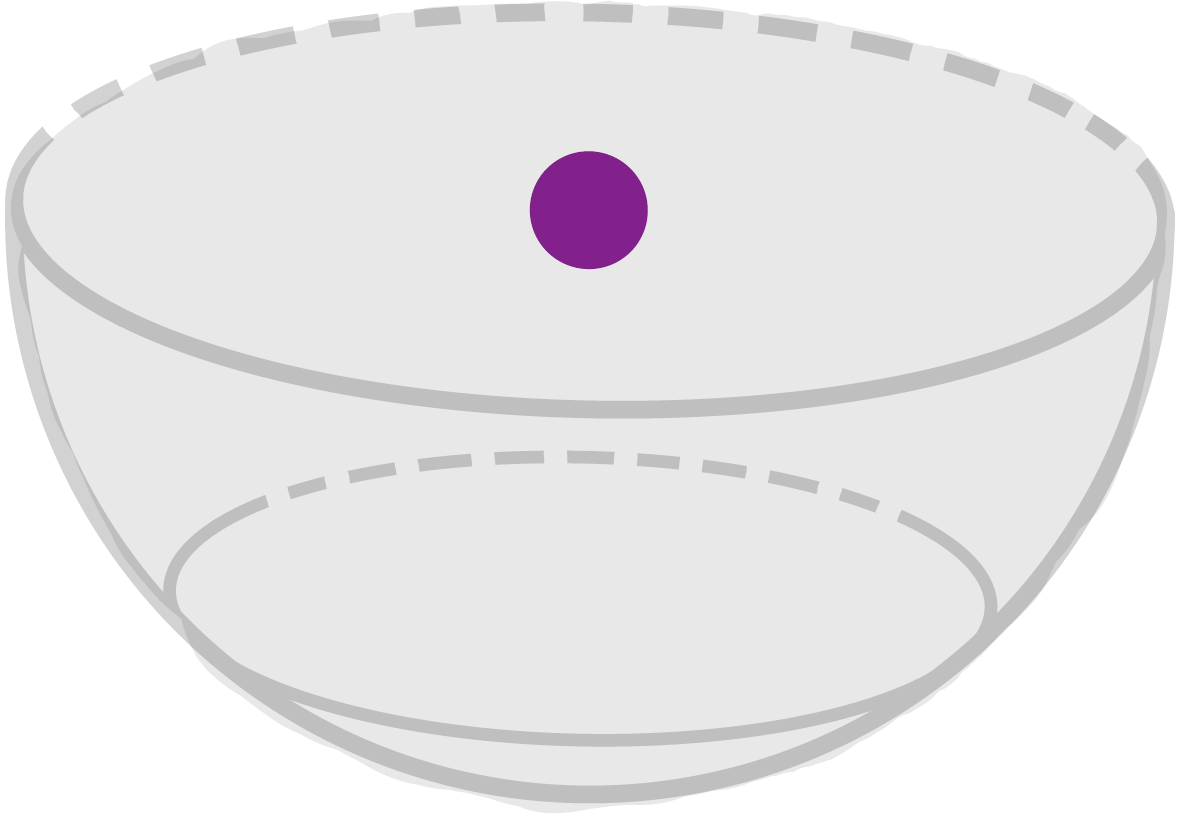}}}}

\newcommand{\leftcurve}{\mathord{\vcenter{\hbox{\includegraphics[height=10ex]{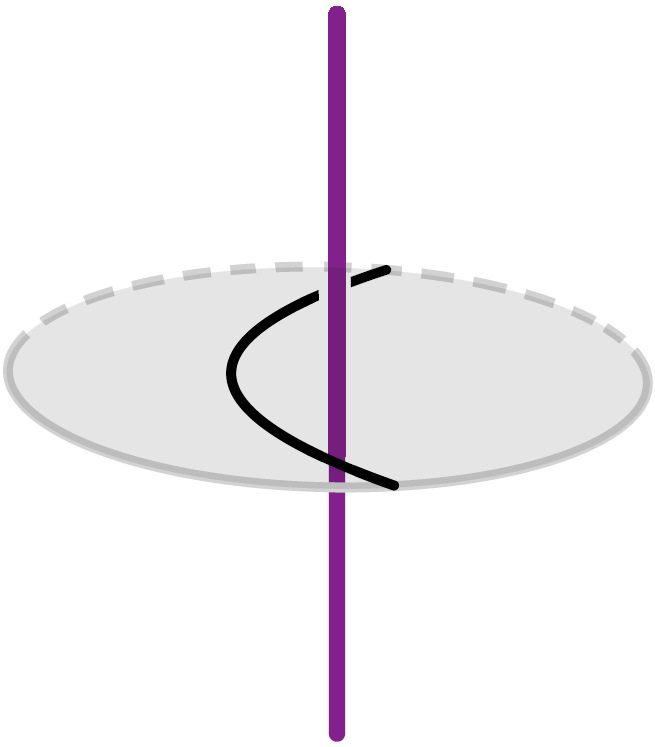}}}}}
\newcommand{\rightcurve}{\mathord{\vcenter{\hbox{\includegraphics[height=10ex]{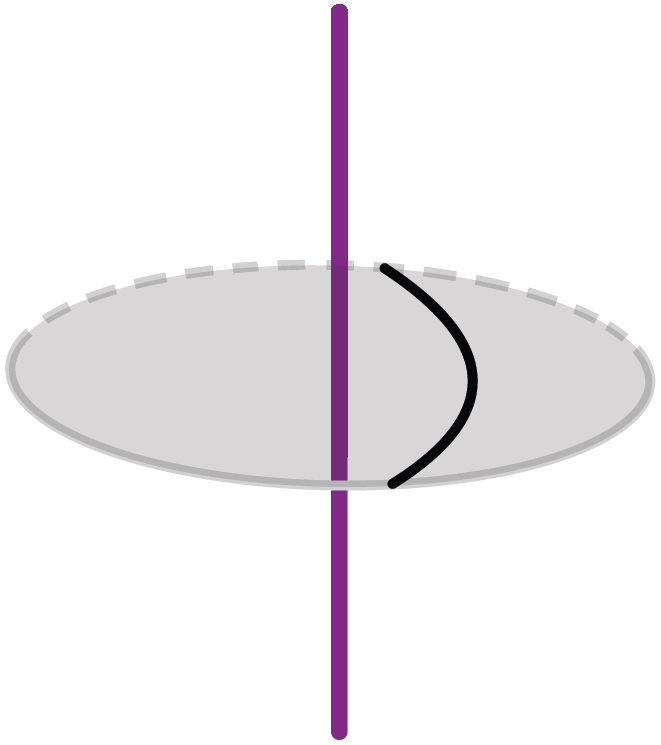}}}}}

\newcommand{\leftvertexblue}{\mathord{\vcenter{\hbox{\includegraphics[height=10ex]{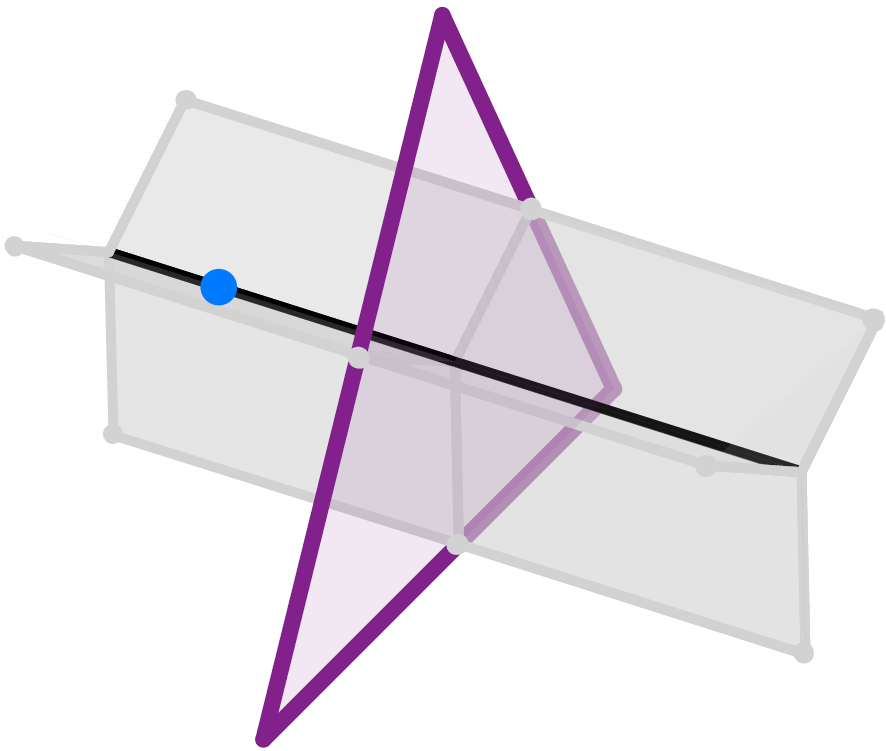}}}}}
\newcommand{\rightvertexblue}{\mathord{\vcenter{\hbox{\includegraphics[height=10ex]{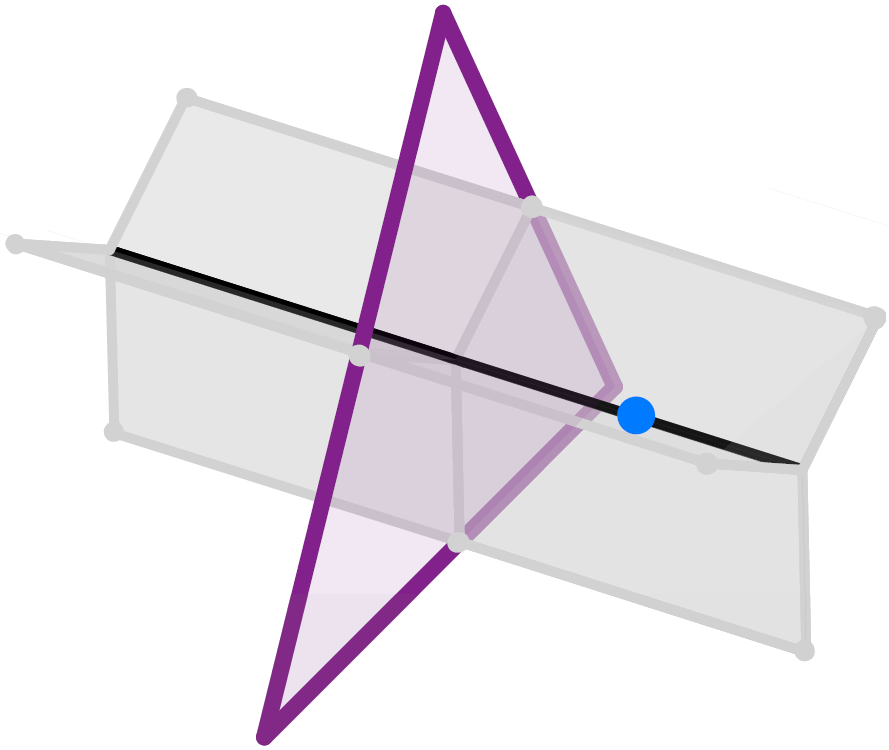}}}}}
\newcommand{\removablesummand}{\mathord{\vcenter{\hbox{\includegraphics[height=20ex]{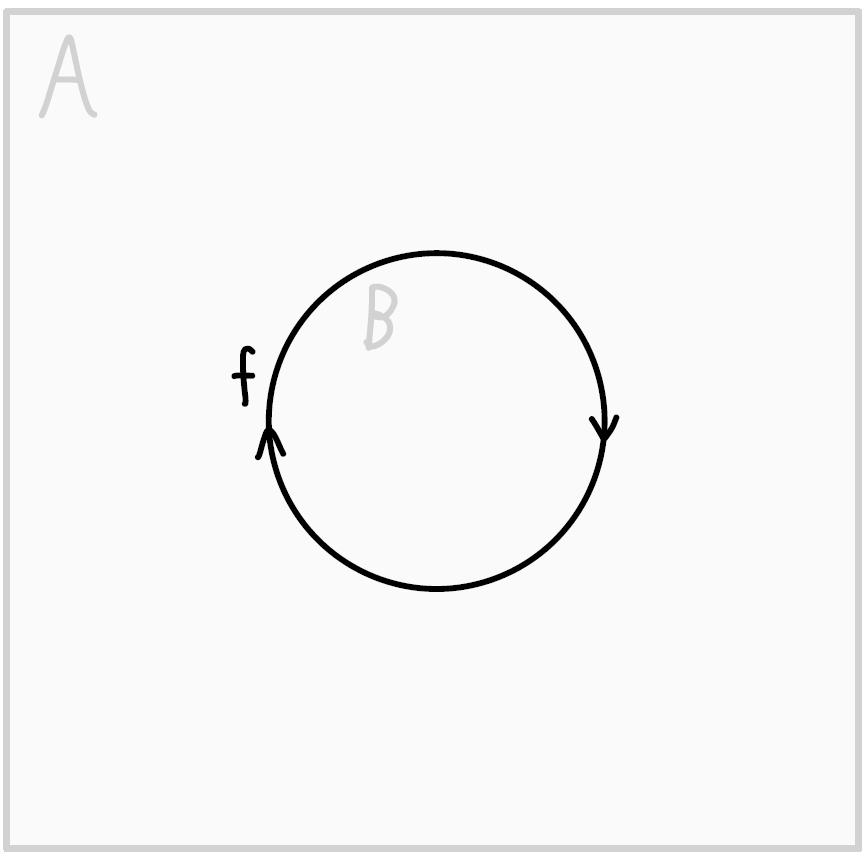}}}}}
\newcommand{\rhoiota}{\mathord{\vcenter{\hbox{\includegraphics[height=20ex]{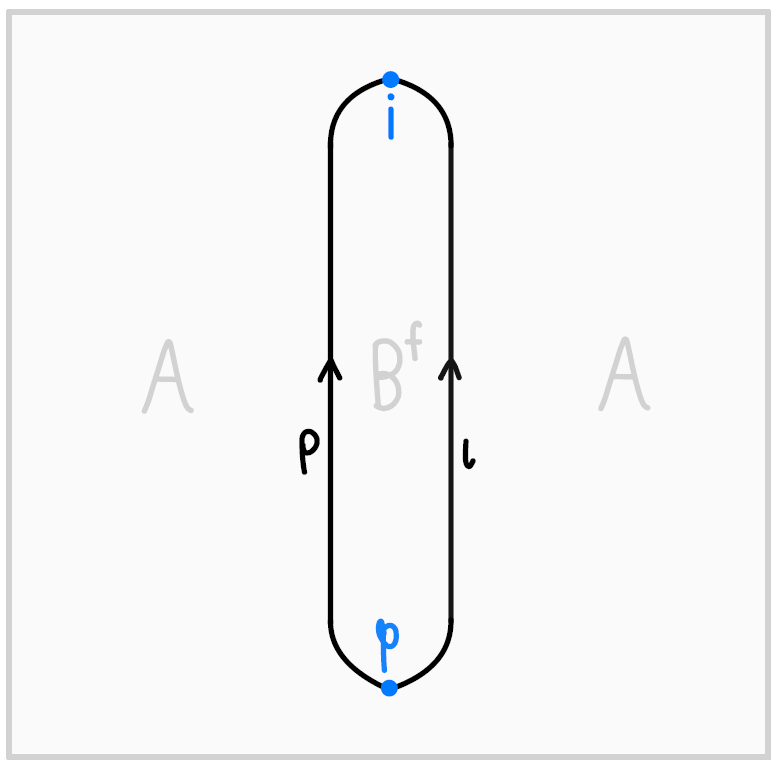}}}}}
\newcommand{\removablekirby}{\mathord{\vcenter{\hbox{\includegraphics[height=20ex]{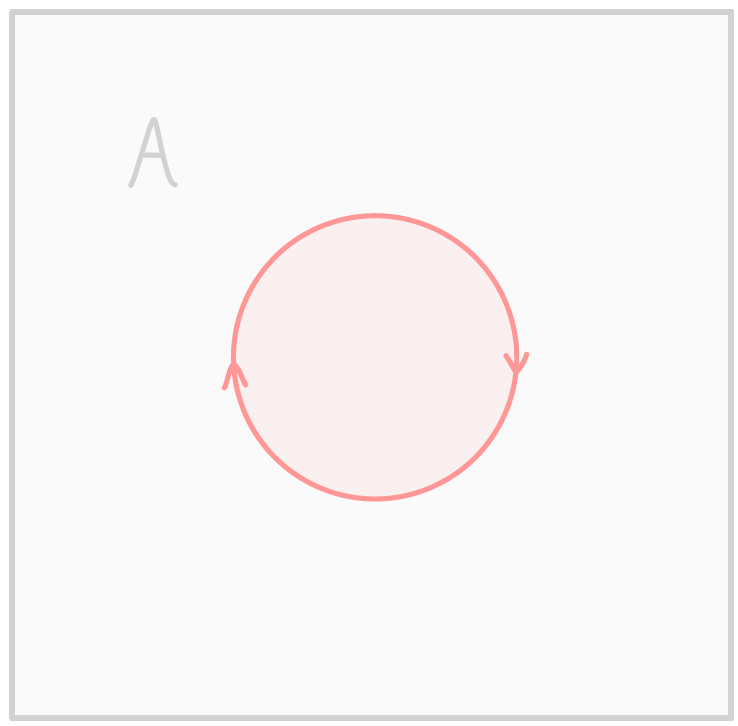}}}}}
\newcommand{\removablesphere}{\mathord{\vcenter{\hbox{\includegraphics[height=20ex]{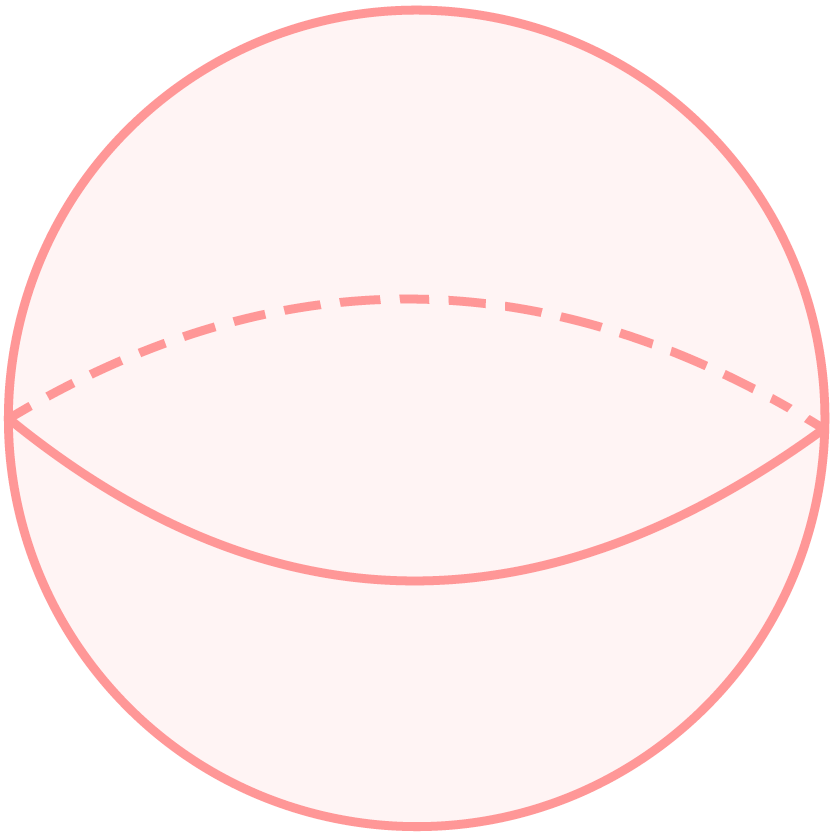}}}}}
\newcommand{\Asheet}{\mathord{\vcenter{\hbox{\includegraphics[height=20ex]{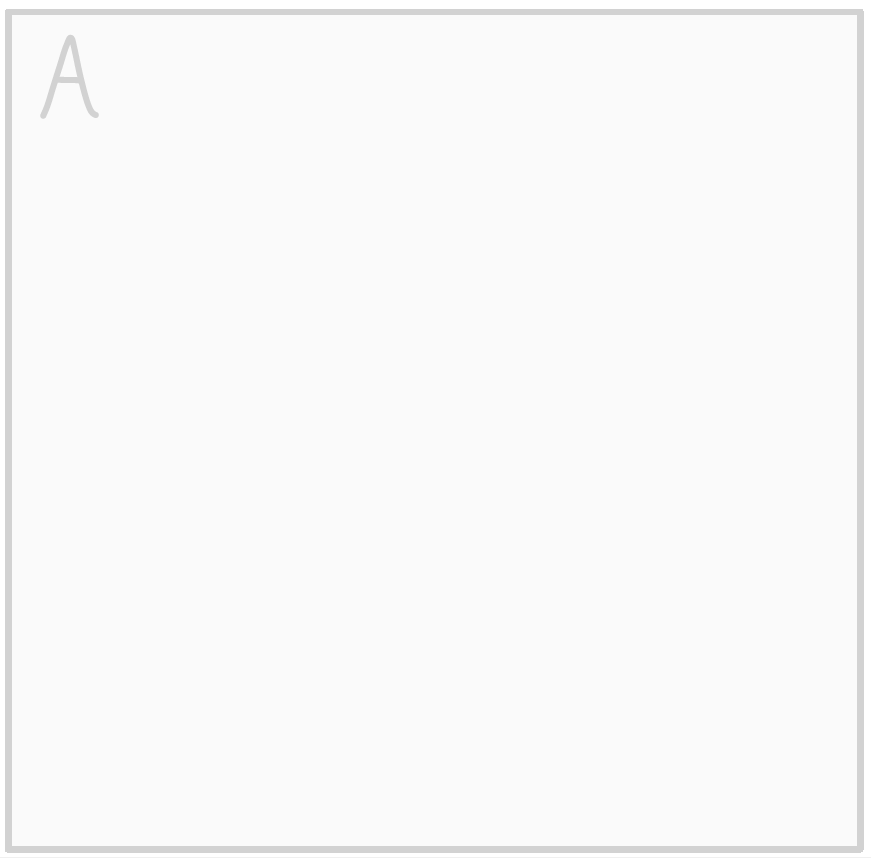}}}}}
\newcommand{\twoloopsvariant}{\mathord{\vcenter{\hbox{\includegraphics[height=20ex]{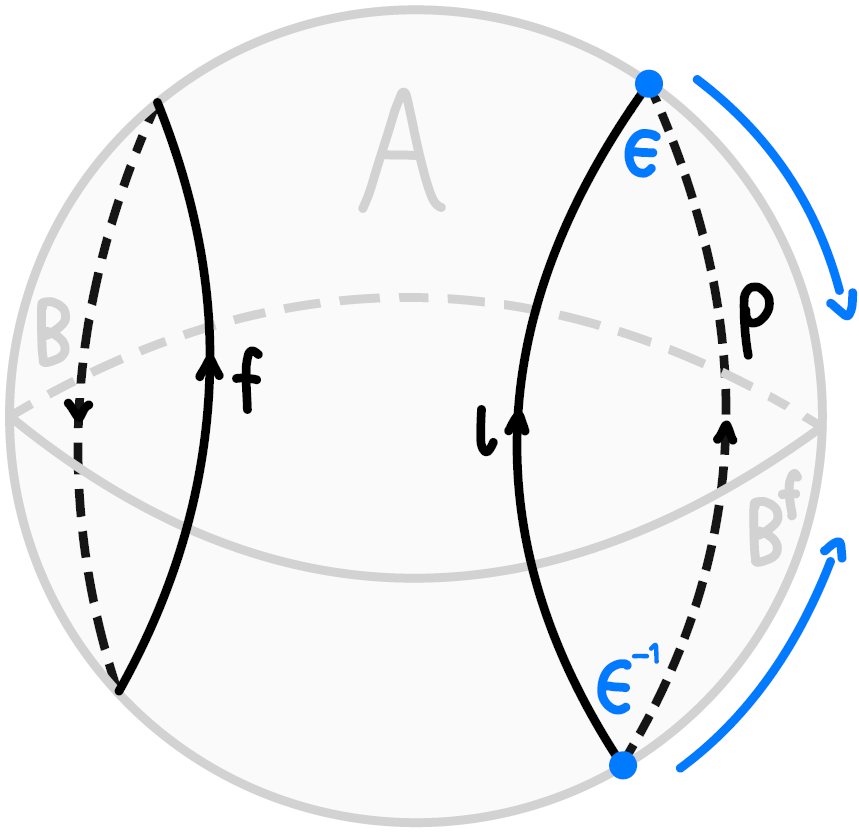}}}}}
\newcommand{\afterslitting}{\mathord{\vcenter{\hbox{\includegraphics[height=30ex]{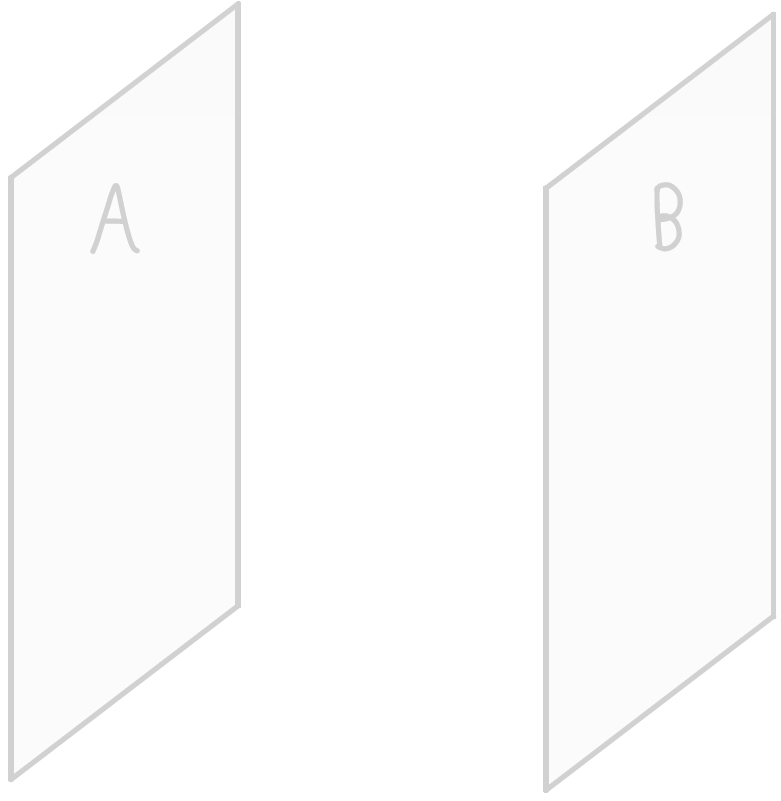}}}}}
\newcommand{\dimfvariant}{\mathord{\vcenter{\hbox{\includegraphics[height=20ex]{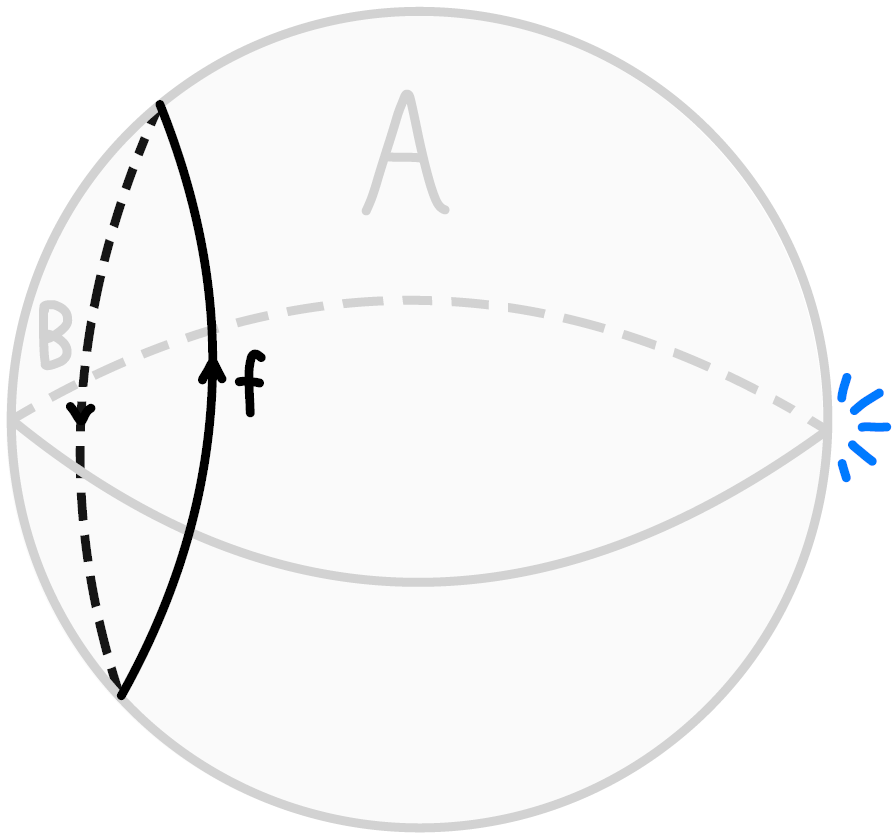}}}}}

\newcommand{\doubleloop}{\mathord{\vcenter{\hbox{\includegraphics[height=20ex]{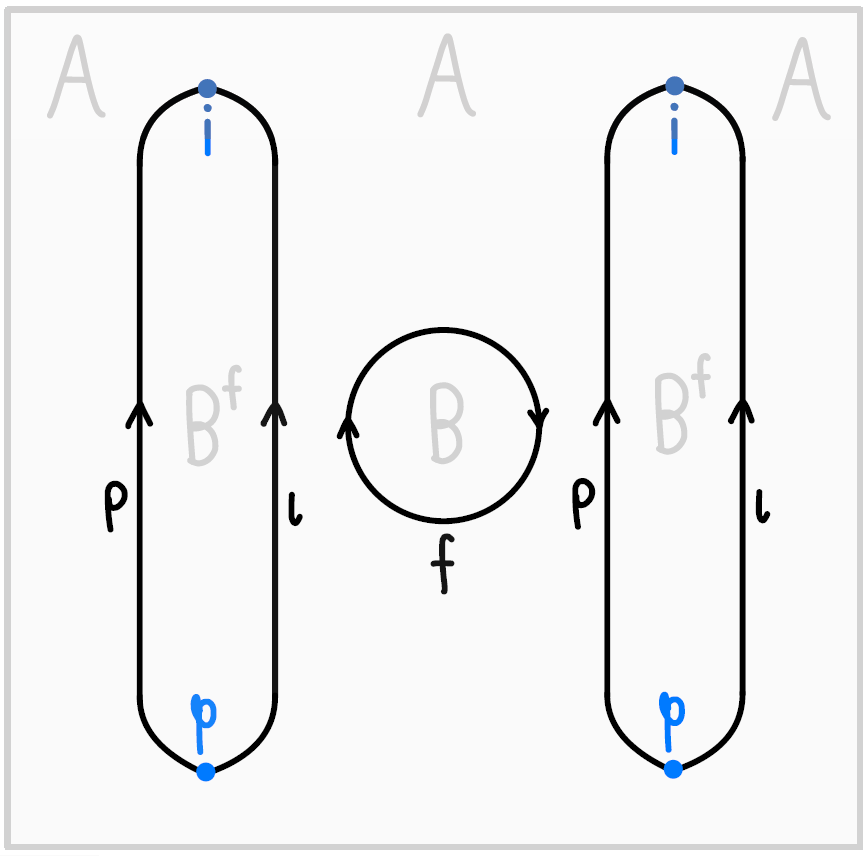}}}}}
\newcommand{\kirbyleftcurve}{\mathord{\vcenter{\hbox{\includegraphics[height=20ex]{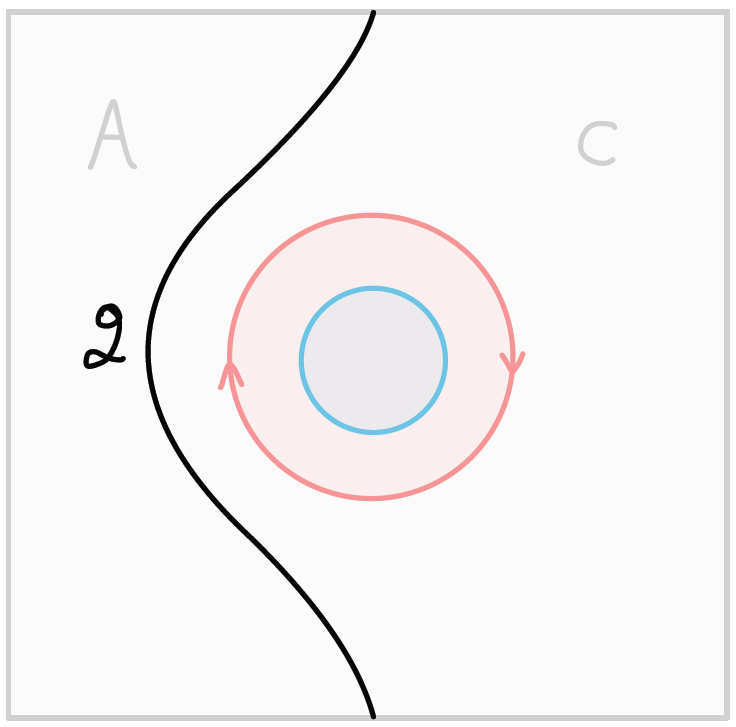}}}}}
\newcommand{\kirbyrightcurve}{\mathord{\vcenter{\hbox{\includegraphics[height=20ex]{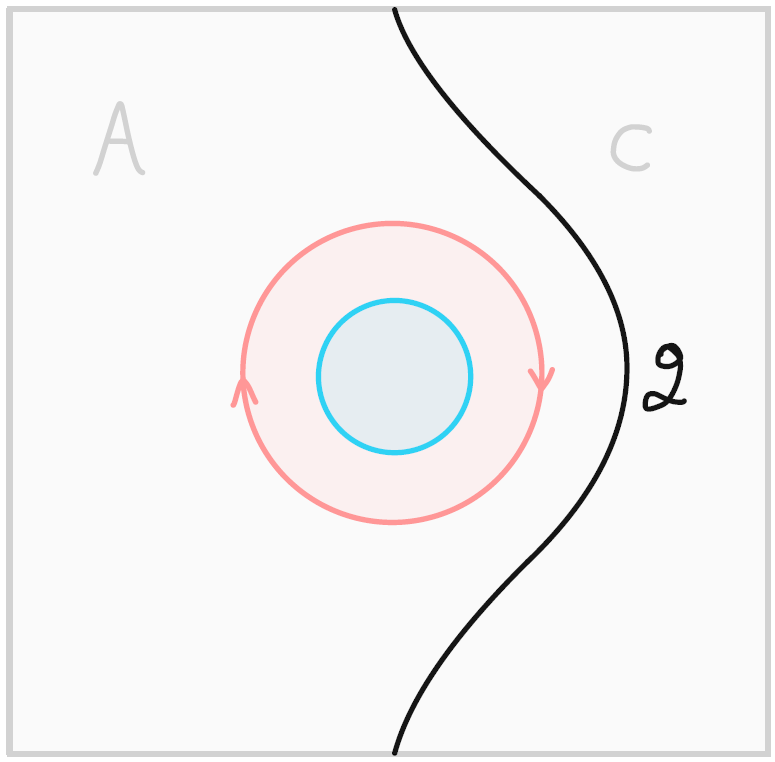}}}}}
\newcommand{\leftcollapse}{\mathord{\vcenter{\hbox{\includegraphics[height=20ex]{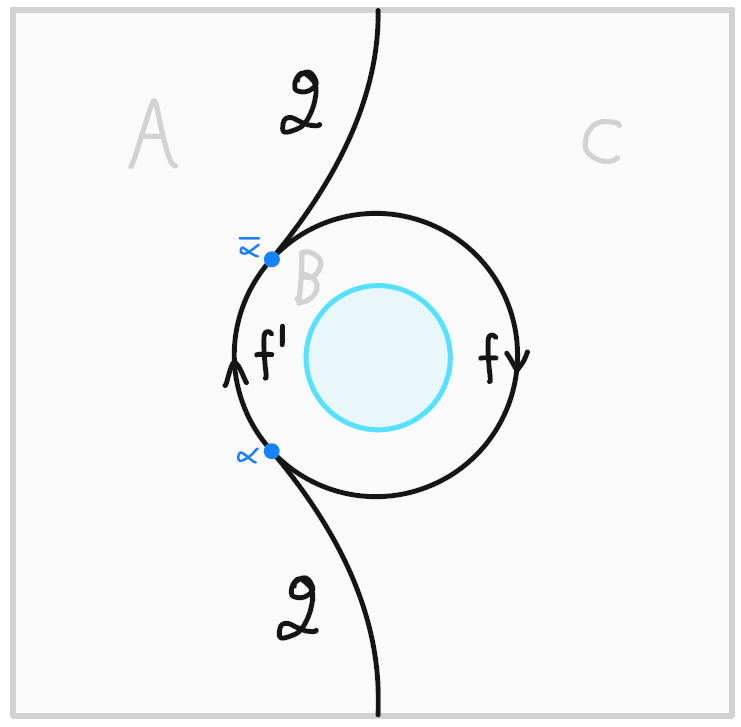}}}}}
\newcommand{\rightcollapse}{\mathord{\vcenter{\hbox{\includegraphics[height=20ex]{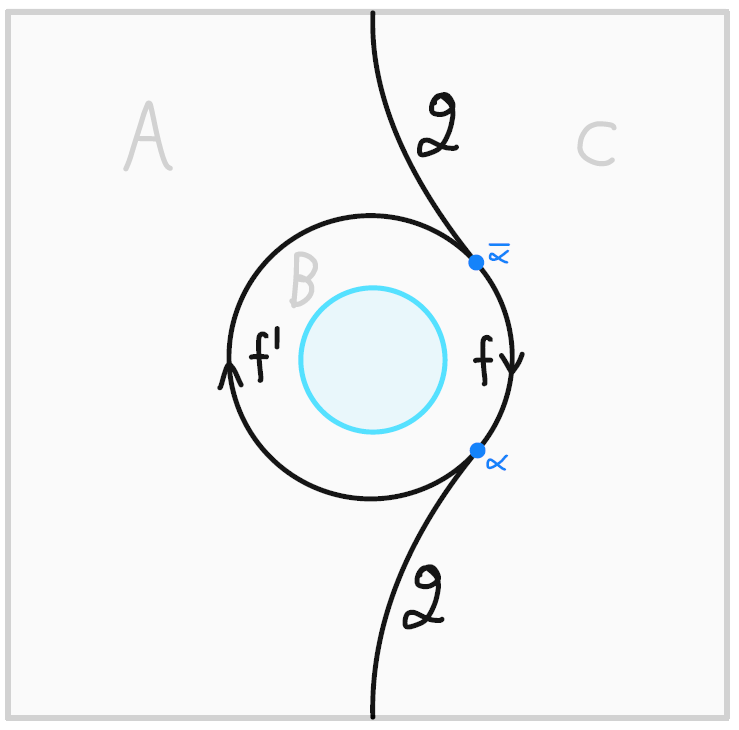}}}}}
\newcommand{\leftmerge}{\mathord{\vcenter{\hbox{\includegraphics[height=30ex]{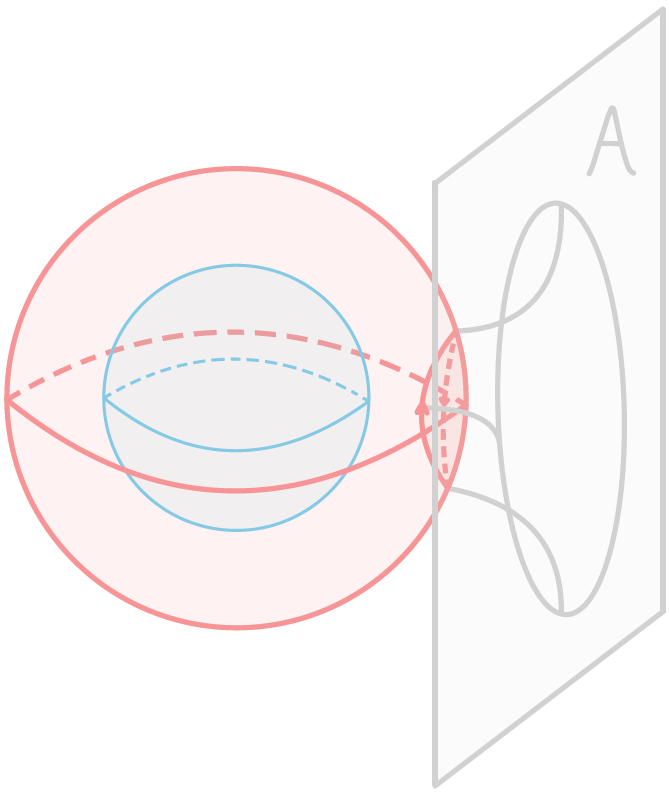}}}}}
\newcommand{\rightmerge}{\mathord{\vcenter{\hbox{\includegraphics[height=30ex]{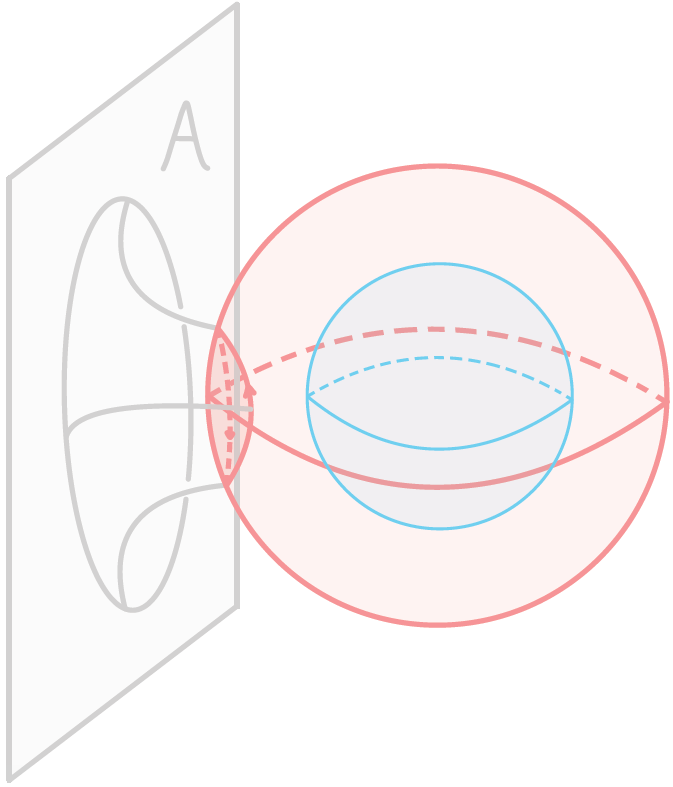}}}}}
\newcommand{\leftsphere}{\mathord{\vcenter{\hbox{\includegraphics[height=30ex]{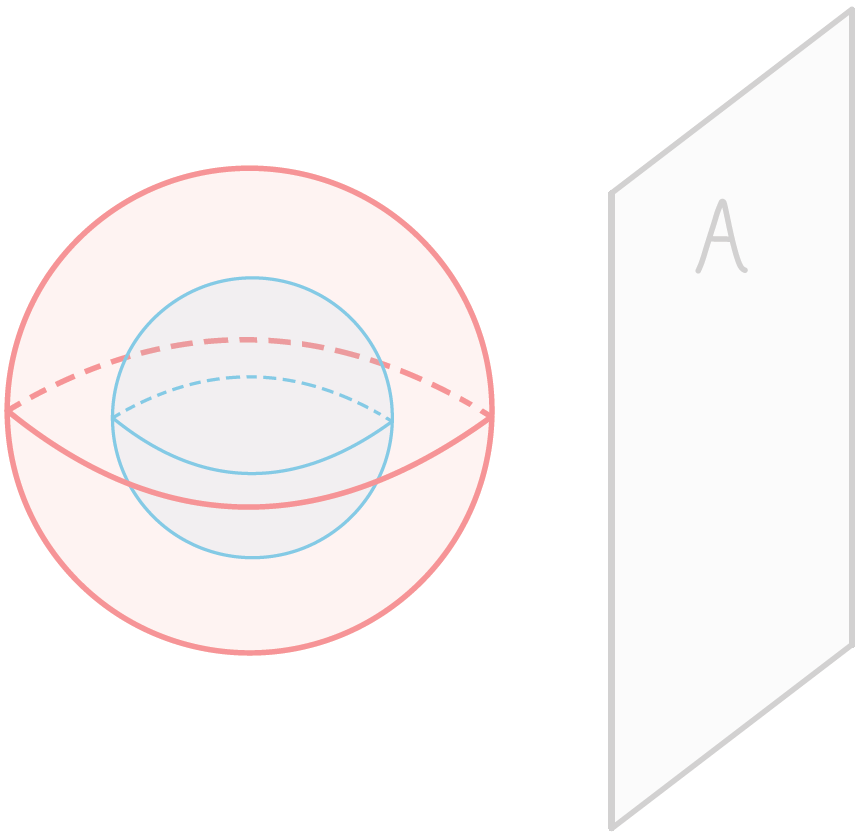}}}}}
\newcommand{\rightsphere}{\mathord{\vcenter{\hbox{\includegraphics[height=30ex]{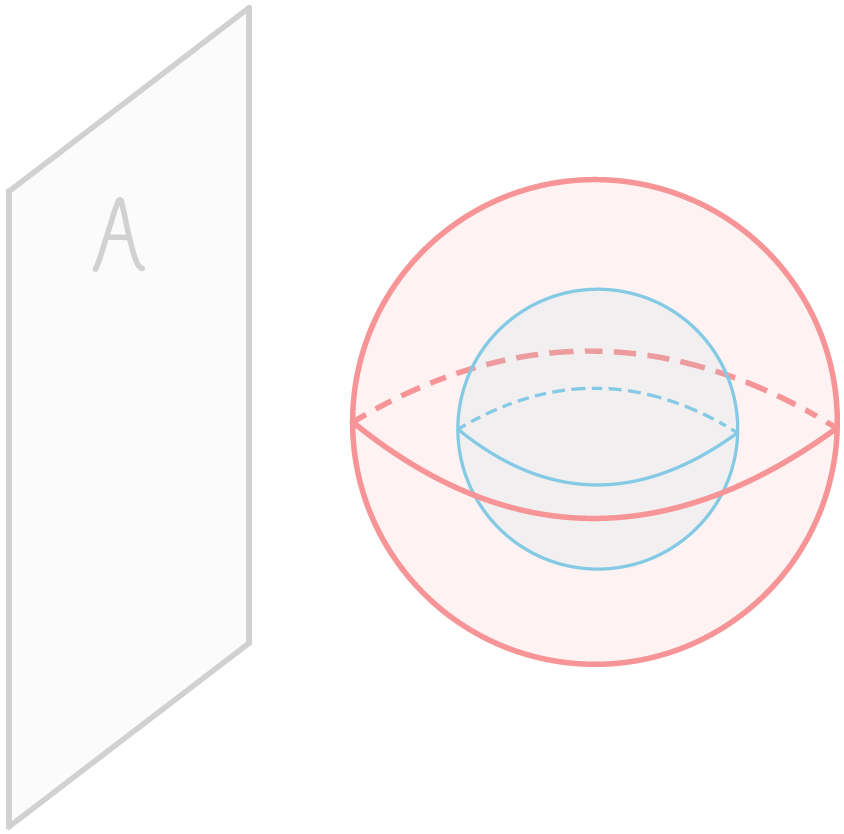}}}}}
\newcommand{\nestedloops}{\mathord{\vcenter{\hbox{\includegraphics[height=15ex]{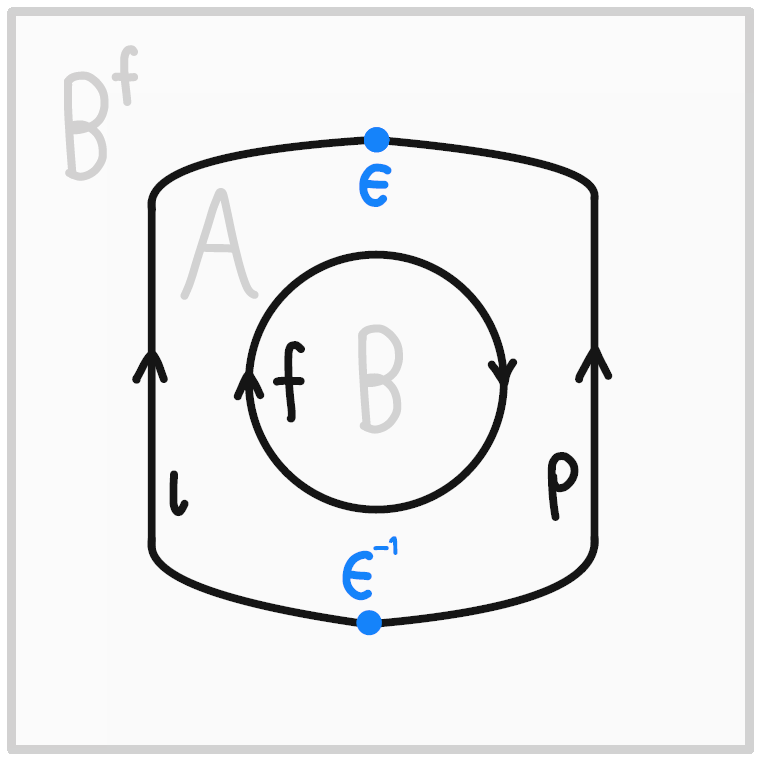}}}}}
\newcommand{\poppingtwoendomorphism}{\mathord{\vcenter{\hbox{\includegraphics[height=20ex]{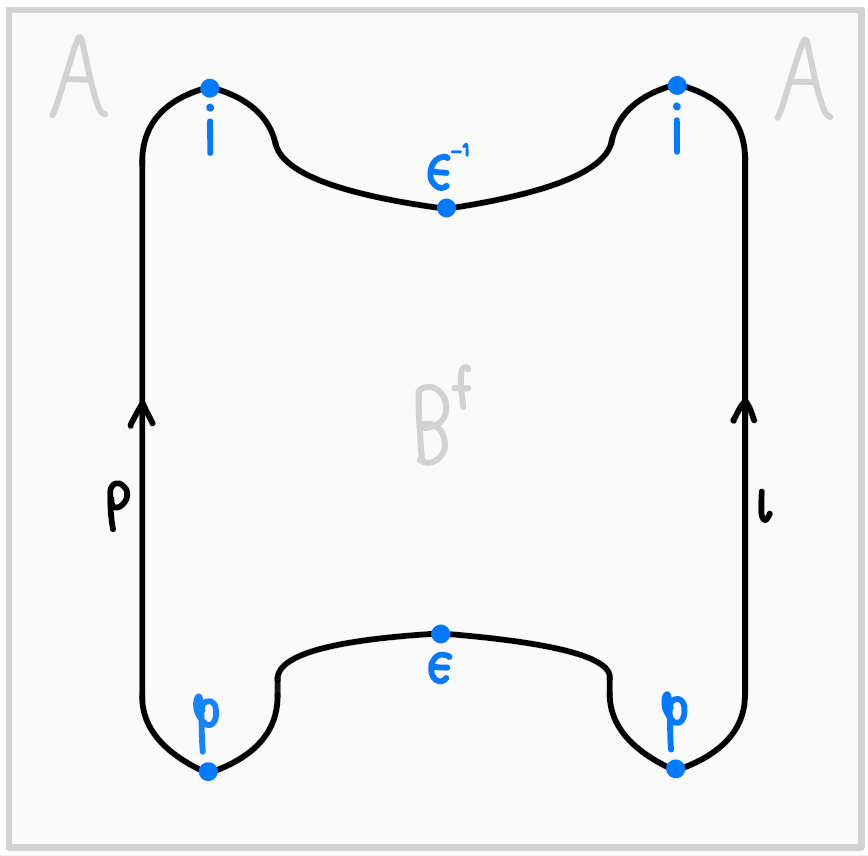}}}}}
\newcommand{\saddleiso}{\mathord{\vcenter{\hbox{\includegraphics[height=20ex]{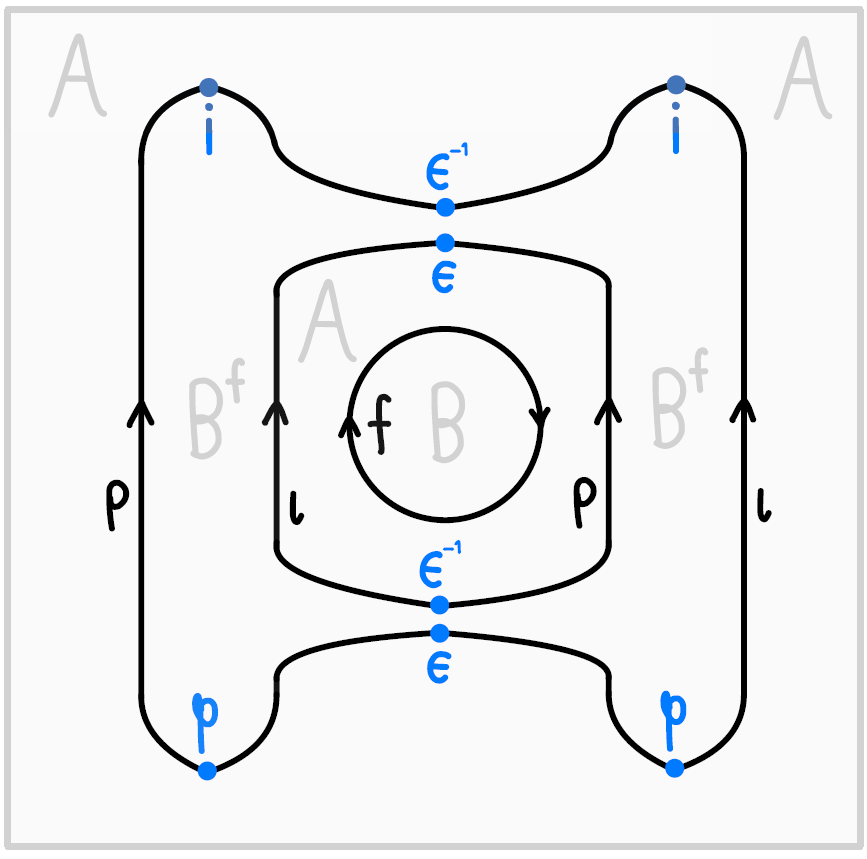}}}}}
\newcommand{\saddleisotwo}{\mathord{\vcenter{\hbox{\includegraphics[height=20ex]{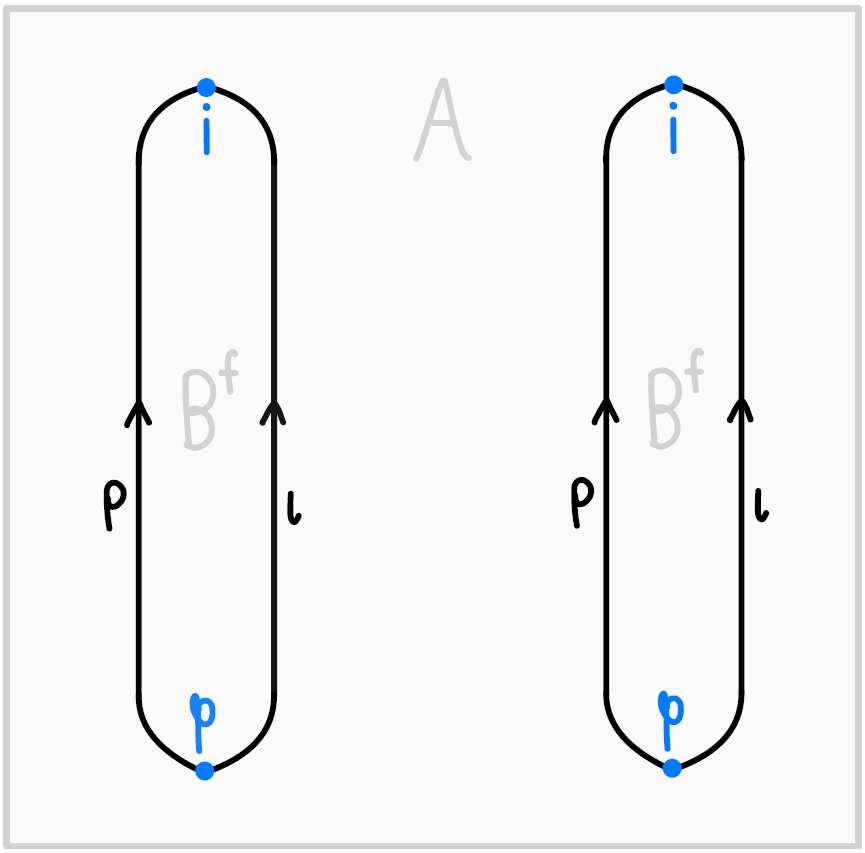}}}}}
\newcommand{\summedleftcurve}{\mathord{\vcenter{\hbox{\includegraphics[height=20ex]{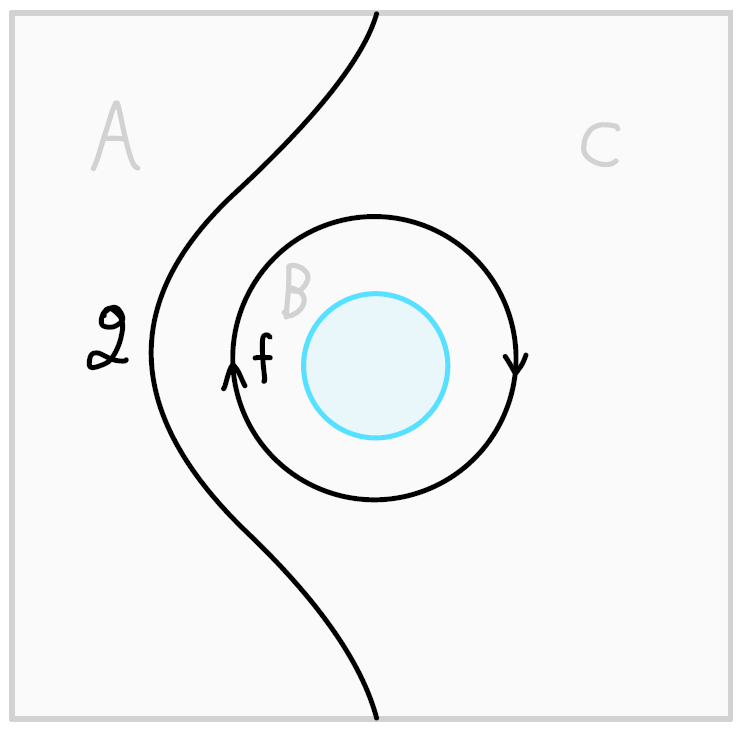}}}}}
\newcommand{\summedrightcurve}{\mathord{\vcenter{\hbox{\includegraphics[height=20ex]{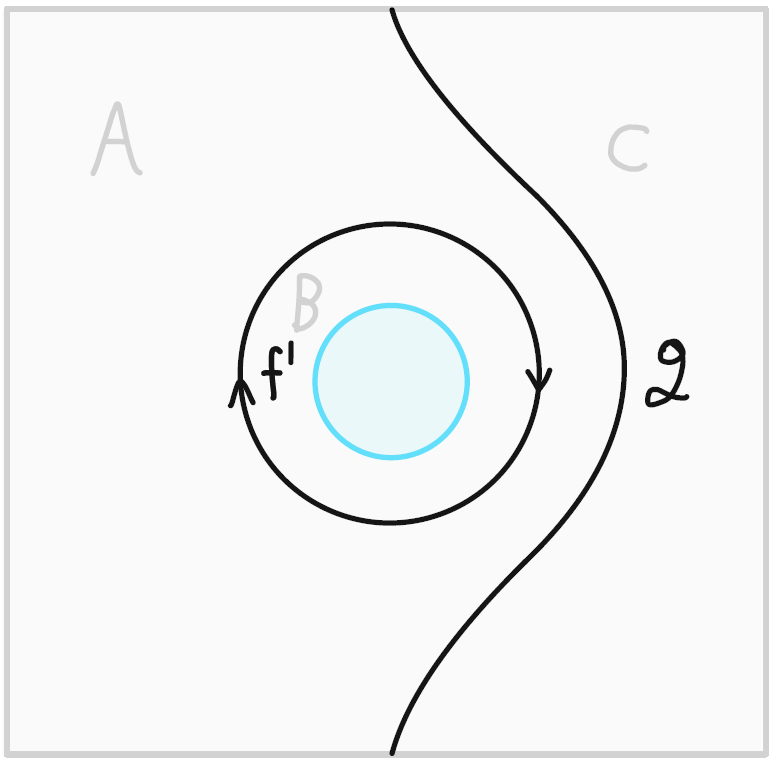}}}}}
\newcommand{\throat}{\mathord{\vcenter{\hbox{\includegraphics[height=30ex]{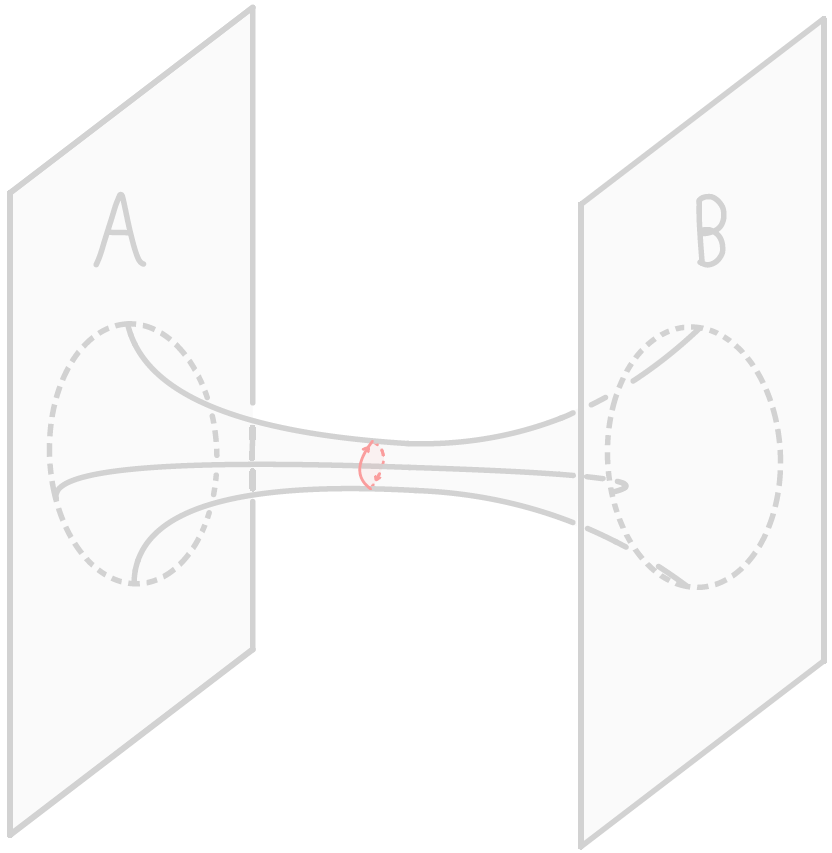}}}}}

\newcommand{\bj}{\mathord{\vcenter{\hbox{\includegraphics[height=20ex]{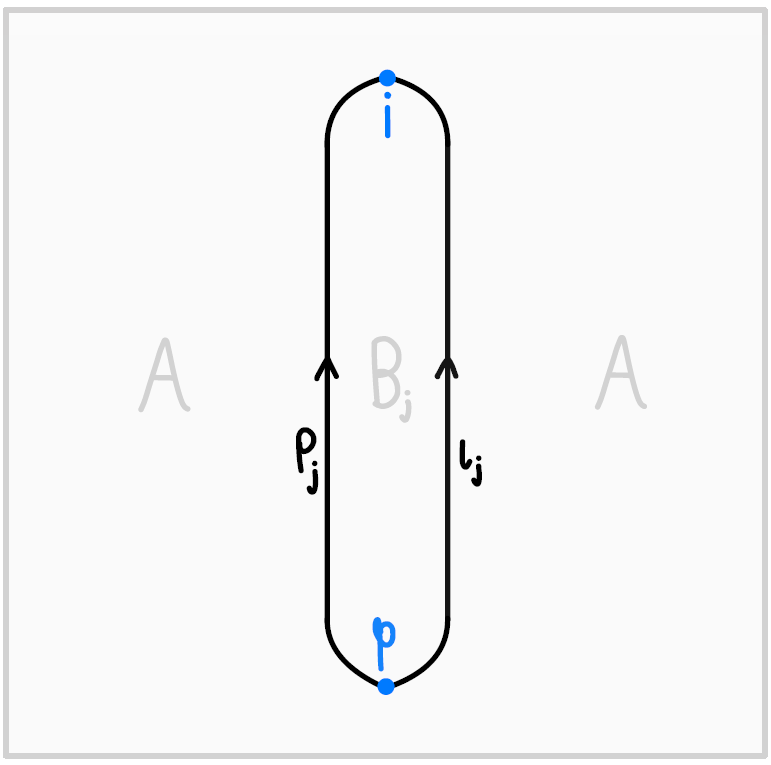}}}}}

\newcommand{\smallbj}{\mathord{\vcenter{\hbox{\includegraphics[height=15ex]{figures/symbols/Bj.png}}}}}
\newcommand{\smallremovablekirby}{\mathord{\vcenter{\hbox{\includegraphics[height=15ex]{figures/symbols/removable_kirby.png}}}}}
\newcommand{\smallAsheet}{\mathord{\vcenter{\hbox{\includegraphics[height=15ex]{figures/symbols/A-sheet.png}}}}}
\newcommand{\ehwazvariation}{\mathord{\vcenter{\hbox{\includegraphics[height=25ex]{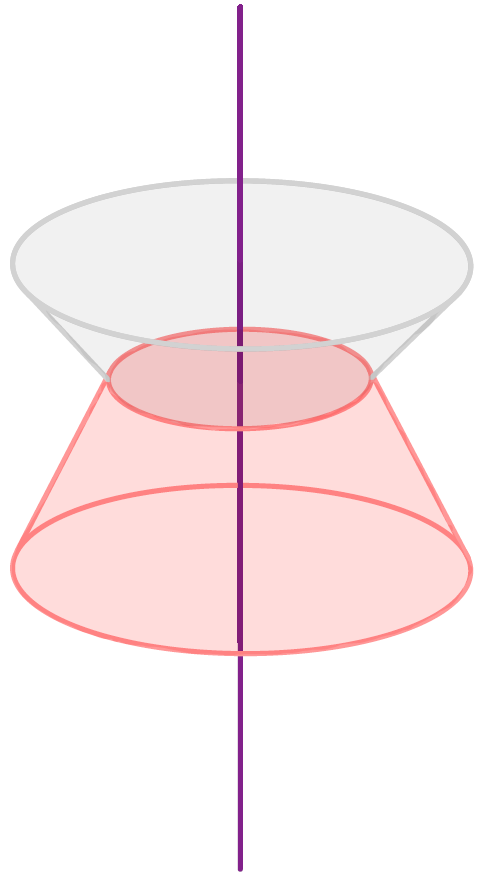}}}}}
\newcommand{\postehwazvariation}{\mathord{\vcenter{\hbox{\includegraphics[height=15ex]{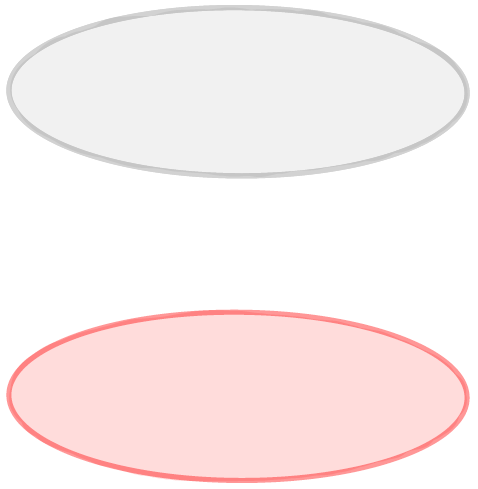}}}}}

\newcommand{\composetriangle}{\mathord{\vcenter{\hbox{\includegraphics[height=30ex]{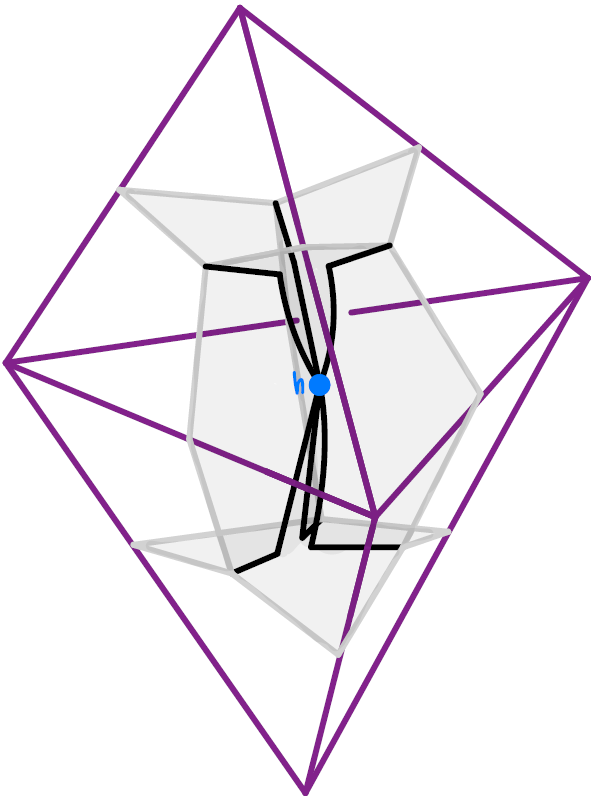}}}}}
\newcommand{\splittingtriangle}{\mathord{\vcenter{\hbox{\includegraphics[height=30ex]{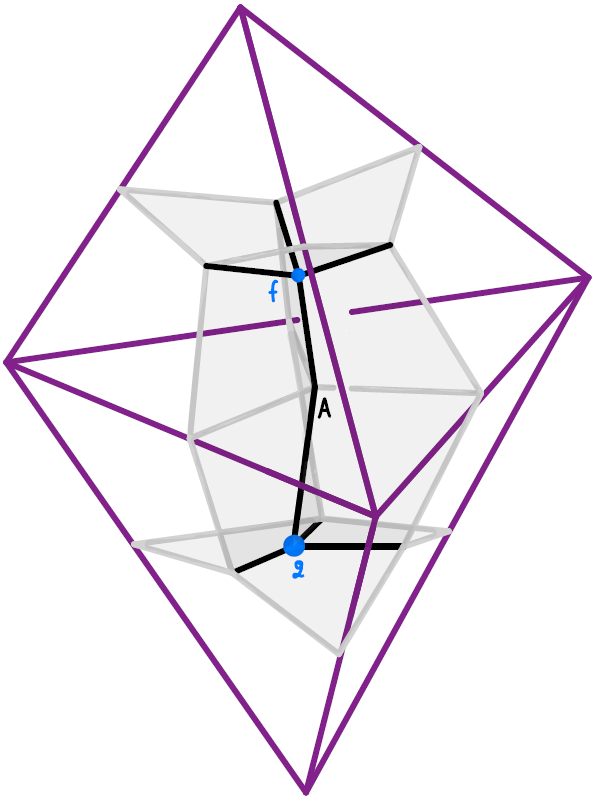}}}}}
\newcommand{\mempty}{\mathord{\vcenter{\hbox{\includegraphics[height=25ex]{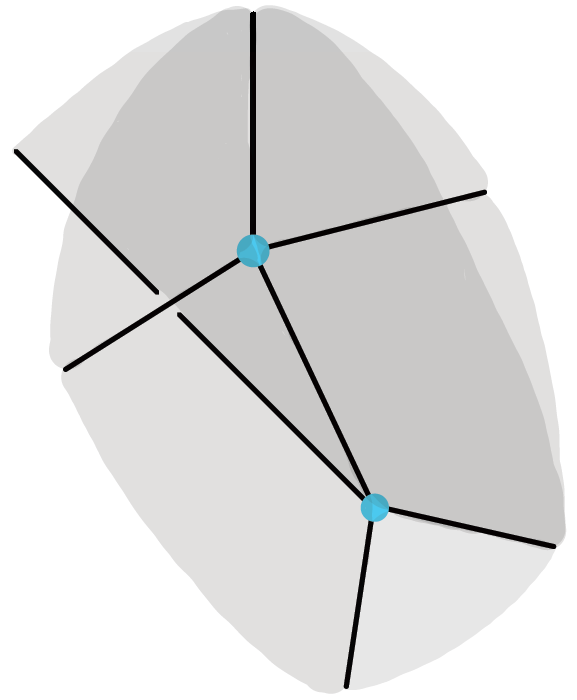}}}}}
\newcommand{\memball}{\mathord{\vcenter{\hbox{\includegraphics[height=25ex]{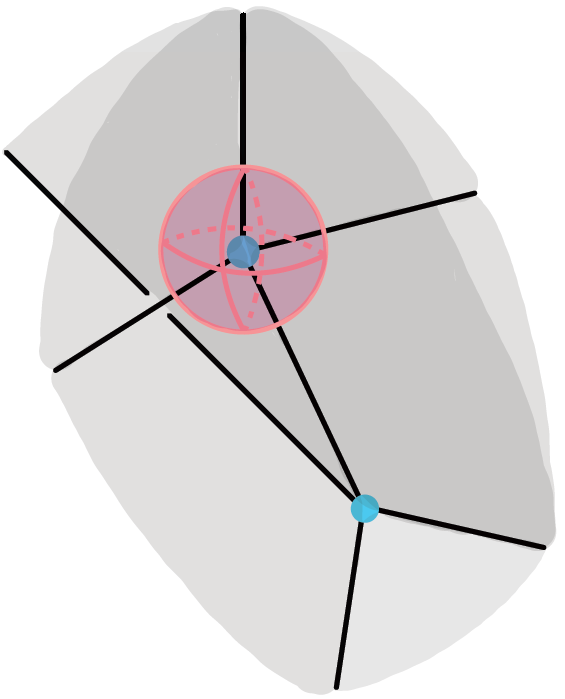}}}}}
\newcommand{\memballone}{\mathord{\vcenter{\hbox{\includegraphics[height=25ex]{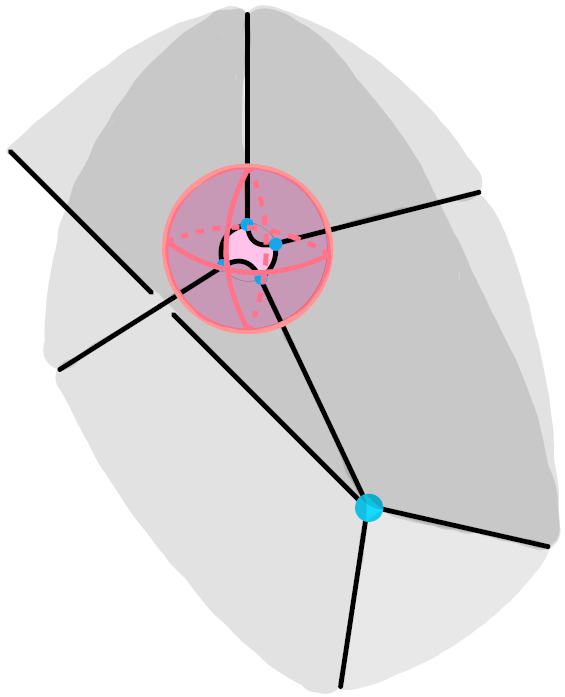}}}}}
\newcommand{\memballtwo}{\mathord{\vcenter{\hbox{\includegraphics[height=25ex]{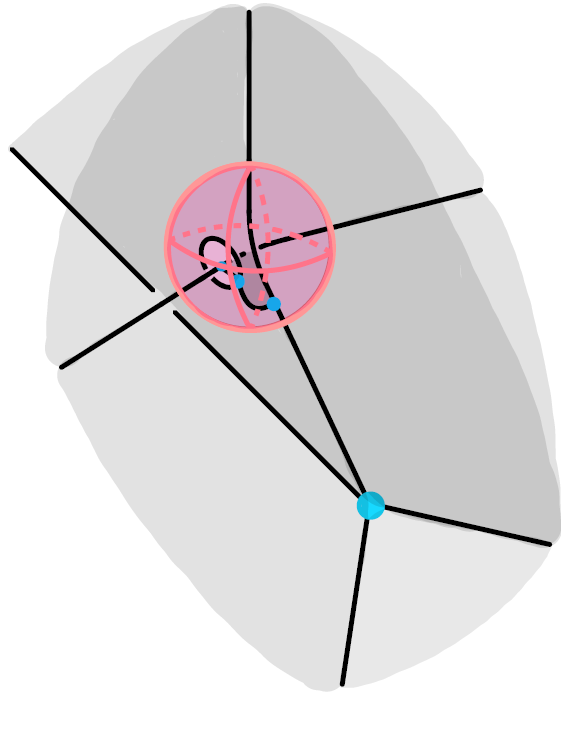}}}}}
\newcommand{\ballone}{\mathord{\vcenter{\hbox{\includegraphics[height=10ex]{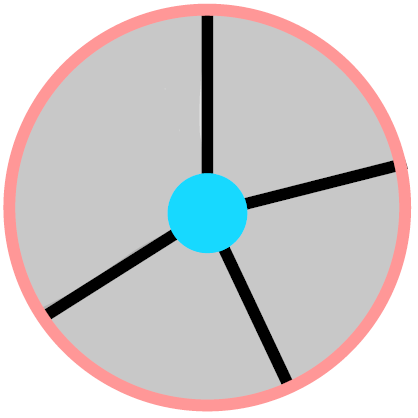}}}}}
\newcommand{\balltwo}{\mathord{\vcenter{\hbox{\includegraphics[height=10ex]{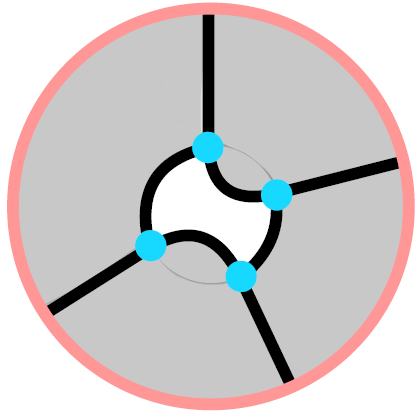}}}}}
\newcommand{\ballthree}{\mathord{\vcenter{\hbox{\includegraphics[height=10ex]{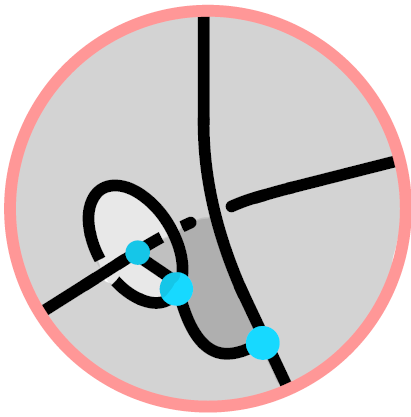}}}}}

\newcommand{\localkirbyleftcurve}{\mathord{\vcenter{\hbox{\includegraphics[height=20ex]{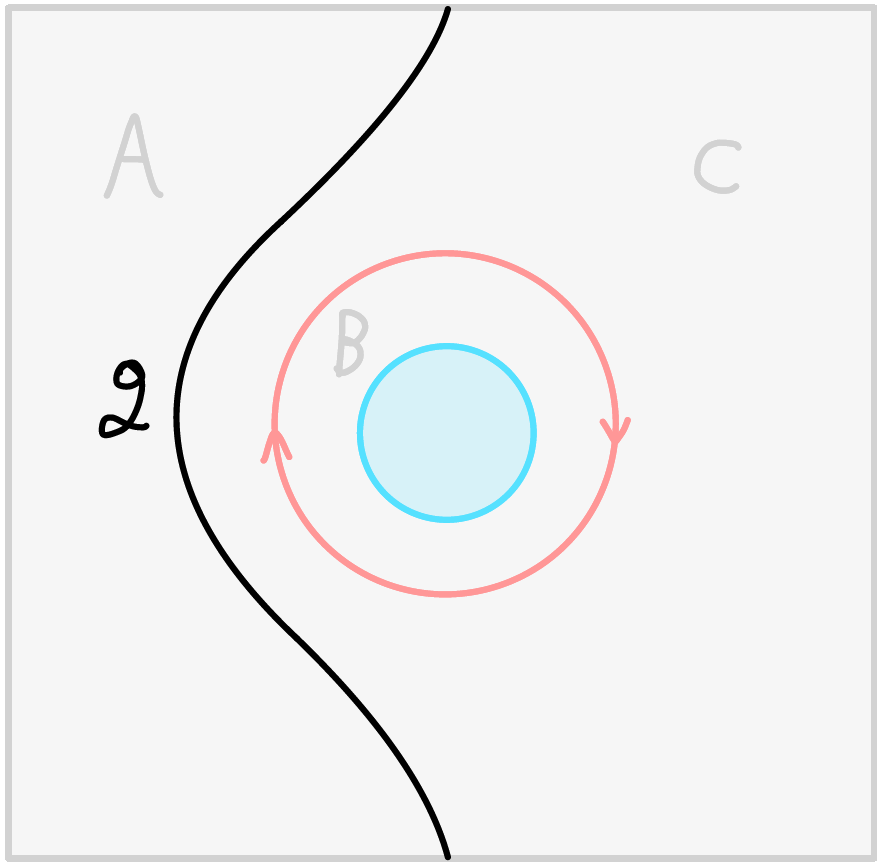}}}}}
\newcommand{\localkirbyrightcurve}{\mathord{\vcenter{\hbox{\includegraphics[height=20ex]{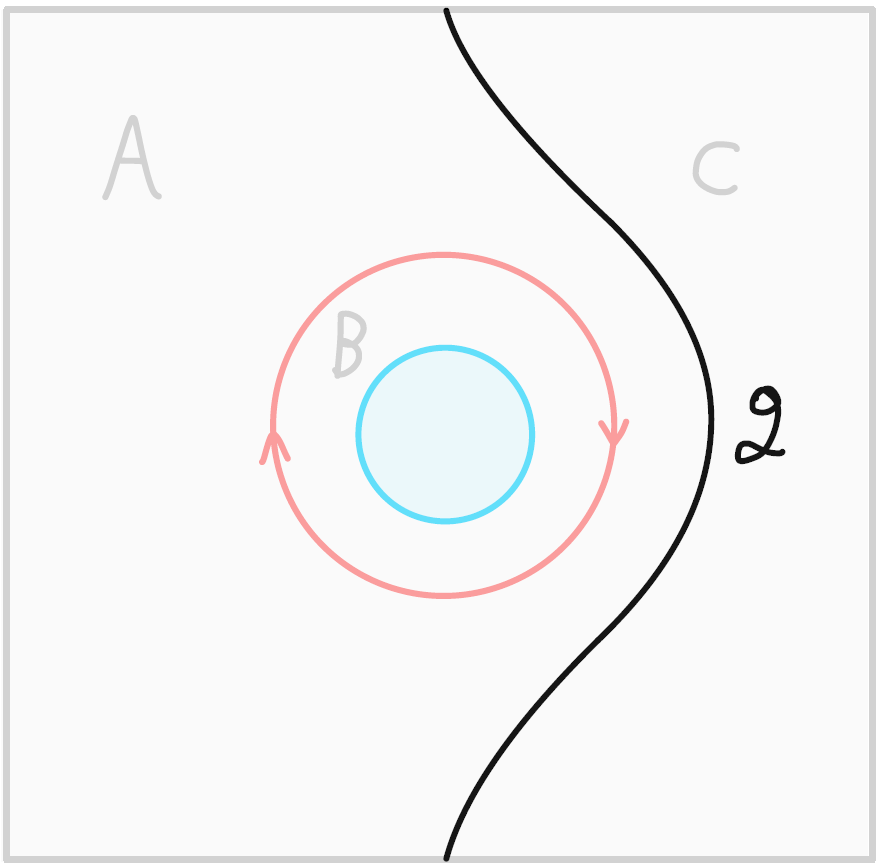}}}}}
\newcommand{\AAsheets}{\mathord{\vcenter{\hbox{\includegraphics[height=20ex]{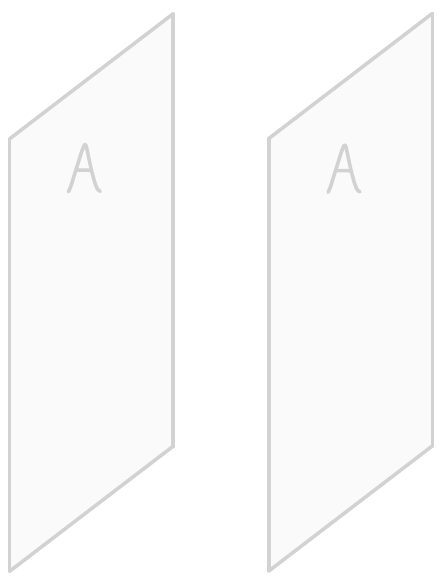}}}}}
\newcommand{\ABsheets}{\mathord{\vcenter{\hbox{\includegraphics[height=20ex]{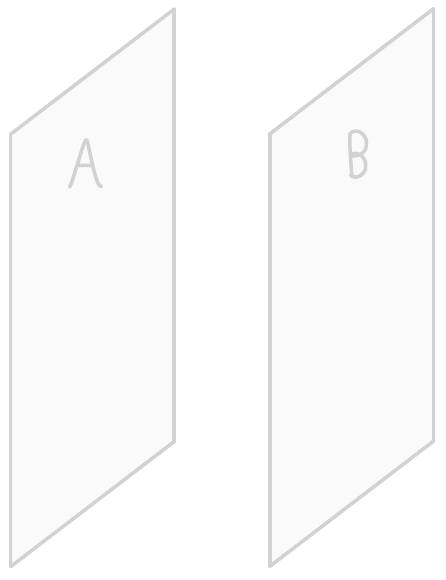}}}}}
\newcommand{\KirKirsheets}{\mathord{\vcenter{\hbox{\includegraphics[height=20ex]{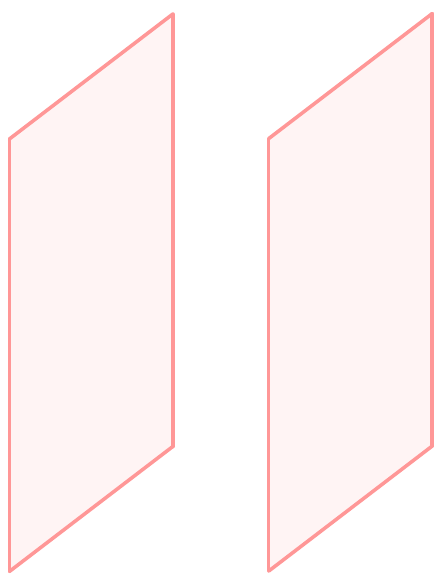}}}}}
\newcommand{\memepsilon}{\mathord{\vcenter{\hbox{\includegraphics[height=13ex]{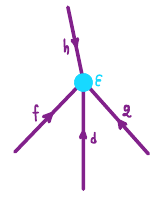}}}}}
\newcommand{\epsilondelta}{\mathord{\vcenter{\hbox{\includegraphics[height=15ex]{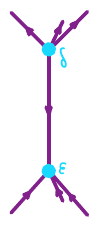}}}}}
\newcommand{\memdelta}{\mathord{\vcenter{\hbox{\includegraphics[height=10ex]{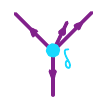}}}}}
\newcommand{\noepsilon}{\mathord{\vcenter{\hbox{\includegraphics[height=15ex]{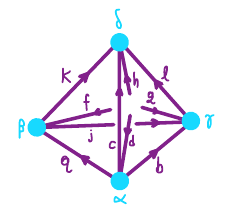}}}}}
\newcommand{\noepsilondelta}{\mathord{\vcenter{\hbox{\includegraphics[height=13ex]{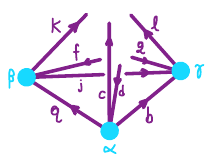}}}}}
\newcommand{\nodelta}{\mathord{\vcenter{\hbox{\includegraphics[height=13ex]{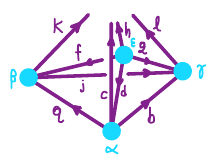}}}}}
\newcommand{\merge}{\mathord{\vcenter{\hbox{\includegraphics[height=25ex]{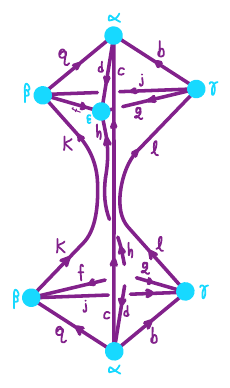}}}}}
\newcommand{\gluebeta}{\mathord{\vcenter{\hbox{\includegraphics[height=25ex]{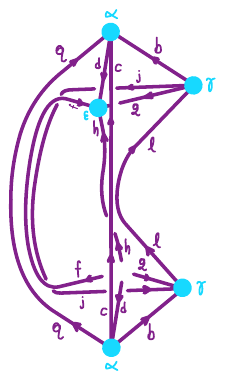}}}}}
\newcommand{\gluegamma}{\mathord{\vcenter{\hbox{\includegraphics[height=25ex]{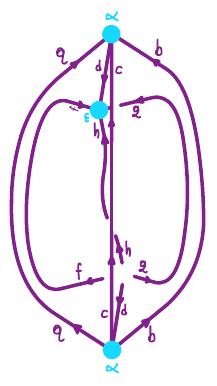}}}}}
\newcommand{\gluealpha}{\mathord{\vcenter{\hbox{\includegraphics[height=25ex]{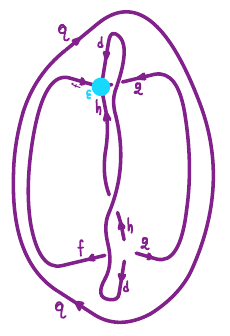}}}}}
\newcommand{\glueI}{\mathord{\vcenter{\hbox{\includegraphics[height=25ex]{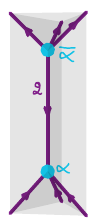}}}}}
\newcommand{\glueII}{\mathord{\vcenter{\hbox{\includegraphics[height=25ex]{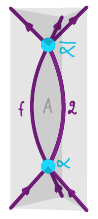}}}}}
\newcommand{\glueres}{\mathord{\vcenter{\hbox{\includegraphics[height=25ex]{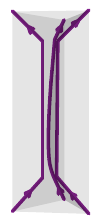}}}}}

\title{4D Topological Quantum Field Theory, part II: TQFTs from (Fusion) 2-Categories, their Membrane-Net Interpretation, and Higher Kirby Colors}

\author[a]{Lily Reisch}
\affiliation[a]{School of Physics and Astronomy, Tel Aviv University, Israel}
\emailAdd{lilyrsch@gmail.com}

\abstract{

\emph{This paper is the second in a six part series of independent papers.}

We construct four-dimensional topological quantum field theories from spherical fusion 2-categories, extending the Douglas–Reutter state-sum invariants of closed four-manifolds to state spaces and cobordism maps satisfying Atiyah’s TQFT axioms. We introduce a higher-dimensional generalization of the Levin–Wen (or Kirillov) string-net construction in terms of membrane-nets: decorated foams in three-manifolds, considered up to isotopy and local relations determined by the fusion 2-category in a skein-theoretic fashion. We prove that the membrane-net spaces are canonically isomorphic to the physical state spaces obtained from the lattice state-sum construction, providing a triangulation-independent description of the theory.
A key ingredient in the proof is the introduction of two higher-categorical analogues of Kirby colors: a global Kirby color associated with objects of the fusion 2-category and local Kirby colors associated with its 1-morphisms. We establish generalized "handle-slide" and "handle-removal" identities obeyed by the categorified Kirby colors and use them to express the lattice cylinder projector diagrammatically, thereby proving the equivalence of the lattice and membrane-net descriptions.
Our construction extends the correspondence between lattice topological models and continuum string-net spaces from three to four spacetime dimensions and provides new graphical tools for the study of higher-categorical TQFTs.
}

\keywords{Topological Field Theories, Topological States of Matter, Anyons, Higher Categories, Tensor networks, Skein theory}

\begin{document}
\maketitle

\section{Introduction}

This paper is the second part of a six part series of independent papers that together tell a cohesive story. The six papers can be read independently, and we provide in \ref{roadmap} a summary of the general motivation, roadmap and goals. As for the introduction relevant to this specific paper, see \ref{intro to 2nd paper}.

\subsection{The paper series and roadmap}\label{roadmap}

\subsubsection{General motivation}
In the last few decades, topological quantum field theories (TQFTs) rose to prominence as an important class of theories arising in many physical and mathematical contexts, becoming an active area of studies. On the physical side, a TQFT is any kind of QFT that can be defined on a spacetime manifold without an a priori choice of geometry, a property sometimes referred to as "general covariance", "background independence" or "diffeomorphism invariance". In particular, any QFT reducing to general relativity in the classical limit ought to have this property, hence QFT approaches to the problem of quantizing gravity should fit into the larger framework of topological quantum field theory\footnote{Note that one might have to consider "partially defined" TQFTs, meaning theories whose path integral is well defined only on certain classes of manifold.}. Other applications include the theories describing topological order in condensed matter physics, the SymTFT paradigm in which the generalized symmetries of any QFT are captured by the physics of a TQFT one dimension higher, and topological quantum error correction which encodes quantum information in the vacuum sector of lattice models for which that sector is a TQFT. Thus, the study of TQFT can be seen either as a study of quantum gravity, the study of generalized symmetries, the study of topological order or the study of topological quantum error correction, all of which fit into a single physical framework.

On the mathematical side, TQFTs can be seen as a nexus where low dimensional topology, homotopy theory, higher category theory and many branches of algebra meet. Concretely, a (fully extended/local) $n$-dimensional TQFT can be defined as a collection of invariants of manifolds in all dimensions up to $n$, beginning with numerical invariants of $n$-manifolds (physically coming from path integrals) and going up the categorification ladder as  we descend in dimension, assigning vector spaces to $(n-1)$-manifolds, (linear) 1-categories to $(n-2)$-manifolds and so forth. These invariants interact in many ways: for instance, an $n$-manifold with boundary yields a distinct number for each element in the vector space assigned to its boundary, an $(n-1)$-manifold with boundary yields a categorified representation of its boundary category, and so forth. From all of this structure we can extract representations of mapping class groups, invariants of isotopy classes of embedded manifolds (such as knot invariants) and many other algebraic invariants of interest to topologists and algebraists alike. By performing kinds of categorified summations over shared boundaries we can compute the behavior of these invariants under gluing and surgery operations, so besides yielding more sophisticated invariants the many layers of structure in a TQFT makes these invariants much more computable.

From the point of view of homotopy theory and higher category theory, TQFTs are functors from the $n$-category (or $(\infty,n)$-category, depending on one's taste) of cobordisms into symmetric monoidal $n$-categories, with the target $n$-categories of physical interest being $n$-fold categorifications of the category of vector spaces, hence TQFTs can be seen as "higher representations" of the cobordism category. The category of cobordisms has a remarkably simple description first conjectured in \cite{BD95} and sketched in detail in \cite{Lur09} as the freely generated symmetric monoidal category on a single fully dualizable object, a celebrated result known as the "cobordism hypothesis". Since the target $n$-category $\mathcal{D}$ is often of a concrete algebraic nature, this allows for the abstract classification of TQFTs in various dimensions by identifying the fully dualizable objects of $\mathcal{D}$\footnote{Depending on tangential structures, one may also need to consider homotopy fixed points of certain canonical group actions.}.

\subsubsection{Dimension 3}
Focusing on specific dimensions\footnote{Wherever we say "$d$ dimensions", we mean $d$ spacetime dimensions.}, TQFTs in dimension 1 are essentially trivial (classified by finite dimensional vector spaces). The classification of TQFTs in dimension 2 can be arrived at without much difficulty (using the classification of surfaces, Morse theory, or via other approaches) and they are classified by commutative frobenius algebras. This should not come as a surprise, as (compact) surfaces (possibly non-orientable and with boundary) have a simple topological classification, and the spatial ("1-0") part of these theories is 1 dimensional which allows only for disjoint unions of arcs and circles. On the other hand, in dimension 3 there is much to do, which can be traced back to the richness of 3D topology. Beginning with Witten's famous paper linking TQFT to invariants of 3-manifolds and knots, the study of 3D TQFTs blossomed into a fairly mature field, with the two main explicit constructions of TQFTs being the Reshetikhin-Turaev (RT) and Turaev-Viro-Barrett-Westbury (TVBW) models. These two very general constructions take as input an algebraic structure in the form of a 1-category (modular categories in the former case and spherical fusion categories in the latter) and explicitly reconstruct all of the physical data of a TQFT (state spaces, time evolutions, boundary conditions, observables, etc) from the algebraic data.

It is known that the RT and TVBW constructions describe "all 3D TQFTs", depending on one's precise assumptions. If one assumes a once-extended theory (a certain mild locality condition required by the physics), then \cite{BDSPV15} shows that all such theories are recovered by the RT construction. If one demands a full extension (a stronger locality condition expected from a physical theory), then TQFTs are abstractly classified by fully dualizable objects of 3Vect, an appropriate tricategory of "3-vector spaces", and similarly fully local $n$-dimensional TQFTs should be classified by "$n$-vector spaces". Following \cite{BD95} (and motivated by \cite{Walker06}), $3$-vector spaces should be a well behaved class of linear 2-categories, for instance cocomplete ones. Full dualizability of linear $n$-categories structures is a highly constraining condition which should manifest as a kind of categorified semisimplicity, discussed extensively for linear 1-categories ("2-vector spaces") in \cite{BDSPV15}. In the latter case (of linear 1-categories) it is known that the bicategory of "2-vector spaces" agrees with another good candidate for $2\text{Vect}$, namely the Morita bicategory of algebras, bimodules and bimodule homomorphisms. Returning to the 3D case (which is a categorification of the above) it is known (see \cite{DR18}) that semisimple 2-categories reduce in a similar Morita fashion to (spherical) fusion 1-categories, which indeed form a Morita 3-category in which they are fully dualizable (see \cite{DSPS20}). While this is the main candidate for the target category appropriate for local 3-dimensional physics (or twice extended 4-dimensional physics), it has yet to be proved whether fully dualizable 2-categories must be semisimple in the sense of \cite{DR18} or perhaps satisfy a slightly weaker condition. Working with this Morita 3-category, it was shown in \cite{Walker06,MW12}'s framework (i.e. using disclike 3-categories) that the corresponding theories are indeed those described by the TVBW construction. In the traditional framework of tricategories it is not known yet, but a very close result is known - namely, if we take the endomorphism bicategory of a spherical fusion category in their Morita 3-category we obtain its Drinfeld center (which is a modular category by \cite{Mue03}), and the corresponding once-extended theory described by the RT construction indeed coincides with the TVBW construction (see \cite{BK10},\cite{Bal10a},\cite{Bal10b}).

The TVBW construction should then be an explicit construction of all fully local 3D theories, with the three main results being the abstract full dualizablity result of \cite{DSPS20} (which focuses on the evaluation at the point), the explicit state sum formula described in \cite{TV92},\cite{BW96} (which describes the evaluation on a 3-manifold) and the construction of an actual TQFT connecting the two by \cite{BK10} (which describes the core data at all dimensions up to 3 and their interactions, with less emphasis on the 0-dimensional part).

\subsubsection{Dimension 4}
As for dimension 4, much less is known and the field is quite young. Our goal in this series of papers is to begin closing some of that gap. Ideally, we would like to have both an abstract classification of TQFTs in terms of some concrete algebraic structure as well as an explicit construction of the corresponding physical theories. Recall that we have two good candidate definitions for the target 4-category 4Vect of a physical TQFT: the 4-category of cocomplete linear 3-categories, and the Morita 4-category of fusion 2-categories, are expected to have equivalent fully dualizable subcategories (hence should describe the same topological theories). While the former notion and a kind of full dualizability condition on 3-categories are yet to be defined and studied, the latter has been studied by \cite{Dec24}, and it was indeed shown that every fusion 2-category is fully dualizable, with the spherical ones conjectured to yield oriented fully extended TQFTs whose top dimensional part (the path integral) reduces to the Douglas-Reutter state sum invariant on closed 4-manifolds.

In analogy with dimension 3, the Douglas-Reutter invariant is the analog of the TVBW invariant, and can be shown to be the universal\footnote{The universal state sum of \cite{Walker21} reduces to the Douglas-Reutter state sum only when the input 3-category has a single object, in which case its monoidal endomorphism 2-category is the input to the corresponding Douglas-Reutter state sum. However, in Walker's formalism one sums over equivalence classes of objects and objects are considered equivalent as soon as there is a nonzero morphism between them, hence on each connected component the object assigned to vertices/0-handles is uniform and one obtains the Douglas-Reutter state sum. Thus, the Douglas-Reutter state sum is universal up to a direct sum of TQFTs.} semisimple construction in dimension 4\footnote{Note that while many state-sum constructions are known in dimension 4 (\cite{CY93},\cite{CKY97},\cite{Mac99},\cite{Cui19},\cite{BB18},...), as described in \cite{DR18} they are all special cases of the Douglas-Reutter invariant, except possibly the dichromatic state sum of \cite{BB18}. The latter was shown to be a special case of \cite{MMT25} in the case that the target of their pivotal functor is modular (and not just ribbon fusion), and we expect the results of papers III and IV in our paper series to allow the direct comparison of the two; \cite{MMT25} ask whether their invariant is a special case of the Douglas-Reutter invariant (and indeed it is expected to be due to the latter's universality), hence the embedding of \cite{MMT25} in our theory should resolve the two conjectures, which we expect to be a short application of papers III and IV.} (describing the path integral itself, which is the evaluation on closed 4-manifolds), \cite{Dec24}'s full dualizability result of fusion 2-categories is the analog of \cite{DSPS20}'s result (Concerning the abstract existence of certain TQFTs using their evaluation at the point), with an explicit TQFT construction missing from the literature. The next logical step is then to explicitly construct 4D TQFTs from spherical fusion 2-categories and study their physics, filling in particular the role of Balsam and Kirillov's work in closing up the 3D story. This is one of our main goals in this series, with further goals filling other missing gaps that were important in the 3D literature.

\subsubsection{Roadmap}

This paper is part of a six part series of independent papers that together tell a cohesive story, ordered logically rather than chronologically. Fix a fusion 2-category $\mathcal{C}$. In order, the six papers discuss the following aspects of this story:
\begin{enumerate}
    \item \textbf{Paper I (\cite{ReischI})} introduces a collection of invariants of pairs consisting of a closed 4-manifold together with a knotted surfaces embedded in the 4-manifold. The collection is indexed by a choice of half-braiding on $\mathcal{C}$ for each connected component of the surface link, hence controlled by the Drinfeld 2-center $\mathcal{Z(C)}$. The invariants are given by a generalization of the Douglas-Reutter state sum formula, imposing different local rules on the decorations and weights assigned to simplices in the vicinity of the knotted surface. Along the way, we reprove the Douglas-Reutter state sum's invariance using purely diagrammatic methods. Conceptually, this paper constructs the path integral in the presence of certain surface operators for the theories discussed in the series.
    \item \textbf{Paper II} extends the Douglas-Reutter state sum invariant to a functorial field theory in the sense of Atiyah, which amounts to introducing state spaces assigned to 3-manifolds and time evolution operators assigned to 4-cobordisms, extracted from a geometry-sensitive theory as its topological sector, or equivalently the ground state of a certain lattice model. We then provide an alternative, continuum description of the state spaces using a skein theory of surfaces ("membrane-net"), which is manifestly topological and highly reminiscent of non-invertible gauge theory, or from the lattice model point of view is the 4D version of Levin-Wen's string nets. We then develop a pair of categorifications of the Kirby color familiar from fusion 1-category, and use them to establish a canonical isomorphism between the former and latter descriptions of the state spaces, which is the paper's main theorem.
    \item \textbf{Paper III (\cite{ReischIII})} computes the time evolution operators of paper II for elementary 4-cobordisms using the state sum, then extracts equivalent description in terms of the membrane-net spaces, completing the alternative membrane-net description of the TQFTs developed in paper II.
    \item \textbf{Paper IV (\cite{ReischIV})} extends both of the TQFTs of previous papers to a twice-extended TQFT (i.e. a trifunctor $\text{Bord}_{4,3,2,1}\to3\text{Vect}$) and shows that the two are equivalent as twice-extended TQFTs. The extensions are very explicit, with one being manifestly finite and amenable to computer implementation, with the second being manifestly topological. This establishes the locality of the field theories, and through a higher state-operator corresponds allows one to obtain all surface, line and point defects/operators and compute all of their correlation functions.
    \item \textbf{Paper V (\cite{ReischV})} constructs an equivalence of 2-categories between the 2-category assigned to the circle by the TQFTs from paper IV and the Drinfeld center $\mathcal{Z(C)}$.
    \item \textbf{Paper VI (\cite{ReischVI})} brings it all together. We show that the equivalence of paper V is an equivalence of braided 2-categories, then use it to obtain a description of the theories from papers II,III,IV in terms of a skein theory of surfaces in 4 dimensions. In particular, one sees in the process that the defects constructed in paper I cover all surface defects obtained through the higher state-operator correspondence of IV, hence provide a state-sum description for all of their correlation functions which allows one to compute them by hand or by computer implementation.
\end{enumerate}

\subsection{Introduction to paper II}\label{intro to 2nd paper}

This paper focuses on the 4-3 part of the theories discussed in this series of papers. Fix a spherical (pre)fusion 2-category $\mathcal{C}$ as the input to the construction.

Our first result in this paper is the construction of the state spaces and time evolution maps associated with the Douglas-Reutter
state sum. To every closed oriented three-manifold $M$, we
assign a finite-dimensional vector space $\mathcal H(M)$, and to every four-dimensional cobordism $C:M_1\to M_2$ (4-cobordism for short), a linear map
\begin{equation}
U(C):
\mathcal H(M_1)
\longrightarrow
\mathcal H(M_2),
\end{equation}
satisfying Atiyah's TQFT axioms. The construction begins with
a triangulation-dependent lattice state space and uses the path
integral on a cylinder to define a projector onto its
topological subspace. We show that the images of these
projectors for different triangulations are canonically
isomorphic, with the isomorphisms induced by corresponding
cylinder cobordisms. The resulting state spaces and evolution
maps are independent of the auxiliary triangulations. This
provides an explicit realization of the general principle that
a topological theory can be extracted from a
geometry-dependent construction by restricting to the subspace
invariant under changes of the auxiliary geometric structure.

Our central result is a continuum description of these state
spaces in terms of \emph{membrane-nets}, providing a
higher-dimensional generalization of the Levin--Wen string-net
construction. A membrane-net is represented by a decorated
foam embedded in a three-manifold. Its two-dimensional sheets
are labeled by objects of $\mathcal C$, the lines along which
sheets meet by 1-morphisms, and their junction points by
2-morphisms. We consider formal linear combinations of such
configurations, up to isotopies and local relations determined
by the graphical calculus of the fusion 2-category. The
resulting quotient defines a skein-theoretic vector space
$\mathcal H^{\mathrm{mem}}(M)$, intrinsically associated with
the three-manifold and independent of any triangulation. For the special case $\mathcal{C}=\text{Mod}(\mathcal{B})$ with $\mathcal{B}$ a ribbon fusion 1-category, we expect this invariant to reduce to the well-known skein modules of ribbon fusion (and in particular modular) categories. $\mathcal{H}^\text{mem}(M)$ is in essence an explicit realization of \cite{Walker06}'s far-reaching framework, applied to the setting of 3+1D TQFTs; our main contribution in this paper is in identifying this space with the lattice construction $\mathcal{H}(M)$.

\emph{\textbf{Important distinction: comparison with \cite{XLLC21}'s membrane-nets} Note that we use "membrane-nets" to describe a continuum, skein-theoretic type of construction. \cite{XLLC21} on the other hand simply use this term to describe a dual lattice construction, whose continuum description should be our membrane-nets. Our use of the term is analogous to the relation of string-nets \cite{Kir11} to the Levin-Wen model \cite{LW05}.}

We prove that the membrane-net space is canonically isomorphic
to the state space obtained from the extension of the state-sum construction:
\begin{equation}
\mathcal H(M)
\cong
\mathcal H^{\mathrm{mem}}(M)
\label{eq:membrane_net_equivalence}
\end{equation}
where the isomorphism commutes with changes of the auxiliary triangulation used to realize the $\mathcal{H}(M)$ spaces. The two descriptions have complementary properties. The
lattice state space is finite-dimensional by construction,
but initially depends on a triangulation. The membrane-net
space is manifestly topological, but is defined as a quotient
of an infinite-dimensional vector space of decorated foams.
Their equivalence establishes the finite dimensionality of
the membrane-net space and identifies it with the physical
state space of the corresponding four-dimensional TQFT.

A principal mathematical ingredient in the proof is the
introduction of \emph{categorified Kirby colors}. In the
graphical calculus of fusion (1-)categories, Kirby colors are
distinguished weighted sums of simple objects satisfying
identities that permit further topological manipulations, similar in spirit to handle slides. The structure of a fusion
2-category leads naturally to two distinct categorifications.
The \emph{global Kirby color} is a weighted superposition of
membranes labeled by simple objects of $\mathcal C$, while
the \emph{local Kirby colors} are weighted superpositions of
junction lines labeled by simple 1-morphisms between specified
objects\footnote{Note that global and local here refer to the category-theoretic terminology and not to globality and locality in the physical sense.}. We establish generalized handle slide-like identities
and relations governing the addition and removal of these
special membrane-net components. These identities extend the graphical
techniques associated with ordinary Kirby colors to the
setting of fusion 2-categories and we expect them to be of interest
independently of their applications in this paper series.

The higher Kirby colors provide the main tool for proving
the membrane-net equivalence, via a \cite{Kir11} type of argument, though the 4-dimensional version of this argument turns out to be much more involved and interesting. We first identify the lattice
state space with a space of decorated foams constrained by
the triangulation (thus defining a larger space of membrane-nets, since we quotient only by relations that respect these constraints). We then describe how the constraints
associated with the two-simplices, edges, and vertices can
be removed successively by certain local relations. We prove three "transparency lemmas", which relate the corresponding quotients with canonical projectors defined using the local and
global Kirby colors, with their handle slide-like properties screening away the constraints placed by the simplices. We explicitly obtain a factorization of the cylinder projector into these diagrammatic projectors, each associated with a corresponding simplex in the triangulation. The image of the cylinder projector is then canonically isomorphic to the membrane-net space $\mathcal{H}^\text{mem}(M)$, establishing the claimed equivalence. The higher Kirby colors are useful for many independent purposes - in particular, in future papers we will use them to define invariants of knotted surfaces in 4-manifolds from handle decompositions (papers III,VI and hopefully also in future work beyond this series).

The construction also admits a natural interpretation in
the language of lattice topological models. In the
Levin--Wen construction, the ground-state space is selected
by local commuting projectors, and its continuum description
is given by string-nets. Our construction provides a
higher-dimensional analogue at the level of the state
spaces and their defining projectors, with membrane-nets
replacing string-nets. The resulting framework relates
the algebraic data of a fusion 2-category to topological
states on spatial three-manifolds, and our local projectors assemble into a collection of Hamiltonians for a certain lattice model on the triangulation whose vacuum sector is our TQFT's space of states. In a special case, this family should specialize to \cite{XLLC21}'s model, which they argue describes all general 3D topological orders with gapped boundary. Thus, our central result should yield a continuum description of all such topological orders. While we introduce the corresponding lattice models and their Hamiltonians (as well as the aforementioned continuum description), the detailed study of their excitations and possible applications to topological quantum error correction lies beyond the present work.

More generally, the membrane-net construction can be viewed
as a higher-categorical analogue of imposing gauge
invariance (and gauge constraints). The local relations identify configurations that differ by transformations determined by the fusion
2-category, while the lattice projector selects the
corresponding invariant states. The equivalence between
the two descriptions thus relates the projection onto the
topological sector of lattice gauge theories to a skein-theoretic
quotient defined directly in the continuum, where the latter can be interpreted as equivalence classes of "higher and non invertible principal bundles", with the decorated foams encoding specific trivializations). We intend to explore this point of view in
future work, and in particular extract a geometric notion of gauge theory for higher and non-invertible symmetries.

\subsection{Organization}

The paper is organized as follows. In Section \ref{review}, we review
the relevant structure of fusion 2-categories and the
Douglas-Reutter state sum, emphasizing its interpretation
as a discrete path integral. In Section \ref{from state sum to TQFT}, we construct
the state spaces and cobordism evolution maps and establish
their independence of the triangulation. Section \ref{membranenets}
introduces membrane-nets and develops the higher Kirby
colors and their graphical identities. We then use these
identities to prove the equivalence between the membrane-net
and state-sum state spaces. We conclude with a tensor network lattice model and general 3D topological orders in \ref{tensor network}.

\section{Background: fusion 2-categories and their state sums}
\label{review}

Topological quantum field theories admit a description in terms of extended topological structures and their composition laws. In three spacetime dimensions, fusion categories provide the algebraic framework underlying the Turaev-Viro state sum and the Levin-Wen string-net construction. In four spacetime dimensions, the corresponding structure is richer: in addition to surface operators, one must account for their junction lines and the local operators at which these junctions meet. Fusion 2-categories provide a natural algebraic framework for organizing these structures and their composition.

A fusion 2-category consists of objects, 1-morphisms, and 2-morphisms, together with composition, a monoidal product, duality, and suitable finiteness and semisimplicity conditions. These data admit a geometric interpretation in terms of labeled surfaces, their junction lines, and junction operators. The graphical calculus translates the categorical operations and identities into local manipulations of these geometric configurations. With an appropriate spherical structure, the categorical data can be used to construct the Douglas-Reutter state sum, which defines an invariant of closed oriented four-manifolds.

In this section, we review the elements of fusion 2-category theory needed for our construction, emphasizing their geometric interpretation. We then describe the Douglas-Reutter state sum as a discrete path integral, in which categorical labels are assigned to simplices of a triangulation and local amplitudes are combined into a partition function. More precisely, this is a globalization of local path integral, with the numerical values associated to simplices representing a finite set of paths integral values from which the path integral on any 4-manifold can be obtained, since all the different geometries of a ball are equivalent in a topological field theory. This state-sum formulation will provide the starting point for the construction of state spaces and cobordism maps.

\subsection{Fusion 2-categories}

\begin{figure}[H]
\includegraphics[width=0.3\textwidth]{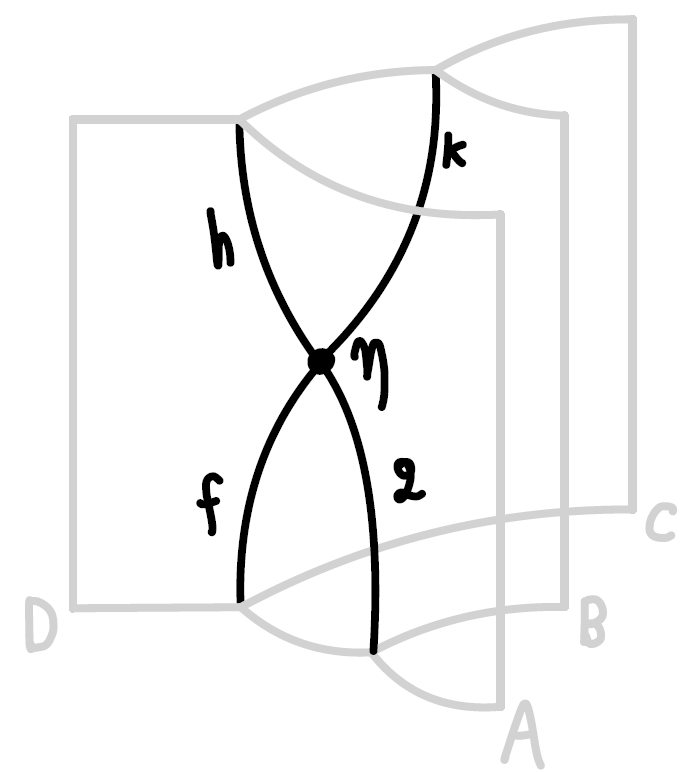}
\centering{}
\caption{Objects, 1-morphisms, and a 2-morphism in a fusion 2-category.
Shaded \textbf{sheets} are \textbf{objects} \(A,B,C,D\). The \textbf{strands} where sheets meet are
\textbf{1-morphisms}; vertical stacking denotes the $\cdot$ composition, horizontal gluing
denotes the $\circ$ composition and juxtaposition in the direction perpendicular to the page denotes the monoidal product $\Box$. The right boundary is the object \(A\!\Box\!B\!\Box\!C\)
and the left boundary is \(D\), so both composites \(f\circ g\) and \(h\circ k\) lie in
\(\mathrm{Hom}(A\!\Box\! B \!\Box\!C,\,D)\). The \textbf{coupon} labeled \(\eta\) is a
\textbf{2-morphism}
\(\eta: f\circ g \Rightarrow h\circ k\) in that Hom 1-category.}
\label{2category}
\end{figure}

Fusion 2-categories have three algebraic layers, each with their own operations, shown in Fig. \ref{2category}.

\subsubsection{objects = spatial boundary conditions}
First, a fusion 2-category $\mathcal{C}$ has a set of "objects" $\text{ob}\mathcal{C}$ which we think of as \emph{string types} or as \emph{spatial boundary conditions} for strings. We can combine two objects $A,B$ by putting them next to each other along some coordinate and this topological operation is algebraically encoded by the "monoidal product" $A\Box B$ \footnote{An example are
1-form symmetry surfaces in four dimensions, where the labels are group elements $A,B$, and
$U_A(\Sigma)\;\square\; U_B(\Sigma) \;=\; U_{AB}(\Sigma)$.}. The $\Box$ operation is associative (which is necessary if we are to think of it as physically putting them next to each other) but not necessarily commutative (order matters). This operation also has a unit which we denote by $1$, the empty boundary condition.
\subsubsection{1-morphisms = string configurations}
The second layer consists of a set of "1-morphisms" between every pair of objects $A,B$ which we denote by $\text{Hom}_\mathcal{C}(A,B)$ and think of as some \emph{spatial configuration of strings} stretching from one boundary condition $A$ to another boundary condition $B$. If we adopt the string type intuition then we think of a string touching a boundary point labeled by $A$ as "a string of type $A$". We will sometimes write the statement that $f\in\text{Hom}_\mathcal{C}(A,B)$ as $f:A\to B$ (invoking the usual notation for functions) and say that "$f$ is a 1-morphism from $A$ to $B$". The 1-morphisms can also depict strings that intersect in interior vertices (or "junctions"), not just strings ending on the boundaries. For 1-morphisms there are two kinds of operations. Given two 1-morphisms $f:A\to B,g:C\to D$ we can put them side by side along the direction of the monoidal product to obtain a 1-morphism:
\begin{equation}
f\Box g:A\Box C\to B\Box D \ .
\end{equation}
If $B=C$ then we can also "compose" the two 1-morphisms by gluing along their shared boundary which we think of as coupling the two sets of strings to obtain a larger string $f\circ g$. These two operations are also associative. For each object $A\in\text{ob}\mathcal{C}$ there is unit 1-morphism $1_A:A\to A$ which we think of as the straight string stretching from one copy of $A$ to the other $A$ and this 1-morphism is a unit w.r.t the $\circ$ operation (at least in topological contexts where simply stretching something doesn't do anything physical). The data mentioned so far defines a \emph{monoidal 1-category}.\footnote{Only if we are in a strict fusion 2-category. For the purpose of this exposition we omit bicategory coherences.}
\subsubsection{2-morphisms = string processes}
The third layer is a kind of "morphisms between morphisms". For every pair of 1-morphisms with the same boundary conditions $f,g:A\to B$ there is a set of "2-morphisms" between them $\text{Hom}_{\text{Hom}_\mathcal{C}(A,B)}(f,g)$\footnote{The notation $\text{Hom}_{\text{Hom}}$ stresses the fact that $\text{Hom}_\mathcal{C}(A,B)$ is a 1-category with objects being 1-morphisms $f:A\to B$. The 1-morphisms inside that 1-category are then the 2-morphisms $\alpha:f\Rightarrow g$, and following the notation from the previous section on 1-morphisms we see where the $\text{Hom}_{\text{Hom}}$ comes from.} which we think of as \emph{processes of strings} in which $f$ turns into $g$ as time progresses while preserving the boundary conditions. We will often write the statement that $\alpha$ is a 2-morphism between $f$ and $g$ as $\alpha:f\Rightarrow g$. 2-morphisms have three kinds of operations - we can put any two of them $\alpha:f\Rightarrow g,\beta:e\Rightarrow h$ side by side along the direction of the monoidal product to form $\alpha\Box \beta$ regardless of the source/target of the 1-morphisms, we can glue $\alpha,\beta$ along the 1-morphism direction if the target object of $f,g$ matches the source object of $e,h$ to obtain a 2-morphism:
\begin{equation}
\beta\circ\alpha:e\circ f\Rightarrow h\circ g \ ,    
\end{equation}
and we can glue two of them along the temporal coordinate if $g=e$ to obtain a 2-morphism $\beta\cdot \alpha:f\Rightarrow h$. The first operation describes two unrelated processes happening side by side, the second describes a process in which a coupled system of strings undergoes two simultaneous independent processes in two of its regions (separated by the shared boundary) and the third operation describes a longer process in which $\alpha$ happens to $f$ turning it into $g$ then $\beta$ occurs on $g$ (turning it into $h$). These are known respectively as the "monoidal product", "horizontal composition" and "vertical composition". Finally, for each 1-morphism $f$ there should be an "identity 2-morphism" $1_f:f\Rightarrow f$ which is a unit w.r.t $\cdot$ - this describes the process in which nothing happens to the string represented by $f$.

We will sometimes denote the set of objects by $\mathcal{C}^0$ and the sets of $k$-morphisms by $\mathcal{C}^k$. The $\circ,\cdot$ define the "2-category" structure of $\mathcal{C}$ and the $\Box$ operation is known as a "monoidal structure". A fusion 2-category has two more additional kinds of structure - linearity and duality.

\subsubsection{Linear 2-category = superpositions}
In quantum mechanical contexts it is natural to equip the higher categories with a "linear" structure to describe superpositions. For 2-categories, linearity means that the 2-morphism spaces come with the structure of a vector space (so we can take linear combinations of processes), the 1-morphisms have a $\oplus$ operation and the objects have a more abstract $\boxplus$ operation. Often we will take a kind of representation of the linear 2-category and then the 2-morphisms will actually be linear maps and the 1-morphism spaces will be their domain/range, in which case $\oplus$ will become the familiar direct sum - this is analogous to the abstract structure of a Lie group that we often represent using vector spaces and linear maps but have many inequivalent representations with the same underlying structure. More formally we define a direct sum of two 1-morphisms $f_1,f_2:A\to B$ to be a 1-morphism $f_1\oplus f_2:A\to B$ with canonical "projection" and "inclusion" 2-morphisms:
\begin{equation}
p_j:f_1\oplus f_2\Rightarrow f_j,\quad i_j:f_j\Rightarrow f_1\oplus f_2 \ 
\end{equation}

satisfying:
    \begin{itemize}
        \item $i_1\cdot p_1+i_2\cdot p_2=1_{f_1\oplus f_2}$
        \item $p_j\cdot i_j=1_{f_j}$
        \item $p_1\cdot i_2=p_2\cdot i_1=0.$
    \end{itemize}
Note that we just used the linear structure on the 2-morphisms to define this operation. Similarly, we define the direct sum of two objects $A_1,A_2$ to be an object $A_1\boxplus A_2$ together with canonical "projection" and "inclusion" 1-morphisms:
\begin{equation}
\rho_j:A_1\boxplus A_2\to A_j,\quad \iota_j:A_j\to A_1\boxplus A_2 \ 
\end{equation}
satisfying:
\begin{itemize}
        \item $\iota_1\circ \rho_1\oplus \iota_2\circ\rho_2=1_{A_1\boxplus A_2}$
        \item $\rho_j\circ \iota_j\cong1_{A_j}$
        \item $\rho_1\circ \iota_2=\rho_2\circ \iota_1=0.$
    \end{itemize}

\subsubsection{2-category with duals = CPT}
Fusion 2-categories are also "2-categories with duals", meaning they come with a few important involutive "duality" operations. First, there is an operation taking an object $A$ into its "object dual" $A^\#$ s.t. $A^{\#\#}=A$, thought of as the \emph{anti-string} of the corresponding string type. This introduces a kind of \emph{charge conjugation} operation into our algebraic structure. This structure can be used to track orientations - strings oriented from source to target and visually labeled by $A$ will correspond to the object $A$, while strings oriented from target to source and visually labeled by $A$ will correspond to the object $A^\#$. Since strings can generically bend in such a way that there are local minima/maxima in the 1-morphism direction, we should have some canonical 1-morphisms:
\begin{equation}
e_A:A\Box A^\#\to 1,i_A:1\to A^\#\Box A \ ,   
\end{equation}
known as the "evaluation" and "coevaluation" 1-morphisms and interpreted as bends in the 1-morphism direction. For the latter reason they are often called "folds". We should also have canonical 2-morphisms:
\begin{equation}
C_A:(e_A\Box 1_A)\circ (1_A\Box i_A)\Rightarrow 1_A, D_A:1_{A^\#}\Rightarrow \left(1_{A^\#}\Box e_A\right)\circ \left(i_A\Box1_{A^\#}\right) \ ,    
\end{equation}
known as the "cusps" and describing a process in which a minimum-maximum pair cancels out. We can use this kind of charge conjugation structure to construct "left and right mates" for 1-morphisms using the folds, taking a 1-morphism $f:A\to B$ into $f^\#:B^\#\to A^\#$ and $^\#f:B^\#\to A^\#$ respectively. We can use the cusps interpolate between the two mates. Similarly, we can construct "left and right mates" of 2-morphisms, taking $\alpha:f\Rightarrow g$ into: $\alpha^\#:g^\#\Rightarrow f^\#$ and $^\#\alpha:\ ^\#g\Rightarrow\ ^\#f$.
The second kind of involution is known as the "1-morphism dual" (or "adjoint"), taking a 1-morphism $f:A\to B$ into $f^*:B\to A$ and satisfying $f^{**}=f$. Notice that it just swaps the boundary conditions but doesn't change them. This can be thought of as a reflection in the 1-morphism direction, a kind of \emph{parity} transformation. Similarly to the object dual, this operation induces further structure. Recall that we decompose the 1-morphisms into junctions of strings taking one set of string types into the other. These junctions behave like particles coupled to each other by the strings and over time they trace curves in spacetime. We can use the 1-morphism dual to keep track of orientations - when a junction is oriented from initial to final configuration is visually labeled by $f$ we will treat it as a junction of type $f$ and when it points in the opposite temporal direction while still being visually labeled by $f$ we will treat it as a junction of type $f^*$. We can bend the curve traced by a junction $f:A\to B$ to get canonical 2-morphisms:
\begin{equation}
i_f:1_A\Rightarrow f^*\circ f,e_f:f\circ f^*\Rightarrow1_B \ ,    
\end{equation}
known as the "unit" and "counit" respectively, and they describe pair creation/annihilation processes for the particles coupled by the strings. Sometimes they will be denoted as $\eta_f,\epsilon_f$ to distinguish them from the analogous 1-morphisms for $\#$, though we prefer this notation since it highlights their similar role. As before we can also form the "left and right mates" of a 2-morphism $\alpha$ which we denote by $^*\alpha,\alpha^*$.

We could also add a "2-morphism dual" that takes a 2-morphism $\alpha:f\Rightarrow g$ into $\alpha^\dagger:g\Rightarrow f$ and describes a time-reversal operation. Though anyonic string models typically have this kind of structure, fusion 2-categories don't always come equipped with 2-morphism duals and indeed we will not need it for constructing TQFTs out of fusion 2-categories.

Finally, all of the above structure obeys some key axioms ensuring a notion of semisimplicity and isotopy invariance:
\subsubsection{Semisimplicity = decomposition into irreps}
The linear structure is of a more algebraic nature than the rest of the structure (which can be described purely graphically). This has to do with the fact that we want the fusion 2-category to eventually encode operator algebras. Typically in physics we deal with "semisimple" theories, meaning we have some set of "irreducible" creatures out of which we can build everything by taking the right notion of direct sums. In the case of 2-morphisms the "sum" operation is just the vector space sum so we can always pick a basis. Given two objects $A,B$ we can decompose the space of 1-morphisms between them $\text{Hom}_\mathcal{C}(A,B)$ into a finite set of "simple" 1-morphisms $f_1,...,f_n$ and the rest of the 1-morphisms will be direct sums of these. This is the abstract version of "decomposing into irreducible representations". At the level of objects, we also have a finite set of "simple" objects whose direct sums give all the other objects. "Simplicity" means three things: simple objects and 1-morphisms cannot be decomposed further, the set of 2-morphisms between two simple 1-morphisms is $\mathbb{C}$ if they are isomorphic or 0 otherwise, and finally the simple objects have simple identities. There is one extra condition defining semisimplicity: the whole 2-category has to be "idempotent complete". This is an important technical condition described in detail in \cite{DR18} but shouldn't worry those that prefer to skip details of this kind - every 2-category can be "completed" to one satisfying this property in an essentially unique way so it doesn't actually add new physical data.

Note that the semisimplicity implies something particularly important - when we bring two string types $A,B$ together into a single object $A\Box B$, we should be able to decompose it into simple objects! \footnote{In a semisimple theory, stacking two simple surfaces decomposes into a sum of simple surfaces:
$A \square B \;\cong\; \bigoplus_{C} N^{C}_{AB}\, C$, where the integers $N^{C}_{AB}$ are the fusion multiplicities.
Choosing a summand $C$ is encoded by a 1-morphism
$f:\; A \square B \longrightarrow C$, which is the topological line living on the fused surface along their seam.
For 1-form symmetry surfaces, the seam carries the canonical 1-morphism implementing the group structure
$A\square B  \to AB$.} Similarly, nearby particles $f_1,f_2$ sitting at the junctions can be composed to yield a single 1-morphism $f_1\circ f_2$ and so we should be able to decompose those into a direct sum of simples as well. These decompositions of pairs are known as the \emph{fusion rules} and as the name suggests they allow us to fuse pairs of surface and line operators into a single one.

\subsubsection{Isotopy invariance = topological theory}
Our graphical language has the property that if we apply any isotopy (that is, a continuous deformation that does not cause the points, lines and surfaces to gain new intersections) that preserves the order of objects, 1-morphisms and 2-morphisms in the directions of $\Box,\circ,\cdot$ then the diagram will represent the same algebraic object - the algebraic structure is simply blind to such deformations. In topological contexts (when the theory or operators are topological or if we are just interested in the statistics of the excitations) something stronger should hold - the graphical calculus should be invariant to \emph{arbitrary} deformations. In fact we have to be a little cautious about what we mean by "arbitrary". When one deals with (potentially exotic) particle statistics in 2+1 dimensions, we have to keep track of the order of the particles in the plane and we just demand that processes with the same initial and final configuration that are related by a continuous deformation that fixes the initial and final configrations will be physically equilvalent - this allows us to treat processes in which particles swap locations as nontrivial but the precise path that each of them took does not matter to us. We will do the same here, namely demand the 2-morphisms (processes) with the same source and target that are related by an isotopy fixing those source and target (and also the boundary conditions!) will be \emph{equal}. This implies a lot of axioms for the fusion 2-categories that come from the topology, for instance if we apply the cusp to create a minimum-maximum pair then we compose with another one to kill those minimum-maximum pair we should get the identity 2-morphisms as the whole surface can be isotoped into a straight one. There are also a lot of axioms we did not mention that tell us how the various structures interact. They all capture exactly what the topological interpretation tells us. For instance, $i_{f\Box g}$ should equal $i_f\Box i_g$. The full axioms are listed explicitly in \cite{DR18}, though all we need to know is that they capture precisely what we need for the graphical calculus to make sense and for isotopies of the 2-morphisms to not do anything. A full treatment of this property is described in \cite{BMS24}.

Note that since we are keeping track of order, the following two compositions should be treated as distinct 1-morphisms for any choice of 1-morphisms $f:A\to B,g:C\to D$:
\begin{eqnarray}
(f\Box1_D)&\circ&(1_A\Box g) \nonumber\\
(1_B\Box g)&\circ&(f\Box1_C) \ .   
\end{eqnarray}

If we define $f\Box g$ as the two string configurations $f,g$ sitting exactly next to each other, then the slightest perturbation will take $f\Box g$ to one of the above two - it is unnatural and ungeneric. We thus choose to \emph{define} $f\Box g$ as putting them both side by side only when one of them is an identity 1-morphism and otherwise define this product as the first of the two perturbations. The process swapping the order of $f,g$ from the first decomposition into the second one is a nontrivial 2-morphism $\phi_{f,g}$ known as the "tensorator" or "interchanger", and corresponds to a nontrivial movie of strings. Its time reversal inverts it due to a straightforward isotopy.

The graphical calculus of monoidal 2-categories with duals is discussed extensively in \cite{DR18}, developed originally in \cite{BMS24} and surveyed shortly in many sources (including \cite{XLLC21}). We will make extensive use of this graphical calculus, relying strongly on the results of \cite{BMS24}, to which we refer readers who wish to familiarize themselves with the many technicalities of Gray category diagrams. In particular, we make strong use of the invariance of the graphical calculus' invariance under 3-ball isotopies (rel boundary).

Now that we explored the definition and structure of fusion 2-categories, let us discuss two last points before moving on to constructing TQFTs out of them.

\subsubsection{Traces and dimensions}\label{quantum dimensions}
Fusion 2-categories admit a kind of "trace" operation that we get for free out of all the previous structure. Given some 2-morphism with the same source and target $\alpha:f\Rightarrow f$ with $f:A\to B$ we can define its trace by first closing $f$ into a sphere using $i_f, e_{f^*}$, getting a closed $f$-labeled circle on an $A$-labeled plane, or algebraically a 2-morphism $i_f\cdot(1_{f^*}\circ\alpha)\cdot e_{f^*}$. Now we close this plane into a sphere using the bottom cup $i_{i_A}$, the top cap $e_{e_A}$ and the folds on either side together with the identity $1_{1_A}$. Together these make up a 2-morphism $\text{Tr}(\alpha):1_1\to1_1$. The vector space of such 2-morphisms is 1-dimensional since $1_1$ is simple and it has a canonical basis element $1_{1_1}$ therefore we can identify $\text{Tr}\alpha$ with a number! As the name implies, this number is invariant under cyclic permutations and satisfies a few properties reminiscent of the usual matrix trace. Note that there is actually one ambiguity in this construction - we could have also done this closing with a reverse orientation which in general can yield a different result (the two are not related by an isotopy in $\mathbb{R}^3$). The extra condition requiring these two traces to coincide is known as \emph{sphericality}.

Given these traces, we can define three important kinds of "dimensions". Let $A$ be an object and $f:B\to C$ be a 1-morphism. We define:

\begin{equation}
    \dim(A)\coloneqq\text{Tr}\left(1_{1_A}\right),\ \dim(f)\coloneqq\text{Tr}\left(1_f\right),\ \dim(\mathcal{C})\coloneqq\sum_{[A]\in\pi_0\mathcal{C}}\frac{1}{\dim(\text{End}(A))}
\end{equation}
where $\dim(\text{End}(A))=\sum_{f\in\text{Irr}(\text{End})}\dim(f)^2$ is the usual dimension of a fusion 1-category (applied on the fusion 1-category $\text{End}(A)$) and $\pi_0\mathcal{C}$ is the set of \emph{connected components} of $\mathcal{C}$ (defined as the quotient of the set of simple objects $\text{Irr}\mathcal{C}$ by the equivalence relation identifying simple objects between which there is a nonzero 1-morphism)\footnote{Note that this is indeed an equivalence relation and $\dim\text{End}A$ is independent of the representative in the equivalence class.}. A useful renormalization of the above is the rescaled dimension:

\begin{equation}
    d(A)\coloneqq\dim(A)\dim(\text{End}A)n(A)
\end{equation}
where $n(A)$ is the number of equivalence classes of simple objects in the connected component of $A$ (under the usual equivalence relation of objects in a 2-category, not under the equivalence relation defining a connected component). We will sometimes shorten $\dim(f)$ to $d(f)$.

We will also make use of the following key formula from \cite{DR18} (appearing there as corollary B.4) which mixes the various dimensions:

\begin{equation}
    \sum_{B,A\xrightarrow{f}B}\frac{\dim(f)^2}{d(B)}=\dim(A)
\end{equation}
where the sum is over simple $B,\ f$ with $A$ a (non necessarily simple) fixed object.

\subsubsection{Summarizing table}

For convenience, we organize all of the above interpretations in a table:
\begin{center}
\begin{tabular}{||c c||} 
 \hline
 algebraic structure/operation & Interpretation\\ [0.5ex] 
 \hline\hline
 object $A $& string type/boundary condition (BC)\\ 
 \hline
 1-morphism $f:A\to B$ & string configuration between the BC $A,B$\\
 \hline
 2-morphism $\alpha:f \Rightarrow g$ & process of strings from $f$ to $g$\\
 \hline
 monoidal product $\Box$ & (ordered) disjoint union\\
 \hline
 horizontal composition $\circ$ & coupling strings along a shared BC\\
 \hline
 vertical composition $\cdot$ & combining processes into a long one\\
 \hline
 $+,\oplus,\boxplus$ & superpositions\\
 \hline
  "object" dual $\#$ & Charge conjugation\\
 \hline
  "1-morphism" dual $*$ & Parity\\
 \hline
  "2-morphism" dual $\dagger$ & Time reversal\\
 \hline
  $e,i$ & folds\\
 \hline
  $C,D$ & cusps\\
 \hline

\end{tabular}
\end{center}

\subsubsection{Gluing rules}

Throughout the paper, we will make extensive use of three cut-and-glue rules that hold in fusion 2-category diagrams, proved in the appendices of \cite{DR18} and reviewed in \cite{ReischI}. We denote by $\overline\alpha$ the dual of alpha w.r.t the $\langle\_,\_\rangle=\text{Tr}(\_\cdot\_)$ pairing, with assumed summations over pairs of a 2-morphism and its dual that appear in the same diagram. Summations over objects and 1-morphisms are done over a fixed set of representatives of the equivalence classes of simple objects and simple 1-morphisms in the corresponding $\text{Hom}$ category. In the graphical calculus, the gluing rules are:

\begin{itemize}
    \item \textbf{Gluing rule I: twining/untwining} For any fixed sources and targets (bottom and top 1-morphism wires and object sheets):
    \begin{equation}
        \sum_{g}\dim(g)\glueI=\glueres.
    \end{equation}
    \item \textbf{Gluing rule II: bubble pop} For any fixed sources and targets (bottom and top 1-morphism wires and object sheets):
    \begin{equation}
        \sum_{f,g,A}\frac{\dim(f)\dim(g)}{d(A)}\glueII=\glueres.
    \end{equation}
    \item \textbf{gluing rule III: trace gluing} For each pair of 2-endomorphisms $\alpha:f\to f,\ \beta:f\to f$ of a simple 1-morphism $f$:
    \begin{equation}
        \frac{\text{Tr}\left(\alpha\right)\text{Tr}\left(\beta\right)}{\dim(f)}=\text{Tr}\left(\alpha\cdot\beta\right)=\langle\alpha,\beta\rangle.
    \end{equation}
\end{itemize}

\subsubsection{State sum}
In principle, the algebra of observables (and their correlators) should contain all of the physical information. We thus expect to be able to reconstruct a TQFT from a fusion n-category. In the case of a 4D TQFT, the only "topologically interesting" operators are surface operators - operators of higher or lower codimension cannot knot, meaning their isotopies are "trivial" in a sense. The relevant category is then a (spherical) fusion 2-category. The most explicit of the general constructions that let us reproduce a TQFT from this kind of data is known as a \emph{state sum model} and allows us to construct spaces of states, a notion of classical fields and a path integral that resemble familiar QFTs. In the 4D case the most general state sum construction was carefully defined by Douglas and Reutter \cite{DR18}\footnote{In fact they only showed how to extract a number from every closed 4-manifold that is computed in a path-integral fashion. We will see that adding state spaces and defining a time evolution operator does not require a lot of work.} and we will later see how to extend it to have most of the other features of a physical QFT, namely including observables and turning the TQFT into a local theory.

Their definition of the path integral for a closed 4-manifold $M$ goes as follows:
\begin{itemize}
    \item First, pick some triangulation $K$ on $M$, stratifying it into a set of 4-simplices $K^4$ intersecting in 3-simplices from $K^3$ which are patched together along the 2-simplices in $K^2$ and so on.
    \item Define a \emph{classical field/state} on the triangulated manifold as an assignment of simple objects to the edges, simple 1-morphisms to the triangles whose domain and range are given by the 3 objects on the boundary of the triangle (with the $\Box$ operation used to combine multiple objects into a single one since we have 3 boundary edges) and finally an assignment of 2-morphisms to each tetrahedron whose domain/range is given by the boundary triangles (again we combine 1-morphisms using $\circ$).
    \item Define the \emph{amplitude} of a classical field configuration to be a product of the amplitudes at each 4-simplex, defined as the trace of the $\cdot$ of all the 2-morphisms in the boundary tetrahedra of the 4-simplex. We also normalize the amplitude by the dimensions of the simple objects and 1-morphisms
    \item Finally, define the partition function of $M$ by summing the amplitudes of all possible classical fields. This sum is not well defined - there are finitely many simple objects and 1-morphisms, but the spaces of 2-morphisms are complex vector spaces so they have an infinite number of elements. The more natural thing would then be to sum over a particular \emph{basis} in each vector space, since this is analogous to the notion of simple objects/1-morphisms as "generating" their corresponding spaces. To pick a canonical basis we can combine $\cdot,\text{Tr}$ into a non-degenerate bilinear form which picks out a class of "nice" bases.
\end{itemize}
Douglas and Reutter proved that the fusion 2-category axioms imply this procedure yields an invariant of the 4-manifold - it does not depend on the choice of triangulation and so we can interpret it as the partition function of the theory. Though they did define a notion of "classical fields/states", they \emph{do} depend on the specific triangulation, but all of the extra auxiliary data disappears under summation. In the next section we will review their construction in full formality then show how it can be extended naturally to a TQFT.

\subsection{The Path Integral}\label{sec:path-integral}
Let $\mathcal{C}$ be a spherical prefusion 2-category and let $K$ be an ordered, oriented combinatorial manifold of dimension~4.  We define a state sum path integral on $K$ by assigning categorical data to its simplices and summing over all such labelings.

\subsubsection{Combinatorial Manifolds}\label{sec:comb-manifolds}

We work throughout with finite simplicial complexes.

\paragraph{Balls, spheres, and manifolds.}
A \emph{combinatorial $n$-ball} is a simplicial complex (PL) isomorphic to the standard $n$-simplex; its boundary is a \emph{combinatorial $(n\!-\!1)$-sphere}.
A \emph{combinatorial $n$-manifold} is a finite simplicial complex in which the link of every vertex is a combinatorial $(n\!-\!1)$-sphere.
A \emph{singular combinatorial $n$-manifold} relaxes this by requiring instead that the link of each vertex is a \emph{combinatorial $(n\!-\!1)$-manifold} (not necessarily a sphere). While the Douglas-Reutter state-sum invariant is well-defined on singular combinatorial closed 4-manifolds, we will mainly restrict our attention to the non-singular case for convenience. We will often deal with \emph{combinatorial manifolds with boundary}, which relax the link condition by allowing the link of a vertex to be either a combinatorial sphere or a combinatorial ball (where the latter is the case of vertices on the boundary and the former is the case of vertices in the interior).

\paragraph{Orientation.}
An \emph{orientation} of a combinatorial $n$-manifold is a choice of orientation on each $n$-simplex such that the orientations induced on any shared $(n\!-\!1)$-face by its two adjacent $n$-simplices are opposite.

\paragraph{Global order and notation.}
An \emph{ordered, oriented} combinatorial $n$-manifold is an oriented combinatorial $n$-manifold together with a global total order $o$ on its vertices (the order need not agree with the orientation).
We write $K^{o}$ for such a triangulation and $K^{o}_{k}$ for its set of $k$-simplices. More often than not we will omit the $o$ from the notation, but we will always assume that we have some fixed order.

The global order $o$ lets us specify faces unambiguously.
For an $n$-simplex $\tau\in K^{o}_{n}$, identify its vertices with $\{0,1,\dots,n\}$ according to the local order induced by $o$.
For any strictly increasing multi-index $[j_{0}\cdots j_{k}]\subset\{0,\dots,n\}$, we denote the corresponding $k$-face by $[j_0\cdots j_k]^\tau$. When the choice of $n$-simplex is clear from context, we will omit the superscript and just write $[j_0\cdots j_k]$.

\paragraph{Standing assumption.}
Unless noted otherwise, all triangulations below are \emph{ordered, oriented combinatorial manifolds} in this sense, and all face maps are taken with respect to the global order $o$.

\subsubsection{$\mathcal{C}$-States}\label{ssec:states-notation}

A \emph{$\mathcal{C}$-state} or \emph{partial state} is a labeling of a fixed triangulated manifold by data taken from a spherical (pre)fusion 2-category $\mathcal{C}$.
Formally, a $\mathcal{C}$-state $\Gamma$ on $K^o$ consists of:
\begin{itemize}
  \item A choice of simple object $\Gamma(e)\in\mathrm{Ob}\,\mathcal{C}$ for each edge $e\in K_1$. When $e$ is given as the $[ij]^\tau$ of some $n$-simplex $\tau$, we will denote $\Gamma([ij])\coloneqq[ij]_\Gamma$ or occasionally just $[ij]$ when it is clear from the context that we are working with the object assigned to the edge and not with the edge itself.
  \item A choice of simple 1-morphism:
  \begin{equation}
    \Gamma(s)\;\in\;
    \text{Hom}_\mathcal{C}\bigl(
      [01]^s_\Gamma\,\Box\,[12]^s_\Gamma\,,\,
      [02]^s_\Gamma
    \bigr) \ .
  \end{equation}
  for each face $s\in K_2$. In the graphical calculus it corresponds to a configuration of strings with "source" BC obtained by putting $[01]^s_\Gamma,[12]^s_\Gamma$ side by side and "target"  BC $[02]^s_\Gamma$.
\end{itemize}
For a 3-simplex $\kappa=[0\,1\,2\,3]$, the four boundary 1-morphisms naturally split into two composable pairs:

\begin{equation}
[(012)3]:=\;[023]\circ \big([012]\Box1_{[23]}\big),~~~~
[0(123)]:=\;[013]\circ\big(1_{[01]} \Box [123]\big) \ .    
\end{equation}
Both are 1-morphisms with the same source ($[01]\Box[12]\Box[23]$) and target ([03]) so we can consider the space of 2-morphisms between them. In the graphical calculus these correspond to a process in which we apply some kind of "associativity" operation between different "factorizations". \footnote{The $[(012)3]$ string-net starts at the triple $[01]\Box[12]\Box[23]$, combines the first two into $[02]$ using the $[012]$ vertex while doing nothing to the $[23]$ on the right, then combines the latter two objects into $[03]$ using the $[023]$ vertex. On the other hand, the $[0(123)]$ string-net starts at $[01]\Box[12]\Box[23]$, combines the last two into $[13]$ using the $[123]$ vertex while doing nothing to the $[01]$ on the left, then combines the latter two objects into $[03]$ using the $[013]$ vertex. The 2-morphisms between $[(012)3]$ and $[0(123)]$ interpolate between the two ways of taking this triple "product" hence we call them "associators".}

We then define the \emph{associator state spaces}
\begin{equation}
  V^+(\Gamma,\kappa)\;=\;\text{Hom}_\mathcal{C}\bigl([012]3\,,\,[0(123)]\bigr),
  \quad
  V^-(\Gamma,\kappa)\;=\;\text{Hom}_\mathcal{C}\bigl([0(123)]\,,\,[012]3\bigr) \ ,
\end{equation}
equipped with the canonical nondegenerate pairing
$\langle\cdot,\cdot\rangle:V^+\otimes V^-\to\Bbbk$ defined by:

\begin{equation}
  \langle\alpha,\beta\rangle\coloneqq\text{Tr}(\alpha\cdot\beta).
\end{equation}
This pairing also induces a canonical isomorphism with the dual $\gamma\to\hat\gamma$

\subsubsection{Classical States And The Action}\label{ssec:action}

For each 3-simplex $\kappa\in K_3$, let
\begin{equation}
  \cup_{\Gamma,\kappa}\;:\;
  \Bbbk\;\longrightarrow\;
  V^+(\Gamma,\kappa)\,\otimes\,V^-(\Gamma,\kappa) \ ,
\end{equation}
be the canonical copairing dual to $\langle\cdot,\cdot\rangle$.  Taking the tensor over all 3-simplices gives
\begin{equation}
  \cup_\Gamma
  \;=\;
  \bigotimes_{\kappa\in K_3}
  \cup_{\Gamma,\kappa}
  (\;1\;)
  \;\in\;
  \bigotimes_{\kappa\in K_3}
  \bigl(V^+(\Gamma,\kappa)\otimes V^-(\Gamma,\kappa)\bigr) \ .
\end{equation}

For a given $\mathcal{C}$-state $\Gamma$, we call $\cup_\Gamma$ the \emph{complete state} or the \emph{classical field}. Implicitly, $\cup_\Gamma$ picks out a particular linear combination of some canonical basis for the associator spaces and we intuitively think of this basis as the set of "simple 2-morphisms"\footnote{Indeed, a basis generates the spaces associated to 3-simplices by taking linear combinations, like the simple objects/1-morphisms generate the spaces associated to 1/2-simplices by taking direct sums.}

Next, for each 4-simplex $\mu\in K_4$, the pentagonator 2-endomorphism defines a linear map:
\begin{equation}
z(\Gamma,\mu)\;:\;
  \bigotimes_{\kappa<\mu}
  V^{\epsilon_0(\kappa)}(\Gamma,\kappa)
  \;\longrightarrow\;\Bbbk .   
\end{equation}
Tensoring over all $\mu\in K_4$ yields
\begin{equation}
  z(\Gamma)\;=\;\bigotimes_{\mu\in K_4}\,z(\Gamma,\mu)
  \;:\;
  \bigotimes_{\kappa\in K_3}
  \bigl(V^+(\Gamma,\kappa)\otimes V^-(\Gamma,\kappa)\bigr)
  \;\longrightarrow\;\Bbbk \ .
\end{equation}
We then define the \emph{amplitude} of the complete state $\Gamma$ by
\begin{equation}
  Z(\Gamma)
  \;=\;
  z(\Gamma)\,(\cup_\Gamma)
  \;\in\;\Bbbk \ .
\end{equation}

\subsubsection{The Partition Function}\label{ssec:partition-function}

Choose a simplicial skeleton $I$ of $C$ (one simple object per equivalence class, one simple 1-morphism per isomorphism class).  For each $\Gamma$ valued in $\Omega$, define normalization factors
\[
  \phi(\mathcal{C})=\dim(\mathcal{C})^{-1},\quad \phi(\Gamma(e))=\left(\dim(\Gamma(e))\dim(\text{End}_\mathcal{C}(\Gamma(e)))n(\Gamma(e))\right)^{-1},\quad
  \phi(\Gamma(s))=\dim\bigl(\Gamma(s)\bigr),
\]
where $n(\Gamma(e))$ counts the equivalence classes of simple objects in the component of $\Gamma(e)$.  Then the \emph{state sum} is
\begin{equation}
  Z(K^o,\Omega)
  \;=\;
  \sum_{\Gamma\colon K^o\to\Omega}
    \biggl[
      \prod_{v\in K_0}\phi\bigl(\mathcal{C}\bigr)\,
      \prod_{e\in K_1}\phi\bigl(\Gamma(e)\bigr)\,
      \prod_{s\in K_2}\phi\bigl(\Gamma(s)\bigr)
    \biggr]
    Z(\Gamma) \ .
\end{equation}
Douglas–Reutter’s proof (and our alternative, purely diagrammatic proof in \cite{ReischI}) shows that $Z(K^o,\Omega)$ depends only on the PL-homeomorphism type of the underlying 4-manifold, and thus furnishes a topological invariant. We think of this number as the partition function.

\paragraph{Remark on physical meanings of the state sum}

Note that the categorical labels assigned to simplices should not
be confused with the excitation content of the theory. They are the microscopic degrees of freedom summed over by the path integral, flux membranes and their junctions, in the same way that the strings of a string-net model are flux lines rather than
anyons. The genuine topological excitations of the resulting $4$D theory are the codimension-$\geq2$ defects, and these are governed by the Drinfeld $2$-center $\mathcal{Z}(\mathcal{C})$, as we discuss in other papers in the series.

The partition function defined above may seem somewhat unfamiliar, but it is a straightforward generalization of very well known theories. For particular choices of $\mathcal{C}$, this recovers the Dijkgraaf-Witten model, also known as finite group gauge theory, with the state sum being a sum over $G$-bundles with the 10j weights reducing to cocyle weights. For other values of $\mathcal{C}$, we can also recover higher gauge theories, as well as Crane-Yetter theories (whose boundary is Reshetikhin-Turaev theory, a vast generalization of Chern-Simons theory).

While the state sum may be interpreted as a discretized path integral, one may see it instead as a globalization of a continuous path integral. They key principle behind it is that given the path integral on the $4$-ball for a collection of appropriate boundary conditions, one may recover the path integral on a general 4-manifold by dividing the latter into pentachora (which are topologically 4-balls) to yield a combinatorial presentation of the 4-manifold, multiplying the path integrals on each pentachoron for fixed boundary conditions on that match along intersections, then summing over said boundary conditions. The complex numbers extracted from the spherical fusion 2-category (quantum dimensions and 10j symbols) are then precisely the finite collection of path integral results on the 4-ball needed for this recipe.

\section{TQFTs from fusion 2-categories}

\subsection{From state sum to TQFT}\label{from state sum to TQFT}

We now turn to showing that the Douglas-Reutter invariant without defects extends to a TQFT in the sense of Atiyah - that is, we should assign vector spaces $\mathcal{H}(M)$ to closed 3-manifolds and assign linear maps $U(C):\mathcal{H}(\partial_1C)\to\mathcal{H}(\partial_2C)$ to 4-dimensional cobordisms $C:\partial_1C\to\partial_2C$ s.t. $U(C)$ for a closed manifold is map $\mathbb{C}\to\mathbb{C}$ whose single matrix element is the Douglas-Reutter invariant. Our main idea will be to construct a (non-topological) QFT from the categorical data, then separate the data in that QFT into a "geometric sector" and a "topological sector", and finally show that they can be explicitly decoupled.

The procedure itself is an application of standard TQFT arguments to a new setting: in the known setting of gauge theory \cite{DW90} the procedure reduces to demanding the flatness of the connection using Gauss law constraints then carefully omitting anomalous representations. Morally, what we do here is take one of the two approaches to quantizing a gauge theory - we first quantize a (lattice-like) theory with "gauge redundancy", then apply constraints at the quantum level that impose "gauge invariance" (reducing the theory to to a small invariant subspace from the original large space of states), which in particular also turns the theory into a topological one (like the gauging in \cite{DW90}, \cite{Kit03}, \cite{LW05} and similar models). In the section on membrane-nets we will follow the opposite approach, first introducing the local constraints and then quantizing\footnote{Though unlike more standard cases, the gauge constraints involve superpositions of classical fields, so we will actually have to work with formal superpositions even at the "classical" level. The main difference is that we are taking a natural quotient rather than restricting to a subspace}.

\subsubsection{Defining the QFT}\label{defining the QFT}

Let $M$ be a closed (ordered, oriented) combinatorial 3-manifold. Recall that we use $\Gamma$ to denote $\mathcal{C}$-states on $M$ (which we sometimes call partial states), $M^i$ to denote the set of $j$-simplices in $M$, $o$ to denote the order on $M^0$, $\epsilon_o(\kappa)$ to denote the relative sign between the orientations induced on a $j$-simplex $\kappa$ by $o$ and by the orientation of $M$, and finally $V^\pm(\Gamma,\kappa)$ to denote the spaces of associators on a 3-simplex $\kappa$ for a given $\mathcal{C}$-state $\Gamma$ on $M$ (in other words the space of ways to complete the local partial state on $\partial\kappa$ to a full state on $\kappa$). When we want to separate the combinatorial structure from the topology in the notation, we will mention both $M$ (as a topological space) and the combinatorial structure $\Delta$ (useful for when we want to vary $\Delta$ for a fixed $M$). "combinatorial structure" and "triangulation" will implicitly refer from now on to both a choice of triangulation and a choice of order $o$ on the vertices. We can then define the vector space of "lattice states":
\begin{equation}
H(M)\coloneqq\bigoplus_{\Gamma} H(M,\Gamma)\coloneqq\bigoplus_{\Gamma}\bigotimes_{\kappa\in M^3}V^{\epsilon_o(\kappa)}(\Gamma,\kappa).
\end{equation}
$H(M)$ is canonically dual to the space associated to the orientation reverse $H(\overline{M})$ using the pairing $\bigoplus_\Gamma\bigotimes_\kappa\langle\cdot,\cdot\rangle_{\Gamma,\kappa}$\footnote{which is \textbf{not} the pairing Douglas and Reutter use for state sums with boundary, but rather the pairing appearing in the path integral construction for closed 4-manifolds that we presented earlier.}. Note that $H(M)$ is sensitive to changes in $M$'s combinatorial structure so we can't naively set $\mathcal{H}(M)=H(M)$, and while the pairing is non-degenerate it is not necessarily positive definite, so we refrain from calling $H(M)$ a "Hilbert" space.

To an (ordered oriented) combinatorial 4-cobordism $C:\partial_1C\to\partial_2C$ we can assign a linear map $\tilde{U}(C):H(\partial_1C)\to H(\partial_2C)$ (where we use the induced order, orientation and combinatorial structure on $\partial_iC$) which we define by describing its matrix elements $\langle b_2|\tilde{U}(C)|b_1\rangle$\footnote{where $\langle b_2|$ is the dual of $|b_2\rangle$ w.r.t the non-degenerate pairing $\langle\cdot,\cdot\rangle$ mentioned before.}, defined on each summand in $H$'s definition (indexed by $\Gamma|_{\partial C}$) as follows:

\begin{enumerate}
    \item Fix summands $H(\partial_iC, \Gamma_i)$ corresponding to partial states $\Gamma_i$ on $\partial_iC$. These summands decompose as tensor products $\bigotimes_{\kappa\in (\partial_iC)^3} V^{\epsilon_o(\kappa)}(\kappa,\Gamma_i)$
    \item The summands $H(\partial_iC, \Gamma_i)$ are spanned by the elements of the form $|b_i\rangle=\bigotimes_{\kappa\in (\partial_iC)^3}|b_{i,\kappa}\rangle\in H(\partial_iC,\Gamma_i)$ (where $|b_{i,\kappa}\rangle\in V^{\epsilon_o(\kappa)}(\kappa,\Gamma_i)$ may vary over some orthonormal basis of the 3-simplex spaces) which correspond to complete/classical states of the boundary 3-manifolds $\partial_i C$.
    \item To define $\langle b_2|U(C)|b_1\rangle$, compute the path integral on $C$ as for closed 4-manifolds except we fix labels on $\partial_iC$ to be $\Gamma_i,\ |b_i\rangle$ and take the square root of the $\phi$ weights of boundary simplices:
        \begin{equation}
            \langle b_2|U(C)|b_1\rangle\coloneqq\sum_{\Gamma\text{ s.t. }\Gamma|_{\partial_iC}=\Gamma_i}
                \biggl[
                  \prod_{v\notin(\partial C)^0}\phi\bigl(\mathcal{C}\bigr)\,
                  \prod_{e\notin(\partial C)^1}\phi\bigl(\Gamma(e)\bigr)\,
                  \prod_{s\notin(\partial C)^2}\phi\bigl(\Gamma(s)\bigr)
                \biggr]
                \tilde{Z}_\Gamma
        \end{equation}
        \begin{equation}\label{sqrt version}
            \tilde{Z}^{\frac12\frac12}_\Gamma \coloneqq \sqrt{
            \prod_{v\in(\partial C)^0}\phi\bigl(\mathcal{C}\bigr)\,
            \prod_{e\in(\partial C)^1}\phi\bigl(\Gamma(e)\bigr)\,
            \prod_{s\in(\partial C)^2}\phi\bigl(\Gamma(s)\bigr)}
            \prod_{\mu\in C^4}z(\Gamma,\mu)(\cup_{\Gamma|_{C\backslash\partial C}}(1)\otimes|b_1\rangle\otimes|b_2\rangle)
        \end{equation}
\end{enumerate}

One can think of the terms in the square root as (exponents of) boundary terms in the action, or just as part of the (exponent of the) usual bulk action with the square root being an artifact of the lattice approximation near the boundary (essentially halving the action of the boundary simplices since they only affect "half a spacetime portion" compared to the interior ones). Due to the local nature of the proof for the invariance of the path integral for closed 4-manifolds (see \cite{ReischI} or \cite{DR18}), we see that an important generalization immediately follows for the 4-manifold with boundary case: $\tilde{U}(C)$ is sensitive to the combinatorial structure of the boundaries $\partial_iC$ but not to the combinatorial structure of the interior $C\backslash\partial C$. By construction, $H,\tilde{U}$ also behave appropriately under composition of cobordisms:

\begin{equation}\label{combinatorial 1-functoriality}
     \tilde{U}(C\circ C')=\tilde{U}(C)\circ\tilde{U}(C')
\end{equation}
and disjoint unions:

\begin{equation}
    H(M\uplus M')=H(M)\otimes H(M'),\quad \tilde{U}(C\uplus C')=\tilde{U}(C)\otimes\tilde{U}(C').
\end{equation}
The $(H,\tilde{U})$ collections thus define the state spaces and time evolution operators (respectively) of a geometry-sensitive QFT\footnote{Or in fancier terms a 1-functor $\text{Bord}^\text{comb}_{4,3}\to\text{Vect}$, where $\text{Bord}^\text{comb}_{4,3}$ is the category of 4-cobordisms between 3-manifolds with combinatorial structure.}.
Note that the square root in \ref{sqrt version} is crucial for \ref{combinatorial 1-functoriality} to hold, since we get boundary terms from both $\tilde{U}(C),\tilde{U}(C')$ and their combined power has to sum to 1 for them to become bulk terms in $\tilde{U}(C\circ C')$. However, we could've also used different powers for the source and target:

\begin{equation}\label{general version}
\begin{split}
            \tilde{Z}^{\alpha\beta}_\Gamma \coloneqq \left(
            \prod_{v\in(\partial_1 C)^0}\phi\bigl(\mathcal{C}\bigr)\,
            \prod_{e\in(\partial_1 C)^1}\phi\bigl(\Gamma(e)\bigr)\,
            \prod_{s\in(\partial_1 C)^2}\phi\bigl(\Gamma(s)\bigr)\right)^\alpha\ \times\\
            \left(
            \prod_{v\in(\partial_2 C)^0}\phi\bigl(\mathcal{C}\bigr)\,
            \prod_{e\in(\partial_2 C)^1}\phi\bigl(\Gamma(e)\bigr)\,
            \prod_{s\in(\partial_2 C)^2}\phi\bigl(\Gamma(s)\bigr)\right)^\beta\ \times\\
            \prod_{\mu\in C^4}z(\Gamma,\mu)\left(\cup_{\Gamma|_{C\backslash\partial C}}(1)\otimes|b_1\rangle\otimes|b_2\rangle\right)
\end{split}
\end{equation}
so long as $\alpha+\beta=1$, defining an equivalent theory up to a (re)normalization of states in the state spaces. In practical calculations, we will work with $\alpha=0,\ \beta=1$. We will sometimes refer to the above as the \emph{normalized 10j action}, as in \cite{DR18}.
\\
\\
\textbf{Remark: pinched cobordisms}

We will also need the notion of a \emph{pinched cobordism} between (triangulated) 3-manifolds $M_1,M_2$, which is a 4-manifold $C$ with boundary $\partial_1C\cup_{\partial_\text{pinch}}\partial_2C$ for some $\partial_\text{pinch}C\subseteq\partial_iC\subseteq C$ (in other words, we allow $\partial_1C$ and $\partial_2C$ to share a subcomplex). Even more generally, $C$ could be a partial gluing of a 4-manifold $C_\text{4}$ to a (triangulated) 3-manifold $M$ along some gluing region $M_\text{glue}\subseteq M$, so that $C:M\to\left(\partial C_4\cup_{M_\text{glue}} M\right)\backslash \text{interior}\left(M_\text{glue}\right)$. As a particular example, $C_4$ could be a single pentachoron implementing a Pachner move on $M$, in which case $M_\text{glue}$ is the collection of input tetrahedra of the Pachner move and $\partial C_4\backslash\text{interior}\left(M_\text{glue}\right)$ is the output tetrahedra of the Pachner move, and both tetrahedra collections share the (2D) subcomplex $\partial M_\text{glue}$, and the $\partial_i C$ share both this (2D) subcomplex as well as the (closure of the) subomplex $M\backslash M_\text{glue}$. In this more general setting, we can still define the state sum given a pair of states on the $\partial_iC$ which agree along their shared subcomplex, which we define by the same formulae except simplices on the shared subcomplex contribute no $\phi$ weights.

\subsubsection{Decoupling geometry and topology}\label{decoupling}
The next goal should be to get rid of the boundary triangulations. To achieve this, we will separate the data in the triangulation-sensitive QFT $(H,\tilde{U})$ into a "pure geometry" part and a "pure topology" part, beginning with the former\footnote{In this specific case, "geometry" means combinatorial structure. However, the exact same argument works for extracting a TQFT from any QFT whose path integral depends only on the boundary geometry, for any notion of geometry.}.

Fix a closed oriented 3-manifold $M$ (without a combinatorial structure). For every pair of ordered combinatorial structures $\Delta_1,\Delta_2$ on $M$ denote by $I_M(\Delta_1,\Delta_2)$ the (ordered oriented) combinatorial 4-cobordism between combinatorial manifolds $(M,\Delta_1)\to(M,\Delta_2)$ whose underlying topology is that of a cylinder $I_M\coloneqq M\times[0,1]$ and whose combinatorial structure restricts to $\Delta_1,\Delta_2$ on $\partial_i I_M\coloneqq M\times\{i-1\}$ respectively, with an arbitrary combinatorial structure on the interior\footnote{Such a cylinder must exist by Pachner's theorem - we can always thicken a manifold into an identity cobordism with the same triangulation on both source and target (we show an explicit construction for this part in the next section, corresponding to the product of the to-be-defined $\berkanan_v$ operators), then we continue the thickening from the target 3-manifold by attaching 4-simplices corresponding to the Pachner moves that interpolate between $\Delta$ and $\Delta'$.}. For each such $\Delta_1,\Delta_2$ pair we can then define the map:

\begin{equation}
    \tilde{U}_{\Delta_2\Delta_1}\coloneqq\tilde{U}(I_M(\Delta_1,\Delta_2)):H(M,\Delta_1)\to H(M,\Delta_2)
\end{equation}
which depends only on $\Delta,\Delta'$ and not on the arbitrary choice of combinatorial structure in the interior of $I_M$.

Note that if $(H,\tilde{U})$ defined an honest TQFT then all of the $H(M,\Delta)$ spaces were identical for fixed $M$, and all the $\tilde{U}(\Delta_1,\Delta_2)$ time-evolution operators were the identity operator\footnote{Or possibly a projector, depending on the convention, though this more general case doesn't actually add anything as we'll soon see - the kernel of the projector would consist of states that are killed by all time evolution operator, hence are unphysical}. Further, given a 4-cobordism $C:\partial_1C\to\partial C_2$ and a pair of combinatorial structures $\Delta_i,\Delta_i'$ on its boundaries $\partial_i C$ we see that the time evolution operator $\tilde{U}(C_{\Delta_1\Delta_2})$ assigned to the $\Delta_i$ triangulations of $\partial_iC$ is related to the time evolution operator $\tilde{U}(C_{\Delta'_1\Delta'_2})$ assigned to the $\Delta'_i$ triangulations by postcomposing and precomposing $\tilde{U}(C_{\Delta_1\Delta_2})$ with the cylinder maps $\tilde{U}_{\Delta'_2\Delta_2},\tilde{U}_{\Delta_1\Delta_1'}$ to obtain $\tilde{U}(C_{\Delta'_1\Delta'_2})$:

\begin{equation} \label{Geotop}
    \tilde{U}(C_{\Delta'_1\Delta'_2})=\tilde{U}_{\Delta'_2\Delta_2}\circ\tilde{U}(C_{\Delta_1\Delta_2})\circ\tilde{U}_{\Delta_1\Delta'_1}
\end{equation}
and similarly in the other direction, where we used the fact that attaching a cylinder to a boundary doesn't change the topology. Thus, all of the geometry dependence is captured by the $\tilde{U}_{\Delta_2\Delta_1}$ operators and they don't interact with any kind of topological information by themselves.

Since the path integral depends only on the combinatorial structure of the boundaries and the composition of cylinders is a cylinder, the $\tilde{U}_{\Delta_2\Delta_1}$ operators satisfy:

\begin{equation}
    \tilde{U}_{\Delta_3\Delta_2}\circ \tilde{U}_{\Delta_2\Delta_1} = \tilde{U}_{\Delta_3\Delta_1}
\end{equation}
and in particular we see that every $\tilde{U}_{\Delta\Delta}$ is a projector onto $H_\Delta\coloneqq\text{Im}\tilde{U}_{\Delta\Delta}$. Further, by applying equation \ref{Geotop} for an arbitrary cobordism $C$ with boundary triangulations $\Delta_i$, we find that $\tilde{U}(C)$ is entirely supported on $H_{\Delta_1}$ and its image lies entirely in $H_{\Delta_2}$ since:

\begin{equation}
    \tilde{U}(C)=\tilde{U}(C)\circ\tilde{U}_{\Delta_1\Delta_1}=\tilde{U}_{\Delta_2\Delta_2}\circ\tilde{U}(C).
\end{equation}
Thus, we can restrict the domain and range of the $\tilde{U}$s to the $H_\Delta$s in a well defined way without losing any information regarding the "topological part" of the theory: states orthogonal to those in $H_\Delta(M)$ lie in the kernel of $\tilde{U}(C)$ for \emph{any} $C:M\to M'$ (so they are annihilated by any kind of time evolution hence are unphysical) and are orthogonal to \emph{all} states in the image of $\tilde{U}(C')$ for \emph{any} $C:M'\to M$ (so time evolution can't result in such a state). Finally, as a special case of \ref{Geotop} we see that:

\begin{equation}
    \tilde{U}_{\Delta_2\Delta_1}\circ \tilde{U}_{\Delta_1\Delta_2} = \tilde{U}_{\Delta_2\Delta_2} \quad \tilde{U}_{\Delta_1\Delta_2}\circ \tilde{U}_{\Delta_2\Delta_1} = \tilde{U}_{\Delta_1\Delta_1}
\end{equation}
so when we restrict the domain and range of the $\tilde{U}$s to the $H_\Delta$ subspaces, the maps:

\begin{equation}
    \tilde{U}_{\Delta_2\Delta_1}:H_{\Delta_1}\to H_{\Delta_2}
\end{equation}
become canonical isomorphisms that interpolate between the spaces associated with the projectors $\tilde{U}_{\Delta\Delta}$ and invert each other, hence restricting to the $H_\Delta$s gets rid of all the geometry dependence and preserves all of the topology dependence - the choice of $\Delta$ is entirely superficial, utilized only for obtaining an explicit presentation of the physical/invariant space of states. More formally, if we define $\mathcal{H}(M)\coloneqq\text{Im}\tilde{U}(I_M(\Delta,\Delta))\subseteq H(M,\Delta)$ (for arbitrary $\Delta$ on every $M$) and define $U(C)\coloneqq\tilde{U}(C)|_{\mathcal{H}(M)}$ then we just found that:

\[\text{$(\mathcal{H},U)$ furnishes a TQFT in the sense of Atiyah.}\]
\\
\\
\textbf{Remark: $U$ of a pinched cobordism}

Note that given a pinched cobordism $C$ (in the most general sense mentioned in the previous remark on pinched cobordisms), we can define the $U(C):\mathcal{H}(\partial_1C)\to\mathcal{H}(\partial_2C)$ maps using the corresponding state sum, which still satisfy $U(C'\cdot C)=U(C')\cdot U(C)$.

\subsubsection{Digression: flat connection on the moduli space of geometries}
Note that this approach gives us a family of "realizations" of the state spaces that is parametrized by the choice of geometry, and the $\tilde{U}_{\Delta_2\Delta_1}$s let us move between these spaces such that if we start at some $\Delta_1$ then move between different $H_{\Delta\_i}$ in an arbitrary "loop":

\begin{equation}
    H_{\Delta_1}\xrightarrow{\tilde{U}_{\Delta_2\Delta_1}} H_{\Delta_2}\xrightarrow{\tilde{U}_{\Delta_3\Delta_2}}\cdots\xrightarrow{\tilde{U}_{\Delta_n\Delta_{n-1}}} H_{\Delta_n}\xrightarrow{\tilde{U}_{\Delta_1\Delta_{n}}}H_{\Delta_1}
\end{equation}
we always get the identity map. This structure is very reminiscent of the flat connection on the moduli space of conformal structures on a surface induced by Chern-Simons theory on its boundary \cite{Wit89}, and the above argument indeed generalizes to that case (where "geometry" will now mean conformal structure instead of combinatorial structure). More mathematically, we can think of the collection of $H_\Delta$s as the objects of a category whose morphisms are the $\tilde{U}_{\Delta_2\Delta_1}$ and this category can be thought of as the fundamental groupoid of some simply connected space, or more strongly as a "contractible space of choices" (when we think of this as an $\infty$-groupoid) that we can safely ignore when by moving from the QFT to the TQFT using the $H_{\Delta}\to\mathcal{H}$ restrictions.

\subsection{The space of membrane nets}
\label{membranenets}
In this section we define an alternative definition for the state spaces of the TQFT, morally following the second approach to gauge theory, namely applying a quotient imposing invariance under local transformations on (superpositions of) "classical fields", representing a kind of non-invertible gauge transformation. The equivalence classes are referred to as \emph{membrane-nets}, inspired by \cite{LW05}'s naming for their lower dimensional \emph{string-nets}. We prove that the resulting state spaces are canonically isomorphic to $\mathcal{H}(M)$ spaces.

The arguments used here are a higher analog of an argument appearing in \cite{LW05} and its more careful refinement in \cite{Kir11}, though the transition from 3D to 4D happens to be significantly more rich and interesting: we have to define two new notions of "vacuum loops" or "Kirby colors" which interact with the higher categorical nature of our construction in many nontrivial ways, and our $\berkanan_v$ projectors (analogous to the $B_v$ operators of \cite{LW05},\cite{Kir11}) act in a significantly less trivial manner. We will later use these $\berkanan_v$ operators to introduce a tensor-network formulation of our TQFT, which is mostly a technical reorganization of our previous construction but should be useful for quantum error correction. The main definition introduces a new concept which we believe will be useful beyond our TQFT for more general settings such as anomalous TQFTs, non-compact TQFTs and further generalizations: a skein theory of surfaces based on monoidal 2-categories. We expect this to be a concrete and explicit manifestation of analogous constructions in the setting of factorization algebras/homology.

\subsubsection{Main definitions and goal}\label{main defs}

The $\mathcal{H}$ spaces admit a skein-theoretic description which is manifestly triangulation independent, constructed without the help of a combinatorial "scaffolding" from the very beginning. The resulting space can be seen as an analog of Levin and Wen's "string-nets", one dimension higher - instead of webs on surfaces related by local relations in disks, this space is defined using "foams" on 3-manifolds related by local relation in balls, which we refer to as \emph{membrane-nets}. A similar space can also be defined on 4-manifolds given a braiding on the input 2-category $\mathcal{C}$ (or equivalently a second compatible monoidal structure), which should yield many interesting invariants related to skein lasagna modules that we postpone to future work.

Let $M$ be a closed 3-manifold. We define the \emph{space of membrane nets} on $M$ as the quotient:

\begin{equation}
    \mathcal{H}^\text{mem}(M)\coloneqq H^\text{mem}(M)/D(M)
\end{equation}
where $H^\text{mem}(M),D(M)$ are defined as follows:

\begin{itemize}
    \item The (infinite dimensional) vector space $H^\text{mem}(M)$ is defined as the set of formal linear combinations of elements of $\text{Foam}(M)$\footnote{Another good notation for $H^\text{mem}(M)$ would be $\mathbb{C}[\text{Foam}(M)]$, the free span on $\text{Foam}(M)$.}, where the latter is the set of oriented dual cell decompositions of $M$ with sheets labeled by simple objects of $\mathcal{C}$, edges labeled by 1-morphisms and vertices labeled by 2-morphisms subject to certain conditions coming from $\mathcal{C}$'s 2-categorical structure, with higher dual cells playing no role (and will often not be included as part of the available data). The 2-morphism vertices come with closed neighborhoods and canonical homeomorphisms of those neighborhoods into the closed cube, generic in the sense that the homeomorphisms take the region around a 2-morphism vertex to generic diagrams as defined in \cite{BMS24}. The 1-,2-morphism labels are then required to match the sources and targets determined by those progressive diagrams. We also demand that every point in the interior of the lines and surfaces has a neighborhood homeomorphic to a 3-cube taking the foam in that neighborhood to a generic diagram representing an identity 2-morphism in the case of lines or an identity 2-morphism of an identity 1-morphism in the case of surfaces. These condition ensure that every ball in $M$ whose boundary intersects a foam generically has a homeomorphism with the unit cube defining a 2-morphism in $\mathcal{C}$ by evaluating a corresponding generic diagram. We will often soften the condition of being dual to a cell decomposition to allow for arbitrary decorated stratifications (again caring only about 0D,1D and 2D components), so long as the result restricts to a diagram of $\mathcal{C}$ in every 3-ball.
    \item The (infinite dimensional) subspace $D(M)\subseteq H^\text{mem}(M)$ is defined as the direct sum $\bigoplus_d D(d)$ over cube embeddings $d:D^3\to M$, where $D(d)$ is defined as the set of linear combinations $c_1\psi_1+\cdots+c_N\psi_n$ of foams $\psi_i$ s.t. the all the $\psi_i$ agree outside the interior of $\text{Im}(d)$, the pullbacks $d^{-1}(\psi_i)$ define generic diagrams, and the evaluations of those diagrams (denoted $\text{ev}_d(\psi_i)$) satisfy $\sum_ic_i\text{ev}_d(\psi_i)=0$. In essence, the space $D(d)$ was defined s.t. the quotient by it identifies (superpositions of) foams that differ by isotopies and local relations coming from equations among 2-morphisms in $\mathcal{C}$\footnote{Those local relations are a higher dimensional analog of skein relations on knots such as those obeyed by Wilson lines Chern-Simons theory (defining the Jones polynomial), where in that case "skein relations" refers to those coming from a ribbon category using Shum's theorem (of which there are infinitely many), rather than the use of the term in the context of knot polynomials (where a single (often qaudratic) skein relation is chosen such that it implies all of the other relations coming from the ribbon category).}.
\end{itemize}
We will call the elements of $D(M)$ \emph{fabric relations} or \emph{surface skein relations} and may sometimes call the space $\mathcal{H}^\text{mem}(M)$ the \emph{surface skein module} or the \emph{fabric module} assigned to $M$. The former space's elements should be thought of as kinds of gauge transformations. The latter space's elements will individually be referred to as membrane-nets, surface skeins or fabrics, and we should think of them as huge equivalence classes of (superpositions of) foams.

As an example for the kind of manipulations that the $D(M)$ quotient allows us to do, suppose we begin with the membrane-net represented by a foam that locally looks like:

\begin{equation}
    \mempty
\end{equation}
and suppose we draw a ball around one of the 2-morphism vertices, say:

\begin{equation}
    \memball.
\end{equation}
We can now interpret the interior of the ball as a 2-morphism in $\mathcal{C}$ between the 1-morphism diagrams on the ball's "southern" and "northern" hemispheres, and suppose the corresponding Hom space is two dimensional, then for every other pair of diagrams (say those shown in \ref{lineardependence}) in the same Hom space we can find some $\alpha,\beta,\gamma\in\mathbb{C}$ for which:

\begin{equation}\label{lineardependence}
    \alpha\ballone=\beta\balltwo+\gamma\ballthree
\end{equation}
and then the $D(M)$ quotient enforces:

\begin{equation}
     \alpha\memball = \beta\memballone + \gamma\memballtwo.
\end{equation}
We can thus perform such local surgeries on foams, allowing us to decompose complicated foams into a superposition of simpler ones. This is the essence of \emph{skein theory}, a collection of ideas and concepts that inspired the above construction. The physically-oriented reader might recognize this kind of argument from \cite{Wit89}.

In passing from $\mathcal{H}(M)$ to $\mathcal{H}^\text{mem}(M)$ we gain manifest topological invariance, but lose manifest finiteness. The goal of this section is to show that the two are canonically isomorphic (i.e. $\mathcal{H}(M)\cong\mathcal{H}^\text{mem}(M)$ in a way independent of the specific realization of $\mathcal{H}(M)$), and use the proof technique to construct a tensor network description of our TQFT, which we expect to provide a wide variety of quantum error correcting codes. The isomorphism between the two spaces will turn out to be the natural one, described explicitly as follows:

\begin{itemize}
    \item Begin with a state $|\psi\rangle\in\mathcal{H}(M)$.
    \item Pick a triangulation $\Delta$ on $M$, realizing $\mathcal{H}(M)$ as a subspace of $H(M,\Delta)$.
    \item Express $|\psi\rangle$ as a linear combination of states $c_1|\psi_1\rangle+\cdots+c_n|\psi_n\rangle$ each of which corresponds to a specific partial/$\mathcal{C}$-state on $(M,\Delta)$ (corresponding to the summands we used for defining $H(M,\Delta)$).
    \item For each $|\psi_i\rangle$, pick a foam whose underlying cell decomposition is Poincare-dual to $\Delta$ and whose labels are such that the label on a cell agrees with the label on the simplex it is dual to. Pass to the membrane-net equivalence class $\psi_i$ that this foam belongs to.
    \item The image of $|\psi\rangle$ under the $\mathcal{H}(M)\to\mathcal{H}^\text{mem}(M)$ isomorphism is defined to be (the equivalence class of) $c_1\psi_1+\cdots+c_n\psi_n$.
\end{itemize}

The above defines a map $\pi_\Delta:H(M,\Delta)\to\mathcal{H}^\text{mem}(M)$ for each $\Delta$ which we claim descends to an isomorphism $\pi_\Delta|_{\text{Im}U_{\Delta\Delta}}:\mathcal{H}(M)\xrightarrow{\cong}\mathcal{H}^\text{mem}(M)$, but even once we verify that these maps are indeed isomorphisms, a priori they could be \emph{distinct} isomorphisms, depending on the specific choice of $\Delta$. For this collection of isomorphisms to be independent of $\Delta$ (making the $\mathcal{H}(M)\cong\mathcal{H}^\text{mem}(M)$ isomorphism canonical), we must demand:
\begin{equation}\label{canonicity}
    \pi_\Delta=\pi_{\Delta'}\circ\tilde{U}_{\Delta'\Delta}
\end{equation}
as maps $H(M,\Delta)\to\mathcal{H}^\text{mem}(M)$. We refer to this property as the \emph{canonicity of the $\pi_\Delta$ maps}.

In the rest of this section we turn to verifying \ref{canonicity} and to proving that the $\pi|_{\text{Im}U_{\Delta\Delta}}$ maps are indeed isomorphisms, defining the new notion of higher Kirby colors in the process.

\subsubsection{$\mathcal{H}^\text{mem}_\Delta$ and the 1st transparency lemma}\label{H^mem_Delta}

Let $\Delta$ be a combinatorial structure on a closed oriented 3-manifold $M$. Recall that a partial/$\mathcal{C}$-state $\Gamma$ on $\Delta$ consists of maps $\gamma:\Delta^1\to\mathcal{C}^0$ and $\beta:\Delta^2\to\mathcal{C}^1$ where the output of $\beta$ is a simple 1-morphism $\beta(s)$ with source/target determined by the simple objects $\gamma(e'),\gamma(e''),\gamma(e''')$ assigned to  the three edges of $\partial s$. Given a partial state $\Gamma=(\gamma,\beta)$, we can define $H^\text{mem}_{\Delta^{0,1,2}}(M,\Gamma)$ to be the subspace spanned by foams that are Poincare-dual to $\Delta$ s.t. their labels match those of the corresponding dual simplices. Explicitly, this amounts to the following conditions on a foam $\psi$:

\begin{enumerate}
    \item $\psi$ does not intersect $\Delta^0$. In other words, it can be thought of as a foam in the punctured 3-manifold $M-\Delta^0$, so we may regard it as an element of $\text{Foam}(M-\Delta^0)\subseteq H^\text{mem}(M-\Delta^0)$.
    \item $\psi$ intersects $\Delta^1,\ \Delta^2$ generically (i.e. the object-sheets of $\psi$ intersect $\Delta^1$ transversaly, the 1-morphism lines in $\psi$ don't intersect $\Delta^1$ and the 1-morphism lines intersect $\Delta^2$ transversally).
    \item Given $e\in\Delta^1$, the intersection $\psi\cap e$ is a single point on the line $e$. That point inherits a label in $\mathcal{C}^0$ from the corresponding object-sheet of $\psi$ and we demand this label to agree with $\Gamma(e)$ or with $\Gamma(e)^\#$, depending on whether the orientation of the corresponding sheet agrees with the orientation of $e$ at the point of intersection\footnote{By "agreement" we mean that a positively-oriented tangent vector to $e$ at the point of intersection $p$ completes a 2-form representing the orientation of the corresponding sheet at $p$ to a 3-form lying in the equivalence class defining $M$'s orientation.} or not (respectively).
    \item Similarly to the previous condition, given some $s\in\Delta^2$, the intersection $\psi\cap s$ is required to contain a single point coming from a 1-morphism line, which then specifies a well defined 1-morphism $f\in\mathcal{C}^1$ by using either the label on the corresponding line or its adjoint (depending on whether the orientations agree or not, respectively). We require this 1-morphism $f$ to equal $\beta(s)$.
\end{enumerate}

After quotienting by the subspace of $D(M)$ consisting of local relations that don't violate these conditions\footnote{More specifically, we allow isotopies that don't violate the conditions at any point during the corresponding path of embeddings (not just at its beginning and end), and we allow all local relations happening in balls that don't intersect $\Delta^0\cup\Delta^1\cup\Delta^2$.}, we get a space which we denote by $\mathcal{H}^\text{mem}_{\Delta^{0,1,2}}(M,\Gamma)$. We can analogously define $\mathcal{H}^\text{mem}_{\Delta^{0,1}}(M,\gamma)$ by dropping all conditions involving the 2-simplices\footnote{or equivalently setting $\Delta^2=\emptyset$, which is no longer a valid triangulation but still makes sense in the above definitions.}, and similarly define $\mathcal{H}^\text{mem}_{\Delta^0}(M)\coloneqq\mathcal{H}^\text{mem}(M-\Delta^0)$.

Since the graphical calculus of monoidal 2-categories with duals represents the corresponding algebraic objects faithfully (see \cite{BMS24}), we can canonically identify the $H(\kappa,\Gamma)\coloneqq V^{\epsilon_o(\kappa)}(\Gamma,\kappa)$ spaces with $\mathcal{H}^\text{mem}_{\Delta^{0,1,2}}(\kappa,\Gamma|_\kappa)$. Under this identification we thus have:

\begin{equation}
    H(M)=\bigoplus_{\Gamma}\mathcal{H}^\text{mem}_{\Delta^{0,1,2}}(\kappa,\Gamma|_\kappa)=\bigoplus_{\gamma,\ \beta}\mathcal{H}^\text{mem}_{\Delta^{0,1,2}}(\kappa,\gamma|_\kappa,\beta|_\kappa).
\end{equation}
For later use, we define:

\begin{equation}
    \mathcal{H}^\text{mem}_{\Delta^{0,1,2}}(M,\gamma)\coloneqq\bigoplus_\beta\mathcal{H}^\text{mem}_{\Delta^{0,1,2}}(M,\gamma,\beta), \quad \mathcal{H}^\text{mem}_{\Delta^{0,1,2}}(M)\coloneqq\bigoplus_{\gamma,\ \beta}\mathcal{H}^\text{mem}_{\Delta^{0,1,2}}(M,\gamma,\beta).
\end{equation}

Now, let $\Delta$ be an arbitrary embedded cell complex in $M$ (not necessarily a triangulation) with 0-,1-,2-simplices and possibly some (irrelevant) 3-simplices. The above list of conditions makes sense for this more general case as well, and $\mathcal{H}^\text{mem}_{\Delta^{0,1}}(M),\mathcal{H}^\text{mem}_{\Delta^0}(M)$ are obtained from $\mathcal{H}^\text{mem}_{\Delta^{0,1,2}}(M)$ by removing the 2-simplices, 1-,2-simplices from $\Delta$ (respectively). Let:

\begin{equation}
    v\in\Delta^0,\ e\in\Delta^1,\ s\in\Delta^2
\end{equation}
and consider the embedded cell complexes $\Delta_v,\ \Delta_e,\ \Delta_s$ obtained by erasing $v,\ e,\ s$ (respectively) from $\Delta$. By a Cerf-theoretic argument\footnote{and similarly to lemma 5.5 in \cite{Kir11}}, we can obtain the spaces $\mathcal{H}^\text{mem}_{\Delta_s^{0,1,2}}(M,\Gamma|_{\Delta_s}),\ \mathcal{H}^\text{mem}_{\Delta_e^{0,1}}(M,\Gamma|_{\Delta_e}),\ \mathcal{H}^\text{mem}_{\Delta_v^0}(M)$ by quotienting by the following relations:

\begin{equation}
    \mathcal{H}^\text{mem}_{\Delta_s^{0,1,2}}(M,\Gamma|_{\Delta_s})=\bigoplus_{\Gamma(s)}\mathcal{H}^\text{mem}_{\Delta^{0,1,2}}(M,\Gamma)/\left\langle \rightvertexblue - \leftvertexblue \right\rangle
\end{equation}

\begin{equation}
    \mathcal{H}^\text{mem}_{\Delta_e^{0,1}}(M,\Gamma|_{\Delta_e})=\bigoplus_{\Gamma(e)}\mathcal{H}^\text{mem}_{\Delta^{0,1}}(M,\Gamma)/\left\langle \rightcurve - \leftcurve \right\rangle
\end{equation}

\begin{equation}
    \mathcal{H}^\text{mem}_{\Delta_v^0}(M)=\mathcal{H}^\text{mem}_{\Delta^0}(M)/\left\langle\shiftedfillupsphere-\shiftedfilldownsphere\right\rangle
\end{equation}
where the purple simplices on the r.h.s are the $s,e,v$ on the l.h.s and the gray sheets, black lines and blue vertices are labeled by arbitrary objects, 1-morphisms and 2-morphisms (respectively), and the $\bigoplus_{\Gamma(s)},\ \bigoplus_{\Gamma(e)}$ summations are over all the possible ways to complete $\Gamma|_{\Delta_s}$ to a partial state on $\Delta$ (so a sum over simple 1-morphisms in the case of $s$ and over simple objects in the $e$ case).

Our strategy will be to show that the $\tilde{U}:H\cong \mathcal{H}^\text{mem}_{\Delta^{0,1,2}}\to\mathcal{H}$ projectors from before factorize into a chain of projectors:

\begin{equation}
    \mathcal{H}^\text{mem}_{\Delta^{0,1,2}}\xrightarrow{\prod_{s\in\Delta^2}q_s}\mathcal{H}^\text{mem}_{\Delta^{0,1}}\xrightarrow{\prod_{e\in\Delta^1}q_e}\mathcal{H}^\text{mem}_{\Delta^{0}}\xrightarrow{\prod_{v\in\Delta^0}q_v}\mathcal{H}^\text{mem}
\end{equation}
where in each step a quotient map $q_\_$ corresponding to the above relations is applied (for instance $q_v$ removes the vertex $v$, $q_{e_1}q_{e_2}$ removes the two edges $e_1,e_2$). This factorization then yields a canonical isomorphism $\mathcal{H}^\text{mem}\cong\text{Im}\tilde{U}=\mathcal{H}$. We will construct operators corresponding to each step in this decomposition, whose main properties will be summarized by three \emph{transparency lemmas}, beginning with the:

\paragraph{Transparency lemma I} \emph{Let $\Delta$ be an embedded oriented cell complex in a closed oriented 3-manifold $M$ and let $s\in\Delta^2$, then there is a canonical isomorphism}:

\begin{equation}
    \mathcal{H}^\text{mem}_{\Delta^{0,1,2}_s}(M)\cong\mathcal{H}^\text{mem}_{\Delta^{0,1,2}}(M)
\end{equation}
\emph{and in particular we can identify $\mathcal{H}^\text{mem}_{\Delta^{0,1,2}}(M)$ with $\mathcal{H}^\text{mem}_{\Delta^{0,1}}(M)$.}

\textbf{proof:} The proof of the 1st transparency lemma is essentially an immediate consequence of lemma 4.3.7 from DR's paper. Consider a pair of 3-simplices $\kappa_1,\kappa_2\in\Delta^3$ in a triangulation $\Delta$ sharing a 2-simplex $s\subseteq\kappa_1\cap\kappa_2$, then given a membrane-net:

\begin{equation}
    \psi\in H(M,\Delta)=\mathcal{H}^\text{mem}_{\Delta^{0,1,2}}
\end{equation}
we can obtain an element of $\mathcal{H}^\text{mem}_{\Delta_s^{0,1,2}}$ by partially composing 2-morphisms along the $\Gamma(s)$ 1-morphism. This is precisely what the:

\begin{equation}
    \text{SC}^\pm:\bigoplus_{\Gamma(s)}V^\pm(\kappa_1,\Gamma)\otimes V^\pm(\kappa_2,\Gamma)\to V^\pm(\kappa_1\uplus_s\kappa_2,\Gamma)
\end{equation}
map defined in \cite{DR18} does. Their lemma 4.3.7 asserts that this map is invertible, meaning we can go in the other direction and uniquely factorize a membrane net in $\mathcal{H}_{\Delta_s^{0,1,2}}$ into a membrane net in $\mathcal{H}_{\Delta^{0,1,2}}$ by splitting the interior of $\kappa_1\cup\kappa_2$ into a sum over pairs of 2-morphism vertices (one in each $\kappa_i$) with a single simple 1-morphism line connecting each pair in the sum:

\begin{equation}
    \composetriangle = \sum_{A,f,g,\ \text{s.t.}\ g\circ_A f = h}\splittingtriangle
\end{equation}
thus the quotient map $\prod_{s\in\Delta^2} q_s$ is an isomorphism. Their proof for this lemma only used local semisimplicity to remove the 2-simplex $s$, valid even when $\Delta$ is not a triangulation with barely any modification. $\blacksquare$

Now that we proved transparency lemma 1, we can freely identify $\mathcal{H}^\text{mem}_{\Delta^{0,1,2}}(M)$ with $\mathcal{H}^\text{mem}_{\Delta^{0,1}}(M)$ and will do so from now on. Unlike the 1st one, the next two transparency lemmas involve nontrivial quotients and projectors, requiring a new tool to remove them: the \emph{local and global Kirby colors}.

\subsubsection{The local and global Kirby colors}
\label{KirbyDef}
For transparency lemmas 2 and 3, it will be helpful to define two generalizations of the "Kirby color" familiar from fusion 1-categories. Specifically, we let an oriented pink wire stitching an $A$-labeled sheet to a $B$-labeled sheet (such that the orientation of the pink wire and the sheets corresponds to a 1-morphism $A\to B)$ stand for a sum over simple labelings $f:A\to B$ of the wire, weighted by $\dim(f)$.
We refer to this superposition of labels as the \emph{local Kirby color}. Similarly, we let a pink sheet stand for a sum over simple objects $A$ weighted by $\frac{1}{d(A)}$. We refer to this superposition of objects as the \emph{global Kirby color}. We will often refer to such a pink wire as a \emph{Kirby wire} and denote its label by $\omega:A\to B$, and we will refer to such a pink sheet as a \emph{Kirby sheet} and denote its label by $\Omega$, i.e:
\begin{equation}
    \omega\coloneqq\sum_{f\in\text{Irr}(\text{Hom}(A,B))}\dim(f)f
\end{equation}
\begin{equation}
    \Omega\coloneqq\sum_{A\in\text{Irr}\mathcal{C}}d(A)^{-1}A.
\end{equation}
These formal superpositions of labels make sense as labels of sheets/wires in a membrane-net, since membrane-net spaces are (certain quotients of) free spans on foams. When multiple wires/sheets are Kirby-colored, we will by default assume a separate weighing and summation for each occurrence. However, sometimes we will wish several sheets/wires to share a summation index and weight, which we will call a \emph{correlated Kirby color}. For instance, a pair of correlated Kirby sheets stand for a summation:

\begin{equation}
    \KirKirsheets\underset{\text{correlated}}\coloneqq\sum_{A\in\text{Irr}\mathcal{C}}d(A)^{-1}\AAsheets
\end{equation}
while a pair of uncorrelated Kirby sheets:

\begin{equation}
    \KirKirsheets\underset{\text{uncorrelated}}\coloneqq\sum_{A,B\in\text{Irr}\mathcal{C}}d(A)^{-1}d(B)^{-1}\ABsheets
\end{equation}

The Kirby colors will be our main tool for removing the effect of $\Delta$'s edges and vertices. To use them, we will first need to prove some of their properties:

\paragraph{Properties of the local and global Kirby colors}\label{KirbProp}
\begin{enumerate}
    \item \emph{Let $A\in\mathcal{C}^0$ be an arbitrary object and let $B\in\mathcal{C}^0$ be a simple object, then for every nonzero simple 1-morphism $f:A\to B$ we have:}
        \begin{equation}
            \removablesummand = \frac{\dim(f)}{\dim(\iota\circ\rho)}\rhoiota
        \end{equation}
        \emph{where $A\xrightarrow{\rho} B^f\xrightarrow{\iota} A$ are some projection/inclusion 1-morphisms from the definition of the $\boxplus$ decomposition of $A$, $B^f$ is some canonical simple summand in $A$'s decomposition uniquely determined by $B,f$ and lying in $B$'s component\footnote{In the sense of \cite{DR18}.}, and $1_A\overset{p}{\Rightarrow}\iota\circ\rho\overset{i}{\Rightarrow}1_A$ are the projection/inclusion 2-morphisms provided by the $\oplus$ decomposition of $1_{1_A}$ into $\iota_j\circ\rho_j$s.
    \item A local Kirby loop with global Kirby interior satisfies}:
        \begin{equation}
            \removablekirby = \Asheet
        \end{equation}
    \item \emph{A local Kirby loop with global Kirby interior satisfies:}
        \begin{equation}
            \kirbyleftcurve = \kirbyrightcurve
        \end{equation}
        \emph{where $A,C\in\mathcal{C}^0$ are arbitrary objects,  $g:A\to C$ is an arbitrary 1-morphism and the cyan region contains arbitrary edges/membrane-nets/etc.}
    \item \emph{Given an $A$-sheet and a $B$-sheet for arbitrary $A,B\in\mathcal{C}^0$ connected by a ``wormhole" with a local Kirby loop drawn on the throat of the ``wormhole'' that bounds a global Kirby disk as drawn below, we can slit the throat:}
        \begin{equation}
            \throat = \afterslitting
        \end{equation}
    \item \emph{Given an $A$-sheet and a nearby global Kirby sphere with some cyan region in its interior containing arbitrary punctures/membrane-nets/etc, we can isotope some disk in the $A$-sheet through the sphere while keeping the disk's boundary fixed}\footnote{note that in the drawing it might look like the disk's boundary moved or as if the $A$-sheet remained fixed and the sphere moved. One should understand the square in the drawing as a subdisk of the disk that is actually being isotoped, so the squares in the images continue to some (fixed) large circle concentric with the (fixed) global Kirby sphere, with only the $A$-subdisk bounded by this large circle being isotoped. The isotoped region then traces a ball that engulfs the Kirby sphere as well as the cyan region.}:
    \begin{equation}
        \leftsphere = \rightsphere
    \end{equation}
    \item \emph{A global Kirby sphere with empty interior can be removed and replaced by a factor of $\dim(\mathcal{C})$}:
    \begin{equation}
        \removablesphere = \dim(\mathcal{C})\cdot1_{1_1}
    \end{equation}
\end{enumerate}

\textbf{proof of property 1:}

Let $A,B,f$ be as in property 1. By semisimplicity and the definition of $\boxplus$ there are 1-morphisms $A\xrightarrow{\rho_j} B_j\xrightarrow{\iota_j} A$ (where each $B_j$ is taken from our fixed set of simple objects) s.t.:

\begin{equation}
    1_A = \bigoplus_{j=1}^n\iota_j\circ\rho_j
\end{equation}
and s.t. there are 2-isomorphisms:

\begin{equation}
    \epsilon_j:\rho_j\circ\iota_j\Rightarrow 1_{B_j}.
\end{equation}
By the definition of $\oplus$ there are 2-morphisms
\begin{equation}
    1_A\overset{p_j}{\Rightarrow}\iota_j\circ\rho_j\overset{i_j}{\Rightarrow}1_A
\end{equation}
s.t. $p_j\cdot i_j=1_{\iota_j\circ\rho_j}$, $\sum_{j=1}^ni_j\cdot p_j=1_{1_A}$ and $\rho_k\circ\iota_j\cong 0$ for $j\neq k$. We thus have:
\begin{equation}
    f=f\circ1_A=\bigoplus_{j=1}^n\left(f\circ\iota_j\circ\rho_j\right).
\end{equation}
The above equation is a direct sum decomposition of $f$, and since $f$ is simple there is a unique $j$ for which $f\circ\iota_j\circ\rho_j$ is nonzero.  We will denote the corresponding $\iota_j,\rho_j,i_j,p_j,\epsilon_j$ by $\iota,\rho,i,p,\epsilon$ and the corresponding $B_j$ by $B^f$. In particular, we get that:
\begin{equation}\label{stuff}
    1_f\circ (i\cdot p)= 1_f
\end{equation}
We can therefore harmlessly horizontally precompose the $f$ loop on the l.h.s of property 1 with two copies of $i\cdot p$, then isotope one of the $i\cdot p$s to the right side of the $f$ loop, getting:

\begin{equation}
    \removablesummand = \doubleloop.
\end{equation}

Both directly above and directly below the loop we have $1_\iota\circ1_\rho\circ1_\iota\circ1_\rho$ so we can replace the middle $1_\rho\circ1_\iota$s by $\epsilon^{-1}\cdot\epsilon$ pairs, getting:

\begin{equation}\label{numerator}
    \removablesummand = \saddleiso = \frac{\text{Tr}\left(\nestedloops\right)}{\dim(B^f)}\poppingtwoendomorphism.
\end{equation}
where the right equality is a consequence of $1_{B^f}$'s simplicity. To compute the spherical trace in the numerator we use its graphical definition (a sphere closure) then pull the $\epsilon$-$\epsilon^{-1}$ pair away towards the other side of the sphere. Note that due to reversed orientations, the other side of the sphere is an $A$ disk containing the vertical composition $1_A\xRightarrow{\left(\epsilon^{-1}\right)^*}\iota\circ\rho\xRightarrow{\epsilon^*} 1_A$ (where $\epsilon^*$ is the mate of $\epsilon$ w.r.t to 1-morphism duality) which don't invert each other in general. However, these mates live in the 1-dimensional spaces\footnote{To see that they are indeed 1-dimensional, we can substitute the direct sum decomposition of $1_A$ to get $\text{Hom}(1_A,\iota\circ\rho)=\bigoplus_k\text{Hom}(\iota_k\circ\rho_k,\iota\circ\rho)\coloneqq \bigoplus_kV_k$. Then, a 2-morphism $\alpha\in V_k$ can be harmlessly precomposed and postcomposed with $1_A=\bigoplus_m\iota_m\circ\rho_m$, yielding a decomposition $\alpha=\sum_{m,n}\alpha_{m,n}=\sum_{m,n}\left(i_m\cdot p_m\right)\circ\alpha\circ\left(i_n\cdot p_n\right)$, whose source $\iota_m\circ\rho_m\circ\iota_k\circ\rho_k\circ\iota_n\circ\rho_n$ is nonzero iff $m=k=n$ and whose target is nonzero iff $m=k_0=n$ (and similarly for $\iota$) where $k_0$ is the value for which $\rho_{k_0}=\rho,\iota_{k_0}=\iota$, hence $V_k=\{0\}$ for all $k$ except $k_0$, in which case $V_{k_0}\cong\text{End}(\iota_{k_0}\circ\rho_{k_0}\circ\iota_{k_0}\circ\rho_{k_0}\circ\iota_{k_0}\circ\rho_{k_0})$ (since only the $\alpha_{k_0k_0}$ components are nonzero). Conjugating by $\epsilon_{k_0}$ yields a canonical isomorphism $V_{k_0}\cong\text{End}(1_{B^f}\circ1_{B^f}\circ1_{B^f})=\text{End}(1_{B^f})=\mathbb{C}$.} $\text{Hom}(1_A,\iota\circ\rho),\ \text{Hom}(\iota\circ\rho,1_A)$ which are spanned by $p,i$ (respectively), hence $\epsilon^*\cdot\left(\epsilon^{-1}\right)^*=a\left(i\cdot p\right)$ for some $a\in\mathbb{C}$. Combined with \ref{stuff}, we get:

\begin{equation}
    \dim\left( \nestedloops \right) = \twoloopsvariant = a\cdot\dimfvariant =a\cdot\dim(f).
\end{equation}
To compute $a$, we can take the trace of its defining equation, getting:

\begin{equation}
    a=\frac{\text{Tr}\left(\epsilon^*\cdot\left(\epsilon^{-1}\right)^*\right)}{\text{Tr}\left(i\cdot p\right)}=\frac{\text{Tr}\left(\epsilon\cdot\epsilon^{-1}\right)}{\text{Tr}\left(p\cdot i\right)}=\frac{\text{Tr}\left(1_{1_{B^f}}\right)}{\text{Tr}\left(1_{\iota\circ\rho}\right)}=\frac{\dim(B^f)}{\dim(\iota\circ\rho)}
\end{equation}
Substituting back into \ref{numerator}, we get:

\begin{equation}
    \removablesummand = \frac{a\dim(f)}{\dim(B^f)}\poppingtwoendomorphism = \frac{\dim(f)}{\dim(\iota\circ\rho)}\saddleisotwo.
\end{equation}

We can now replace the r.h.s by the same expression except we replace the $B^f$ inside the right loop by an arbitrary $B_j$ (and the $\rho,\iota,i,p$ by the corresponding $\rho_j,\iota_j,i_j,p_j$) and sum over $j$ (which we can freely do since only the $B^f$ term survives\footnote{See the proof of proposition 1.2.21 in \cite{DR18}.}), and by the definition of $\oplus$ this sum evaluates to $1_A$ so we can just remove the right loop, getting:

\begin{equation}
    \removablesummand = \frac{\dim(f)}{\dim(\iota\circ\rho)}\rhoiota.
\end{equation}

\textbf{proof of property 2:}

By property 1 and the definition of the Kirby colors:

\begin{equation}\label{originalequation}
    \removablekirby=\sum_{f,B}\frac{\dim(f)}{d(B)}\removablesummand = \sum_{f,B}\frac{\dim^2(f)}{\dim(\iota\circ\rho)d(B)}\rhoiota.
\end{equation}

Let us compute the sum over $f$ for some fixed simple $B$. First note that for all $f$, $B^f$ is in the same component as $B$. Suppose WLOG that in the decomposition of $A$ into simples $A=B_1\boxplus\cdots\boxplus B_n$, the first $m$ terms lie in the same component as $B$ and the rest don't, then the sum over $f$s splits into:

\begin{equation}\label{weighted sum over diagrams}
    \sum_{f:A\to B}\frac{\dim^2(f)}{\dim(\iota\circ\rho)d(B)}\rhoiota = \underset{\substack{1\leq j\leq m, \\ f:A\to B\text{ s.t. }B^f = B_j}}{\sum}
    \frac{\dim^2(f)}{\dim(\iota\circ\rho)d(B)}\rhoiota.
\end{equation}

The r.h.s of the above equation can be simplified by noting that:

\begin{equation}
    \text{Hom}(A,B) = \bigoplus^n_{j=1}\text{Hom}(B_j,B)=\bigoplus^m_{j=1}\text{Hom}(B_j,B)
\end{equation}
where the right equality follows from DR's definition of a component, and the left decomposition of Hom 1-categories comes from the following $\bigoplus$ decomposition of every $f:A\to B$:

\begin{equation}
    \left(1_f\circ\sum_{j=1}^n p_j\right):f \rightleftarrows \bigoplus^n_{j=1} f\circ \iota_j\circ \rho_j:\left(1_f\circ\sum_{j=1}^n i_j\right),
\end{equation}
which in particular implies that every simple $f:A\to B$ with $B^f=B_j$ is a direct sum of some canonically determined simple $f':B_j\to B$ (which is the unique nonzero $f\circ\iota_j$ term in the above decomposition, unique due to $f$'s simplicity) and zero 1-morphisms $0_{k\neq j}:B_k\to B$ (which are the rest of the $f\circ\iota_j\circ\rho_j$s in the above decomposition). We then have:
\begin{equation}
    \dim(f)=\sum_k\dim(f\circ\iota_k\circ\rho_k)=\dim\left(f'\circ\rho_j\right)=\frac{\dim\left(f'\right)\dim\left(\rho_j\right)}{\dim\left(B_j\right)}
\end{equation}
where the last equality is proposition B.1 from \cite{DR18}, and the weighted sum over diagrams in \ref{weighted sum over diagrams} reduces to:

\begin{equation}
    \underset{\substack{1\leq j\leq m, \\ f:A\to B\text{ s.t. }B^f = B_j}}{\sum}
    \frac{\dim^2(f)}{\dim(\iota\circ\rho)d(B)}\rhoiota = \sum_{\substack{1\leq j\leq m, \\ f':B_j\to B}}\frac{\dim^2(f')\dim\left(\rho_j\right)^2}{\dim(\iota\circ\rho)\dim(B_j)^2d(B)}\bj.
\end{equation}
Again using B.1 from \cite{DR18}, we can rewrite the coefficient as:
\begin{equation}
    \frac{\dim^2(f')\dim\left(\rho_j\right)^2}{\dim(\iota\circ\rho)\dim(B_j)^2d(B)} = \frac{\dim^2(f')\dim\left(\rho_j\right)}{\dim(\iota)\dim(B_j)d(B)} 
\end{equation}
and since $\iota=\iota_j$ and $\rho_j$ is dual to $\iota_j$, they have the same dimension, hence:

\begin{equation}
    \frac{\dim^2(f')\dim\left(\rho_j\right)}{\dim(\iota)\dim(B_j)d(B)} = \frac{\dim^2(f')}{\dim(B_j)d(B)} 
\end{equation}

To substitute back into our original equation of interest (\ref{originalequation}), assume WLOG that for every simple $B\in\mathcal{C}^0$ there are integers $m_B,m_B'$ s.t. the summands of $A$ that lie in the same component as $B$ are exactly $B_{m_B},B_{m_B+1},\dots,B_{m_B'}$\footnote{The assumption here was only that the decomposition of $A$ is ordered into blocks corresponding to components, which of course can always be (re)arranged.}, then we get:

\begin{equation}
    \removablekirby = \sum_{\substack{B,\ m_B\leq j\leq m_B', \\ f':B_j\to B}}\frac{\dim^2(f')}{\dim(B_j)d(B)}\bj.
\end{equation}
where the first sum is over simple $B\in\mathcal{C}^0$. We can now employ the following change of summation order. The above sum can be rewritten as:

\begin{equation}\label{Rearrangement}
    \removablekirby = \sum_{\substack{[x]\in\pi_0\mathcal{C},\ m_x\leq j\leq m_x', \\ B\in[x],\ f':B_j\to B}}\frac{\dim^2(f')}{\dim(B_j)d(B)}\bj,
\end{equation}
so we now sum over components $[x]$ and then for every $j$ corresponding to a $B_j\in[x]$ we sum over simple $B\in[x]$ and simple $f:B_j\to B$. The advantage in that reordering is that now for fixed $[x]\in\pi_0\mathcal{C}$ and fixed $m_x\leq j\leq m'_x$ the weighted sum simplifies dramatically:

\begin{equation}\label{ind}
    \left(\sum_{B\in[x],\ f':B_j\to B}\frac{\dim^2(f')}{\dim(B_j)d(B)}\right)\bj = \bj
\end{equation}
where the identity we used is proposition B.4 from \cite{DR18}\footnote{we actually substituted the adjoint of the 1-morphisms into that identity, which doesn't affect its validity since $\dim(f)=\dim(f^*)$.}, and the bracketing is there only to stress the fact that the diagrams in \ref{ind} are $B,f'$-independent. We can then substitute back into \ref{Rearrangement} and use the definition of $\oplus$ to finish the proof:

\begin{equation}
    \smallremovablekirby = \sum_{[x]\in\pi_0\mathcal{C},\ m_x\leq j\leq m_x'}\smallbj = \sum_{j=1}^n\smallbj = \smallAsheet.
\end{equation}

\textbf{proof of property 3:}

Let $A,B,C,g$ be as in property 3, then we can plug in the definition of the local Kirby color:

\begin{equation}
    \localkirbyleftcurve = \sum_{f:C\to B} \dim(f)\summedleftcurve
\end{equation}
where the sum is over simple $f$ as usual. We can then use gluing rule I (specifically its "twining" direction) to insert an $\alpha$-$\overline{\alpha}$ pair (summing over possible $\alpha$s taken from an arbitrary basis of $\text{Hom}(f\circ g, f')$):

\begin{equation}
    \sum_{f:C\to B} \dim(f) \summedleftcurve = \sum_{\substack{f':A\to B, \ f:C\to B,\\ \alpha:f\circ g\Rightarrow f'}} \dim(f)\dim(f') \leftcollapse.
\end{equation}
Isotoping the $\alpha$-$\overline{\alpha}$ pair to the right gives (for every $f,f',\alpha$ separately):

\begin{equation}
    \leftcollapse = \rightcollapse.
\end{equation}
Finally, we can use gluing rule I again (this time its "untwining" direction) to obtain:
\begin{equation}
    \sum_{\substack{f':A\to B, \ f:C\to B,\\ \alpha:f\circ g\Rightarrow f'}} \dim(f)\dim(f') \rightcollapse = \sum_{f':A\to B} \dim(f') \summedrightcurve
\end{equation}
and once again using the definition of the local Kirby color:
\begin{equation}
    \sum_{f':A\to B} \dim(f') \summedrightcurve = \localkirbyrightcurve.
\end{equation}

Finally, by summing over simple $B$ with weight $\frac{1}{d(B)}$ we get:

\begin{equation}
\begin{split}
    \kirbyleftcurve=\sum_B\frac{1}{d(B)}\localkirbyleftcurve\\ =\sum_B\frac{1}{d(B)}\localkirbyrightcurve=\kirbyrightcurve
\end{split}
\end{equation}

\textbf{proof of property 4:}

This is a special case of property 2, obtained by plugging $A\Box B$ into its l.h.s then using the representation of $\Box$ in the graphical calculus of $\mathcal{C}$, as can be seen by projecting the wormhole picture onto a plane aligned with the $A,B$ sheets.

\textbf{proof of property 5:}

We begin by attaching the $A$ sheet to the global Kirby sphere using property 4, inserting a local Kirby loop with global Kirby source and target $\omega:\Omega\to\Omega$:

\begin{equation}
    \leftsphere = \leftmerge
\end{equation}
then we isotope the loop to the other side of the sphere and use property 4 again to sever the connection:

\begin{equation}
    \leftmerge = \rightmerge = \rightsphere.
\end{equation}

\textbf{proof of property 6}:

The global Kirby sphere has empty interior so we can apply a local relation by forming a closed ball around it and evaluating in $\mathcal{C}$, allowing us to replace it by any algebraically equivalent diagram. By definition of the global Kirby color, the evaluation yields a weighted sum of 2-endomorphisms $1_1\Rightarrow 1_1$ indexed by simple objects $A$ and weighted by $\frac{1}{d(A)}$. For a fixed $A$ the 2-endomorphism is precisely the diagram defining the spherical trace of $A$, so this 2-endomorphism is exactly $1_{1_1}\cdot\dim(A)$. Thus, we can replace the Kirby sphere by $1_{1_1}\cdot\sum_A\frac{\dim(A)}{d(A)}$. Plugging in the definition of $d(A)$ we find that the factor multiplying $1_{1_1}$ is:

\begin{equation}
    \sum_A\frac{\dim(A)}{d(A)} = \sum_A\frac{1}{\dim(\text{End}([A]))n([A])} = \sum_{[A]}\frac{1}{\dim(\text{End}([A]))} = \dim(\mathcal{C}).
\end{equation}
In particular, this means that an insertion of a global Kirby sphere with empty interior into another membrane-net simply multiplies it by $\dim(\mathcal{C})$.

\subsubsection{The 2nd and 3rd transparency lemmas and $\ehwaz,\pertho$ projectors}

Given a graph embedded in a closed 3-manifold $\Delta^{0,1}\subseteq M$, an edge $e\in\Delta^1$ and a simple object $A\in\mathcal{C}^0$, we can define an operator:

\begin{equation}
    \ehwaz_{e,A}:\bigoplus_{\gamma:\Delta^1\to\mathcal{C}^0\ \text{s.t. }\gamma(e)=A}\mathcal{H}^\text{mem}_{\Delta^{0,1}}(M,\gamma)\to\mathcal{H}^\text{mem}_{\Delta^{0,1}}(M)
\end{equation}
which takes a foam $\psi$ for which $\psi\cap e$ is an $A$-labeled sheet and conjures on that sheet a small $\omega:A\to \Omega$ loop around $e$. By $\oplus$ing over simple $A$s we get an operator:

\begin{equation}
    \ehwaz_e:\mathcal{H}^\text{mem}_{\Delta^{0,1}}(M)\to \mathcal{H}^\text{mem}_{\Delta^{0,1}}(M).
\end{equation}
Recall that $\mathcal{H}^\text{mem}_{\Delta_e^{0,1}}(M)$ is naturally a quotient of $\mathcal{H}^\text{mem}_{\Delta^{0,1}}(M)$. Let us denote the corresponding quotient map (which simply removes the $e$ intersection condition) by:

\begin{equation}
    q_e: \mathcal{H}^\text{mem}_{\Delta^{0,1}}(M) \to \mathcal{H}^\text{mem}_{\Delta_e^{0,1}}(M) \cong \mathcal{H}^\text{mem}_{\Delta^{0,1}}(M)/\left\langle \rightcurve - \leftcurve \right\rangle.
\end{equation}
We can now state:

\paragraph{Transparency lemma II} \emph{Let $\Delta^{0,1}$ be a graph embedded in a closed oriented 3-manifod $M$ and let $e\in\Delta^1$, then $\ehwaz_e:\mathcal{H}^\text{mem}_{\Delta^{0,1}}(M)\to \mathcal{H}^\text{mem}_{\Delta^{0,1}}(M)$ is a projector onto a subspace which surjects isomorphically onto $\mathcal{H}^\text{mem}_{\Delta_e^{0,1}}(M)$ under the canonical surjection $q_e:\mathcal{H}^\text{mem}_{\Delta^{0,1}}(M)\hookrightarrow \mathcal{H}^\text{mem}_{\Delta_e^{0,1}}(M)$. Further, the various $\ehwaz_e$ projectors commute for different $e$, so $\ehwaz\coloneqq \prod_e\ehwaz_e$ is a projector onto (a subspace canonically isomorphic to) $\mathcal{H}^\text{mem}_{\Delta^0}(M)$.}

\textbf{proof}: The proof essentially mimics the proof of theorem 3.9 from \cite{Kir11}. First, note that $\ehwaz_e$ commutes with $\ehwaz_{e'}$ for $e\neq e'$ by the definition of $\ehwaz_e$.

Now fix a specific $e\in\Delta^1$. To see that $\ehwaz_e$ is a projector, observe that $\ehwaz_e^2$ surrounds $e$ by two concentric Kirby loops, and by property 3 of the Kirby colors we can pull the outer loop away from $e$ and the internal loop. The outer loop then has a global Kirby disk as its interior (and nothing other than that) so we can use property 2 to remove it. We are left with a single local Kirby loop around $e$ with global Kirby interior, precisely the definition of $\ehwaz_e$.

To see that $\text{Im}(\ehwaz_e)$ is isomorphic to $\mathcal{H}^\text{mem}_{\Delta_e^{0,1}}(M)$, note that property 3 of the Kirby colors implies the inclusion:

\begin{equation}
    \text{ker}(q_e) = \left\langle \leftcurve - \rightcurve \right\rangle\subseteq\text{ker}(\ehwaz_e)
\end{equation}
so $\ehwaz_e$ descends to a well defined operator on $\mathcal{H}^\text{mem}_{\Delta_e^{0,1}}(M)$. By property 2, the descendant operator is the identity: $q_e(\ehwaz_e\psi)=q_e(\psi)$. This finishes the proof. $\blacksquare$

As a particularly important case, consider a rod $e\in\Delta^1$ going through a "wormhole throat" as in property 4 of the Kirby color, then by the argument for $\ehwaz_e^2=\ehwaz_e$ we see that:

\begin{equation}\label{wormhole rod}
    \ehwazvariation=\ehwaz_e\ehwazvariation\underset{q_e|_{\text{Im}\ehwaz}}\mapsto \postehwazvariation
\end{equation}
where we used property 4 of the Kirby colors after applying the $q_e|_{\text{Im}\ehwaz}$ isomorphism of transparency lemma II. We may picture this as collapsing the $\omega$ loop along the $\Omega$ disk it bounds, slitting the rod in the process. This picture will useful later.

Let us move on to Transparency lemma 3. Given a finite set of points embedded in a closed 3-manifold $\Delta^0\subseteq M$ and a point $v\in\Delta^0$, we can define an operator:

\begin{equation}
    \pertho_v:\mathcal{H}^\text{mem}_{\Delta^0}(M)\to\mathcal{H}^\text{mem}_{\Delta^0}(M)
\end{equation}
which takes a membrane-net $\psi$ that avoids $v$ and adds a small global Kirby sphere engulfing $v$ (and nothing other than $v$), then normalizes the membrane-net by $\frac{1}{\dim(\mathcal{C})}$. Recall that $\mathcal{H}^\text{mem}_{\Delta_v^0}(M)$ is naturally a quotient of $\mathcal{H}^\text{mem}_{\Delta^0}(M)$. Let us denote the corresponding quotient map (which simply embeds a membrane-net from $M-\Delta^0$ into $M-\Delta_v^0$) by:

\begin{equation}
    q_v: \mathcal{H}^\text{mem}_{\Delta^0}(M) \to \mathcal{H}^\text{mem}_{\Delta_v^0}(M) \cong \mathcal{H}^\text{mem}_{\Delta^0}(M)/\left\langle \shiftedfillupsphere - \shiftedfilldownsphere \right\rangle.
\end{equation}

We can now state:

\paragraph{Transparency lemma III} \emph{Let $\Delta^0$ be a finite set of points embedded in a closed oriented 3-manifold $M$ and let $v\in\Delta^0$, then $\pertho_v:\mathcal{H}^\text{mem}_{\Delta^0}(M)\to\mathcal{H}^\text{mem}_{\Delta^0}(M)$ is a projector onto a subspace which surjects isomorphically onto $\mathcal{H}_{\Delta^0_v}^\text{mem}(M)$ under the canonical surjection $q_v:\mathcal{H}^\text{mem}_{\Delta^0}(M) \to \mathcal{H}^\text{mem}_{\Delta_v^0}(M)$. Further, these projectors commute, so $\pertho\coloneqq\prod_v\pertho$ is a projector onto (a subspace isomorphic to) $\mathcal{H}^\text{mem}(M)$.}

\textbf{proof}: The proof is a straightforward variation of the previous proof, using global Kirby spheres instead of local Kirby loops. First, note that $\pertho_v$ commutes with $\pertho_{v'}$ for $v\neq v'$ by the definition of $\pertho_v$.

Now fix a specific $v\in\Delta^0$. To see that $\pertho_v$ is a projector, observe that $\pertho_v^2$ engulfs $v$ by two concentric global Kirby spheres then normalizes by $\frac{1}{\dim(\mathcal{C})^2}$, and by property 5 of the Kirby colors we can pull the outer sphere through the inner one. The outer sphere then has empty interior so we can replace it by a factor of $\dim(\mathcal{C})$, leaving us exactly with the effect of a single $\pertho_v$.

Finally, to show that $\text{Im}\pertho_v\cong\mathcal{H}^\text{mem}_{\Delta^0_v}(M)$, notice that property 5 of the Kirby colors implies:

\begin{equation}
    \text{ker}(q_v)=\left\langle \shiftedfillupsphere - \shiftedfilldownsphere \right\rangle \subseteq \text{ker}(\pertho_v)
\end{equation}
so $\pertho_v$ descends to a well defined operator on $\mathcal{H}^\text{mem}_{\Delta^0_v}(M)$. The descendant operator simply embeds a membrane-net from $M-\Delta^0$ into $M-\Delta^0_v$ then inserts a global Kirby sphere normalized by $\frac{1}{\dim(\mathcal{C})}$ around the position of the removed vertex $v$. By property 6, the latter sphere insertion and the normalization cancel each other, so the descendant operator is the identity.
\\
\\
\textbf{Remark: $\pertho_v$ in the presence of edges}

While we have defined $\pertho_v$ as a projector $\mathcal{H}^\text{mem}_{\Delta^0}(M)\to\mathcal{H}^\text{mem}_{\Delta^0}(M)$, it can be generalized in a straightforward manner to a projector $\mathcal{H}^\text{mem}_{\Delta^{0,1}}(M)\to\mathcal{H}^\text{mem}_{\Delta^{0,1}}(M)$ so long as no edge of $\Delta^1$ intersects $v\in\Delta^0$. Similarly, $q_v$ can be generalized to the quotient map:
\begin{equation}
    q_v: \mathcal{H}^\text{mem}_{\Delta^{0,1}}(M) \to \mathcal{H}^\text{mem}_{\Delta_v^{0,1}}(M) \cong \mathcal{H}^\text{mem}_{\Delta^{0,1}}(M)/\left\langle \shiftedfillupsphere - \shiftedfilldownsphere \right\rangle
\end{equation}
in this setting (i.e. when no edge ends on $v$). Both generalizations are possible due to the inherent locality of $\pertho_v$ and $q_v$. With these more general definitions, we obtain a variation of transparency lemma III, which states that the image of $\pertho_v$ surjects isomorphically onto $\mathcal{H}^\text{mem}_{\Delta_v^{0,1}}=\text{Im}q_v$ under the canonical surjection $q_v$. When $\Delta^1$ is empty this reduces to our former definition of $\pertho_v$, and $\pertho_v$ commutes with $\ehwaz_e$ for all $e\in\Delta^1$ and $v$ that is disjoint from $\Delta^1$.

\textbf{Remark: implicit uses of the transparency lemmas as an annihlation event}

The transparency lemmas allow us to canonically identify the images of certain projectors\footnote{Somewhat vacuously, the identity projector in the case of transparency lemma I.} with $\mathcal{H}^\text{mem}_{\Delta^{0,1,2}}$ spaces. We will use this identification implicitly throughout the paper, which in the case of transparency lemmas II/III we picture as the corresponding Kirby loop/sphere "collapsing" onto the corresponding edge/vertex (respectively), the two annihilating in the process. An instance of this pictorial interpretation appeared in the special case of \ref{wormhole rod}.

\subsubsection{Canonicity of the $\pi_\Delta$ maps}
As mentioned in \ref{main defs}, we must verify \ref{canonicity} to truly claim a \emph{canonical} isomorphism $\mathcal{H}(M)\cong\mathcal{H}^\text{mem}(M)$. To see that this holds, let $|\psi\rangle\in H(M,\Delta)$ and let $\Delta'$ be a triangulation differing from $\Delta$ by a single Pachner move, meaning there are $k$ adjacent tetrahedra in $\Delta$ which form part of the boundary of a pentachoron $\kappa$ and $\Delta'$ is obtained from $\Delta$ by removing those tetrahedra and gluing along the new boundary the $5-k$ tetrahedra from the complement of the old $k$ tetrahedera in the boundary sphere of the pentachoron. Denote by $\Delta_\text{out}$ the embedded complex in $M$ obtained from $\Delta$ by removing those $k$ tetrahedra and all their internal simplices (not including the lower-dimensional simplices in $\Delta$ along which this collection of tetrahedra is glued to the rest of $\Delta$), or equivalently the embedded complex in $M$ obtained by removing from $\Delta'$ the $5-k$ tetrahedra forming the other side of the pentachoron boundary (again including internal simplices and not including outer ones). We can split $\pi_\Delta,\pi_
{\Delta'}$ into $\pi_\text{out}\circ\pi_{\Delta,\text{in}}$ and $\pi_\text{out}\circ\pi_{\Delta',\text{in}}$, where:
\begin{equation}
    \pi_{\Delta,\text{in}}:H(M,\Delta)\cong\mathcal{H}^\text{mem}_{\Delta^{0,1,2}}(M)\to\mathcal{H}^\text{mem}_{\Delta^{0,1,2}_\text{out}}(M)\leftarrow\mathcal{H}^\text{mem}_{\Delta'^{0,1,2}}(M)\cong H(M,\Delta'):\pi_{\Delta,\text{in}}
\end{equation}
are the canonical quotients (compositions of the corresponding $q_s,q_e,q_v$ maps) and similarly for $\pi_\text{out}:\mathcal{H}^\text{mem}_{\Delta^{0,1,2}_\text{out}}(M)\to\mathcal{H}^\text{mem}(M)$. It suffices to show that $\pi_{\Delta,\text{in}}=\pi_{\Delta',\text{in}}\circ\tilde{U}(\kappa)$ where $\tilde{U}(\kappa)$ is the usual state sum map applied to the 4-cobordism whose interior consists only of $\kappa$, constant on the rest of $M$ (see the remark at the end of \ref{defining the QFT}). These maps act as the identity on the tensor factors of $H(M,\Delta),\ H(M,\Delta')$ coming from tetrahedra in $\Delta_\text{out}$, so it suffices to fix a $\mathcal{C}$-state $\Gamma$ on $\partial\Delta_\text{in}=\partial\Delta'_\text{in}$ and consider the restriction $|\psi_\text{rest}\rangle$ of $|\psi\rangle$ to those tetrahedra, living in the space $H(B^3,\Delta_\text{in},\Gamma)$, obtained as in \ref{defining the QFT} by $\oplus$ing over interior $\mathcal{C}$-states but fixing the boundary $\Gamma$. The maps $\pi_{\Delta,\text{in}},\ \pi_{\Delta',\text{in}}\circ\tilde{U}(\kappa)$ then induce maps on the restrictions $H(B^3,\Delta_\text{in},\Gamma)\to \mathcal{H}^\text{mem}_{\partial\Delta_\text{in}}(B^3)\cong\text{Hom}_\mathcal{C}(1_1,\Gamma)$ (where in the last expression $\Gamma$ is treated as a 1-morphism by partially composing along edges) and it suffices to show that these induced maps agree. There are only 4 cases to check, corresponding to the values $k=1,2,3,4$, and in each case this just a short computation comparing a pair of 2-morphisms in $\mathcal{C}$. We show below the $k=1$ case, which is the most complicated one, with each decreasing value of $k$ removing some aspect of this computation (so we leave these simpler versions of the computation to the reader), with $k=4$ being trivial to check. Note that this implies \ref{canonicity}, since any pair of triangulations is related by a sequence of Pachner moves (see \cite{Pac90}, \cite{Pac91}) and $\tilde{U}_{\Delta_3\Delta_2}\tilde{U}_{\Delta_2\Delta_1}=\tilde{U}_{\Delta_3\Delta_1}$. One also has to treat vertex order changes, which are generated by local transpositions, and such transpositions can be implemented by passing (via Pachner moves) to a triangulation with completely disjoint vertices (say a small generic isotopy of the original one) then returning to the original one, applying the order change when the corresponding vertices are created during $k=1$ moves.

Consider then the $k=1$ case, in which case $|\psi_\text{rest}\rangle$ lives in a single associator space, hence can be depicted (with object sheets suppressed from the pictures) as:

\begin{equation}\label{short}
    \memepsilon,
\end{equation}
a graphical calculus expression in $\mathcal{C}$ that coincides with $\pi_{\Delta,\text{in}}|\psi_\text{rest}\rangle$. On the other hand, $\pi_{\Delta',\text{in}}\circ\tilde{U}(\kappa)|\psi_\text{rest}\rangle$ corresponding to the graphical calculus expression:
\begin{equation}\label{long}
    \sum_{\substack{A_{abj},A_{ack},A_{bc\ell},A_{jk\ell},\\ a,b,c,k,j,\ell,\\ \alpha,\beta,\gamma,\delta}}\frac{d(a)d(b)d(c)d(j)d(k)d(\ell)}{\dim(\mathcal{C})d(A_{abj})d(A_{ack})d(A_{bc\ell})d(A_{jk\ell})}\left\langle\memepsilon,\noepsilon\right\rangle\cdot\noepsilon
\end{equation}
where the sum is over simple objects, 1-morphisms and an arbitrary basis of the correpsonding associator spaces, and repeated 2-morphisms with opposite incoming/outgoing 1-morphism directions correspond to duals w.r.t to the non-degenerate pairings $\langle\_,\_\rangle\coloneqq\text{Tr}(\_\cdot\_)$. As in \ref{short}, we do not display object sheets, using the label $A_{\fehu_1\fehu_2\fehu_3}$ for the sheet bounded by the triangle formed by the 1-morphism wires $\fehu_1,\fehu_2,\fehu_3$.

We will now apply a sequence of identities to reduce \ref{long} to \ref{short}. First, using the cyclicity of the trace we can rewrite:
\begin{equation}
    \left\langle\memepsilon,\noepsilon\right\rangle=\left\langle\epsilondelta,\noepsilondelta\right\rangle=\left\langle\memdelta,\nodelta\right\rangle
\end{equation}
and plugging back into \ref{long} we get:
\begin{equation}\label{long2}
    \sum_{\substack{A_{abj},A_{ack},A_{bc\ell},A_{jk\ell},\\a,b,c,k,j,\ell,\\ \alpha,\beta,\gamma,\delta}}\frac{d(a)d(b)d(c)d(j)d(k)d(\ell)}{\dim(\mathcal{C})d(A_{abj})d(A_{ack})d(A_{bc\ell})d(A_{jk\ell})}\left\langle\memdelta,\nodelta\right\rangle\cdot\noepsilon
\end{equation}
then by the definition of duals w.r.t a pairing together with the multilinearity of compositions in a linear 2-category \ref{long2} can be reduced to:
\textbf{\begin{equation}\label{long3}
    \sum_{\substack{A_{abj},A_{ack},A_{bc\ell},A_{jk\ell},\\ a,b,c,k,j,\ell,\\ \alpha,\beta,\gamma}}\frac{d(a)d(b)d(c)d(j)d(k)d(\ell)}{\dim(\mathcal{C})d(A_{abj})d(A_{ack})d(A_{bc\ell})d(A_{jk\ell})}\cdot\merge.
\end{equation}}
We can now apply gluing rule I along the $k$ wire to get:
\textbf{\begin{equation}\label{long4}
    \sum_{\substack{A_{abj},A_{ack},A_{bc\ell},A_{jk\ell},\\ a,b,c,j,\ell,\\ \alpha,\gamma}}\frac{d(a)d(b)d(c)d(j)d(\ell)}{\dim(\mathcal{C})d(A_{abj})d(A_{ack})d(A_{bc\ell})d(A_{jk\ell})}\cdot\gluebeta
\end{equation}}
then we apply gluing rule II along the $\ell,j$ wires and the $A_{jkl}$ sheet:
\textbf{\begin{equation}\label{long5}
    \sum_{\substack{A_{abj},A_{ack},A_{bc\ell},\\ a,b,c,d,\\ f,g,h,\\ \alpha}}\frac{d(a)d(b)d(c)}{\dim(\mathcal{C})d(A_{abj})d(A_{ack})d(A_{bc\ell})}\cdot\gluegamma
\end{equation}}
then gluing rule II along the $b,c$ wires and the $A_{bc\ell}$ sheet:
\textbf{\begin{equation}\label{long6}
    \sum_{A_{abj},A_{ack},a}\frac{d(a)}{\dim\mathcal{C}d(A_{abj})d(A_{ack})}\cdot\gluealpha.
\end{equation}}
Finally, we use Kirby color property 2 on the $a$ loop, which due to the weighted summation becomes a Kirby loop bounding a pair of Kirby disks, hence can be removed and replaced by a single Kirby sphere. The Kirby sphere engulfs \ref{short} with nothing on its exterior, so by sphericality we contract it to infinity and remove it (using property 6), canceling the $\dim\mathcal{C}$ weight and remaining with \ref{short}.

\subsubsection{The $\text{\jrunes{ᛒ}}_v$ operator and the $\mathcal{H}\to\mathcal{H}^\text{mem}$ isomorphism}

\begin{figure}[H]
        \includegraphics[width=0.5\textwidth]{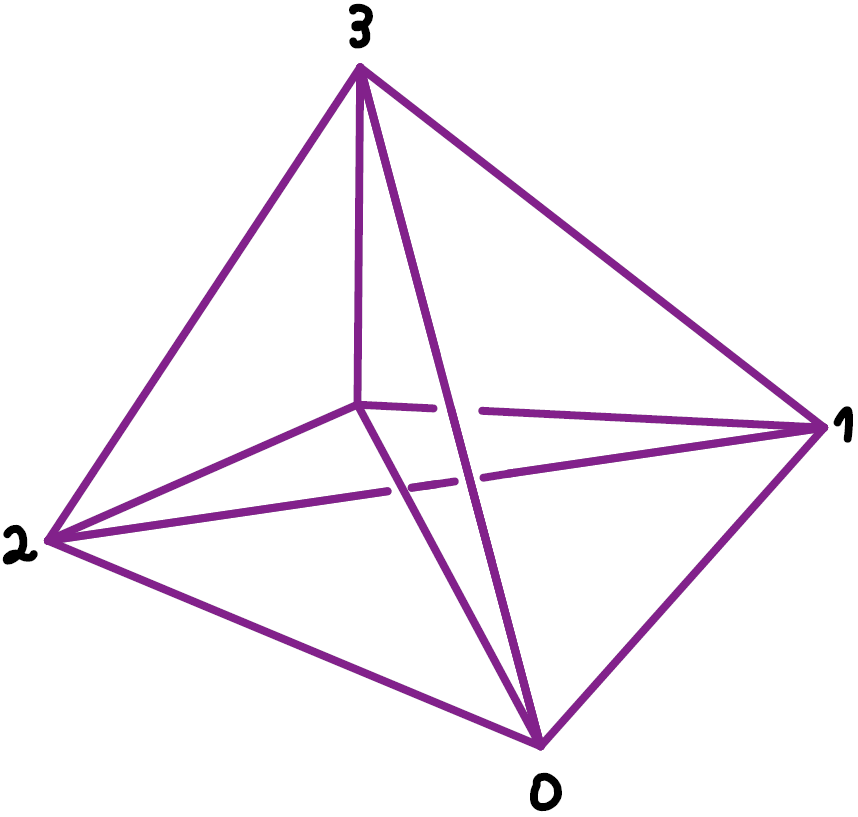}
        \centering{}\caption{The simplest example for the star $\overline{v}\coloneqq\text{St}(v)$ of a vertex $v\in\Delta^0$. The vertex $v$ is the interior vertex, and in general there can be many more edges coming out of it instead of 4 (and correspondingly many more tetrahedra in the interior and triangles/edges/vertices on the boundary $\partial\overline{v}=\text{Link}(v)$). We will denote the particular 3-complex depicted in this figure as $K$.}
        \label{fig:star}
\end{figure}

Now that we have the three transparency lemmas and canonicity at hand, we come to the statement and proof of our main theorem regarding membrane-nets. Given a closed 3-manifold $M$ with combinatorial structure $\Delta$ and a vertex $v\in\Delta^0$, we can define a convenient operator:
\begin{equation}
    \berkanan_v:H(M,\Delta)\to H(M,\Delta)
\end{equation}
as follows. Consider the star of $v$ (denoted $\text{St}(v)$, defined as the (closure of the) set of simplices in $\Delta$ that contain $v$). The simplest example is shown in figure \ref{fig:star}. Now fix a partial state $\Gamma$ on $M$. We will define the action of $\berkanan_v$ on the component\footnote{Where "component" refers to the tensor product decomposition we used for defining $H(M,\Delta,\Gamma)$, tensoring over all 3-simplices.}:
\begin{equation}
    H(\text{St}(v),\Gamma|_{\text{St}(v)})\hookrightarrow H(M,\Delta,\Gamma)\subseteq H(M,\Delta)\ ,
\end{equation}
extend it to the rest of $H(M,\Delta,\Gamma)$ by tensoring with identities on the other $H(\kappa,\Gamma)$s (for $\kappa\in\Delta^3\backslash\text{St}(v)$) and extend it to $H(M,\Delta)$ by $\oplus$ing over $\Gamma$. To define the action on $H(\text{St}(v),\Gamma|_{\text{St}(v)})$, consider the join $\overline{v}\coloneqq\text{St}(v)\star v'$ (where $v'$ is a single new vertex we introduce into the combinatorial 3-ball $\text{St}(v)$, visualized as a vertex deeper in the 4th dimension) and recall that the join of two complexes $A\star B$ is defined as the complex whose $(k+1)$-simplices are disjoint unions of a $(k-m)$-simplex in $A$ and an $m$-simplex in $B$ for any value of $m$, visualized as "coning $A$ over $B$". In the present setting, $\overline{v}=\text{St}(v)\star v'$ is a combinatorial 4-ball whose boundary naturally decomposes into $\text{St}(v)\cup_{\text{Link}(v)} \text{St}(v')$, a gluing of two isomorphic\footnote{Since $\text{St}(v)\cong\text{St}(v')$ as combinatorial manifolds, not just topological manifolds.} combinatorial 3-balls along their shared boundary $\partial\text{St}(v)=\partial\text{St}(v')=\text{Link}(v)$\footnote{Recall that the link of $v$ is defined as the subcomplex of $\text{St}(v)$ consisting of simplices that don't contain $v$.}. We can then define the action of $\berkanan_v$ on a state $|\psi\rangle\in H(\text{St}(v),\Gamma|_{\text{St}(v)})$ by defining the matrix element $\langle\phi|\berkanan_v|\psi\rangle$ for all $|\phi\rangle \in H(\text{St}(v),\tilde{\Gamma})$ where $\Gamma|_{\text{Link}(v')}=\tilde \Gamma$ fixes the partial state only on the boundary, $\oplus$ing over the rest of the labels. We let this matrix element be the normalized 10j action. Equivalently, we define it as the matrix element of $\tilde{U}$ on the 4-ball with boundary classical state fixed to be the gluing of $|\psi\rangle,|\phi\rangle$ along their shared $\tilde{\Gamma}$-labeled boundary.

\begin{figure}[H]
        \includegraphics[width=1\textwidth]{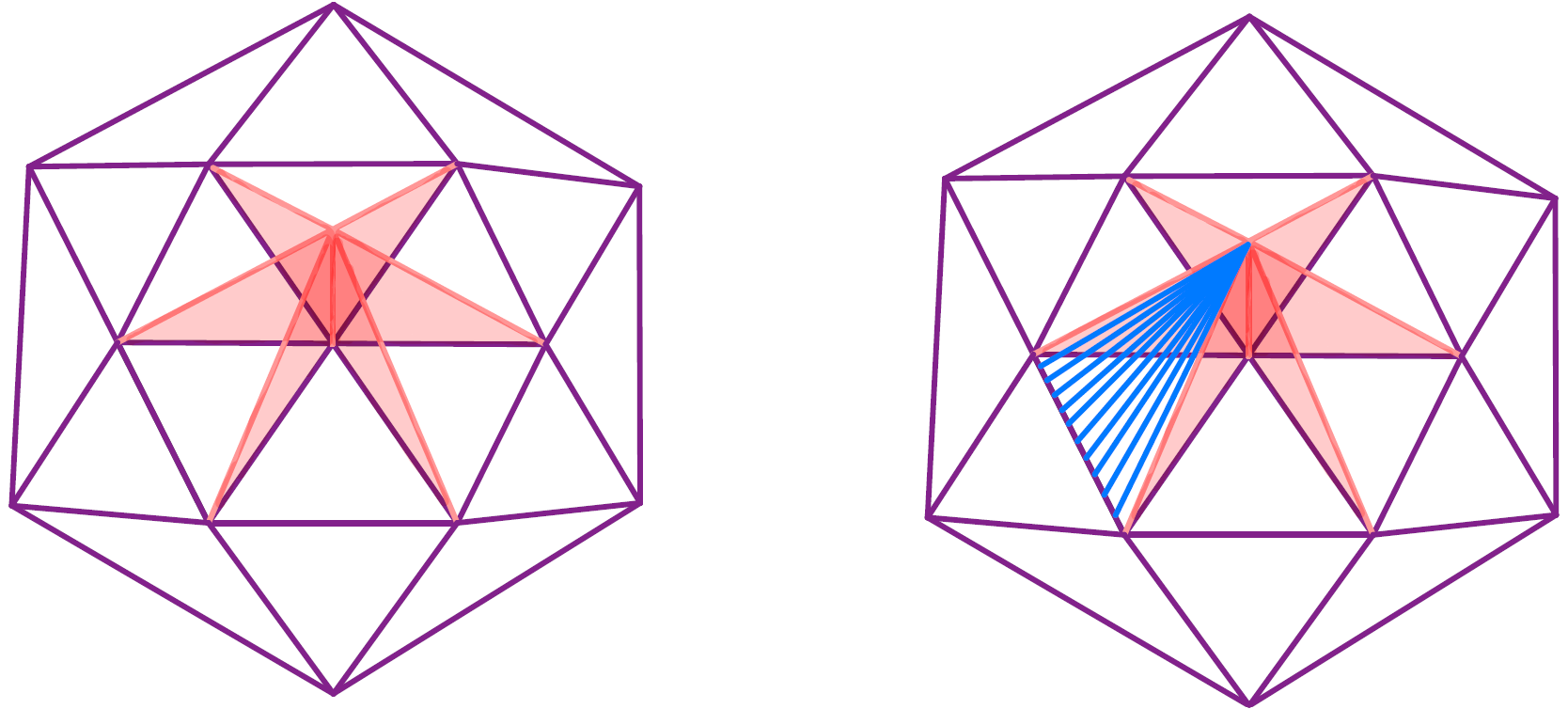}
        \centering{}\caption{Adding a 3D blister to a 2D manifold $M$ at the location of a vertex $v$. We begin with some triangulation of $M$ (in purple, left diagram). We add a vertex $v'$ on top of $v$ (the central vertex in the left hexagon is $v$ and the one slightly above it is $v'$) then cone into $v'$ the star of $v$ (where the latter is the hexagon around $v$), thereby forming a "blister" consisting of six 3-simplices. Describing the simplicial complex structure of the blister in detail, the edges of the blister are the six internal and six external purple edges of the hexagon around $v$ together with the six pink edges (overall 18 edges), the blister's internal 2-simplices are colored in transparent pink, its external 2-simplices are the six triangles making up the base hexagon together with the six triangles that close the top of the blister (one of which is colored in blue lines on the right diagram), and its six 3-simplices each have two pink boundary triangles, a boundary triangle from the original hexagon and a boundary triangle from the bulging hexagon on top of the blister (like the blue triangle).}\label{berkanan_v}
\end{figure}

The $\berkanan_v$ operator can be visualized as adding a 4D "blister" to $M$ popping out from $v$, then path-integrating the (normalized) 10j action inside the blister to move from the original flat 3-disk to the bulging 3-disk. Figure \ref{berkanan_v} shows a version of this 1 dimension down.

\begin{figure}[H]
        \includegraphics[width=1\textwidth]{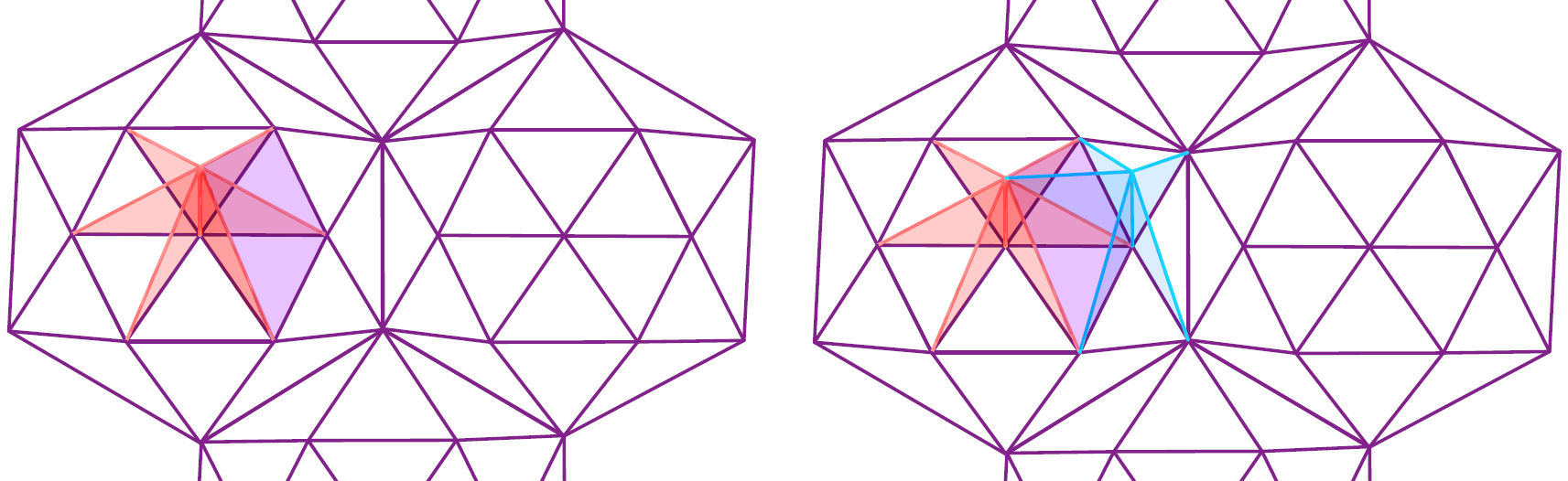}
        \centering{}\caption{Applying a $\berkanan_v$ (red) projector followed by a $\berkanan_u$ (blue) projector for adjacent $v,u$, in the notation of \ref{berkanan_v}. The overall resulting map is defined by computing the state sum of the thickening on the right picture, which is essentially a contraction of two "blisters" like the one from \ref{berkanan_v}, contracted along a pair of shared faces (purple). In figure \ref{berkanan commutation 2} we apply the same pair of operators in the opposite order, getting the same partially-thickened surface with the same source and target triangulations, with the only difference being the interior triangulation, hence we obtain the same linear map upon computing the state-sum.}\label{berkanan commutation 1}
\end{figure}

\begin{figure}[H]
        \includegraphics[width=1\textwidth]{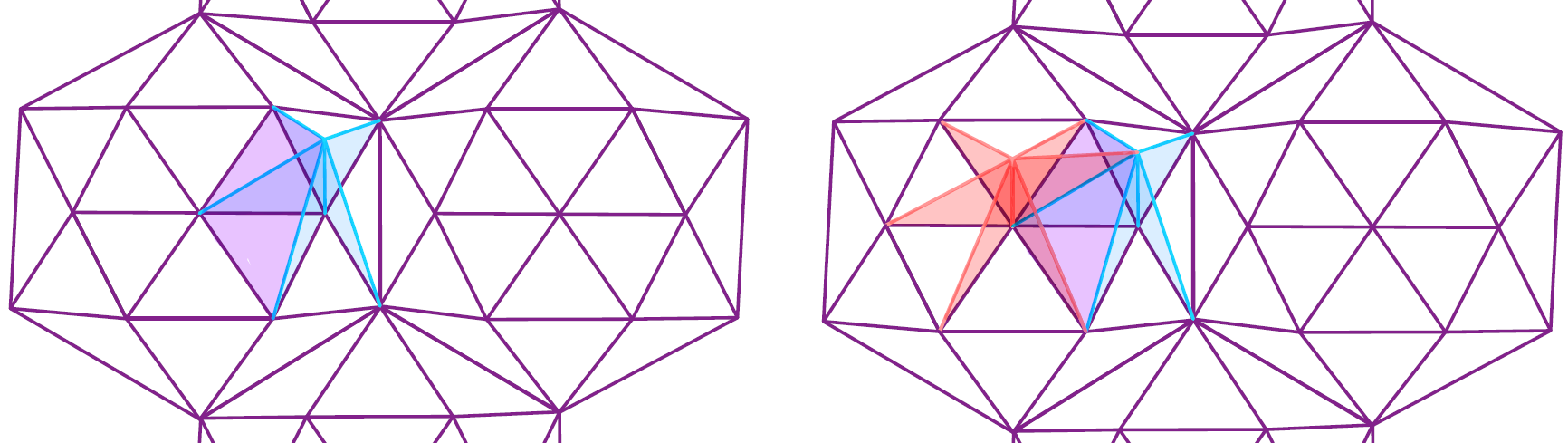}
        \centering{}\caption{Applying a $\berkanan_u$ (blue) projector followed by a $\berkanan_v$ (red) projector for adjacent $v,u$, in the notation of \ref{berkanan_v} and \ref{berkanan commutation 1}.}\label{berkanan commutation 2}
\end{figure}

Due to this description, $\berkanan_v$ is a projector and commutes with $\berkanan_u$ for $u\neq v$ (since the $\tilde{U}$ maps are only sensitive to boundary triangulation, see \ref{berkanan commutation 1}, \ref{berkanan commutation 2} for the only nontrivial case) and the product $\prod_{v\in\Delta^0}\berkanan_v$ corresponds to a thickening of the whole 3-manifold, so it's precisely the $\tilde{U}_{\Delta\Delta}$ projector taking $H(M,\Delta)$ to $\mathcal{H}(M)$. Our goal is thus to show that in terms of membrane-nets the $\berkanan_v$ operator removes the local constraints placed by the triangulation, which we know can be done with $\ehwaz_e,\pertho_p$ projectors, and this goal is captured by the following:

\paragraph{Theorem}\label{Main theorem}

\emph{Let $\Delta$ be a combinatorial structure on a closed 3-manifold $M$. The product $\prod_{v\in\Delta^0}\berkanan_v$ equals the projector $\left(\prod_{p\in\Delta^0}\pertho_p\right)\left(\prod_{e\in\Delta^1}\ehwaz_e\right)$ (where the $\pertho_p$s act on $\text{Im}\prod_e\ehwaz_e$ through its canonical identification with $\mathcal{H}^\text{mem}_{\Delta^0}$, see transparency lemma II), and in particular:}
\begin{equation}
    \mathcal{H}(M)=\text{Im}\tilde{U}_{\Delta\Delta} =\text{Im}\left(\prod_{v\in\Delta^0}\berkanan_v\right) = \text{Im} \left(\prod_{p\in\Delta^0}\pertho_p\right)\left(\prod_{e\in\Delta^1}\ehwaz_e\right) \cong \mathcal{H}^\text{mem}(M)
\end{equation}
We shall prove this theorem in the rest of the section.

\begin{figure}[H]
        \includegraphics[width=0.5\textwidth]{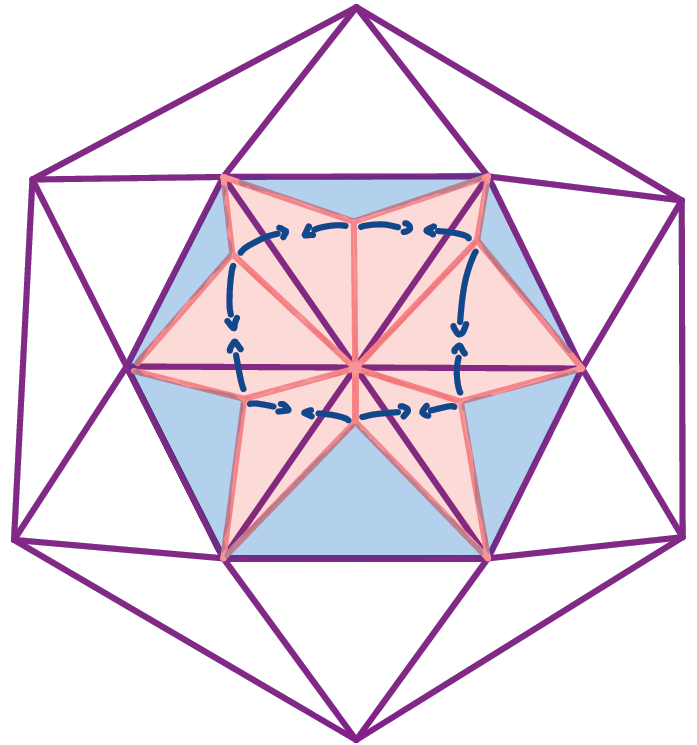}
        \centering{}\caption{The origami decomposition of $\berkanan_v$, or the \emph{polymaw} (due to its "many-toothed maw" resemblance), drawn one dimension down as in the last figure. To illustrate it in a concrete case, we begin with the purple hexagon from figure \ref{berkanan_v}, adding a 3-simplex (tetrahedron) on top of each of its six 2-simplices (triangles). The tetrahedra have a "base" triangle (from the original hexagon), a triangle facing radially out from $v$ (colored in blue), and two triangles facing neighboring tetrahedra (colored in pink). Note that in the 4D picture this will also hold but with three boundary 3-simplices facing the neighbors instead of two 2-simplices. We now glue each pink 2-simplex to the pink 2-simplex next to it as shown in the figure (dark blue arrows indicate the gluing direction), reconstructing the blister by "closing the maw". On the membrane-nets themselves this will correspond to contracting 2-morphism vertices. We are then left with six unglued/uncontracted 2-simplices at the base of the blister and six on top, where the former are regarded as the source of $\berkanan_v$ and the latter as the target.}\label{origami}
\end{figure}

We can compute $\berkanan_v$ by noting that we can equivalently attach on top of every 3-simplex $\kappa\in\text{St}(v)^3$ a 4-simplex $\tilde{\kappa}$ and regard it(s normalized pentagonator) as a map from $H(\kappa,\tilde{\Gamma}|_\kappa)$ to the tensor of the four spaces assigned to each 3-simplex in $\partial\tilde{\kappa}\backslash \kappa$ ($\oplus$ed over the labels on the new edges and faces), then we contract nearby 3-simplices as in the lower dimensional version shown in figure \ref{origami}, leaving a single uncontracted 3-simplex in each $\partial\tilde{\kappa}\backslash \kappa$, and when tensored these uncontracted 3-simplices carry our target/output state.

\begin{figure}[H]
        \includegraphics[width=1\textwidth]{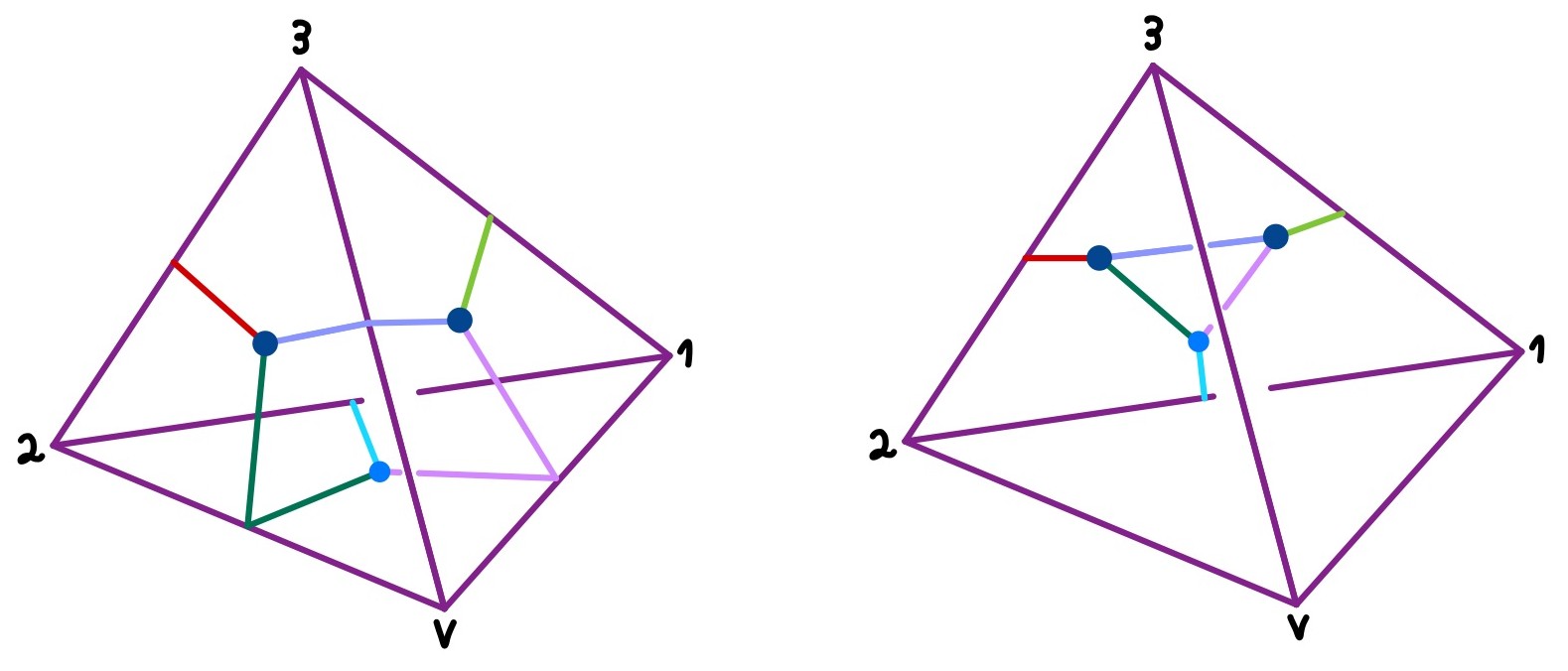}
        \centering{}\caption{This figure illustrates the intermediate step in the origami decompositions, which is the attaching of the 4-simplices. Once again we draw things one dimension down, so in particular we are drawing \emph{string nets} in a \emph{fusion 1-category}, which describes \emph{Turaev-Viro theory} instead of our TQFT, though we will still refer to elements in the drawing as their analogs in our setting. On the left picture, we have drawn the input 2-morphism sitting on the base of the hexagon as a bright blue vertex, and the contracted 2-morphisms as dark blue vertices. Fixing the values of the contracted vertices, the 4-simplex (3-simplex in the figure) takes the input to the partial composition shown on the right picture, sitting in the space assigned to the unglued 3-simplex (the 2-simplex not containing $v$ in the figure). This partial composition can be practically obtained by removing the vertex $v$ and every simplex that shares it in figure (while keeping the nets on top of them), then projecting the membrane-net (string-net in the figure) onto the unglued 3-simplex (2-simplex in the figure). From now on we leave the $2+1$D world in all figures, depicting our setting instead of Turaev-Viro theory.}\label{qua qua}
\end{figure}

We refer to this as the \emph{origami decomposition} of $\berkanan_v$\footnote{Another name suggested to us was "Biblically accurate Qua Qua", referring to the nested and convoluted description of the Cherubim, Ophanim and Seraphim in Isaiah 6 and Ezekiel 1 \cite{Tanakh}, juxtaposed with the popular Israeli origami game "Qua Qua", which produces simpler versions of the above pasting diagrams.}. To compute the state sitting on the unglued 3-simplices (which become the 3-simplices on the top of the blister $\text{St}(v')$ after gluing the rest of the non-source/base 3-simplices) from the state on the base of the blister (i.e. the input state of $\berkanan_v$), note that for a fixed choice of 2-morphisms on the to-be-glued 3-simplices (corresponding to pink 2-simplices in the above figure), the (normalized) pentagonator assigned to a 4-simplex (a 3-simplex popping out of the page in the above figure) can also be regarded as the non-degenerate trace pairing between the 2-morphism on its unglued 3-simplex and the partial composition of the 2-morphism sitting on the 3-simplex $\kappa$ at the base of the blister with the three fixed 2-morphisms on the to-be-contracted 3-simplices on top of $\kappa$, where the partial composition is along the 1-morphisms on the base 2-simplices (base edges in the above figure). This shows that before the summation over the contracted 2-morphisms (with this summation corresponding to the gluing of the origami decomposition), each 4-simplex sends the 2-morphism on its base 3-simplex $|\psi\rangle\in H(\kappa,\Gamma|_\kappa)$ to the same 2-morphism partially composed with all of the to-be-contracted 2-morphisms on top of it, where the latter partial composition sits on the unglued 3-simplex above $\kappa$ (which is the target of the 4-simplex map). The result is illustrated in figure \ref{qua qua}. After contracting the 2-morphism labels associated with the glued 3-simplices and normalizing as in the normalized 10j action, one obtains $\berkanan_v$.

\begin{figure}[H]
        \includegraphics[width=0.9\textwidth]{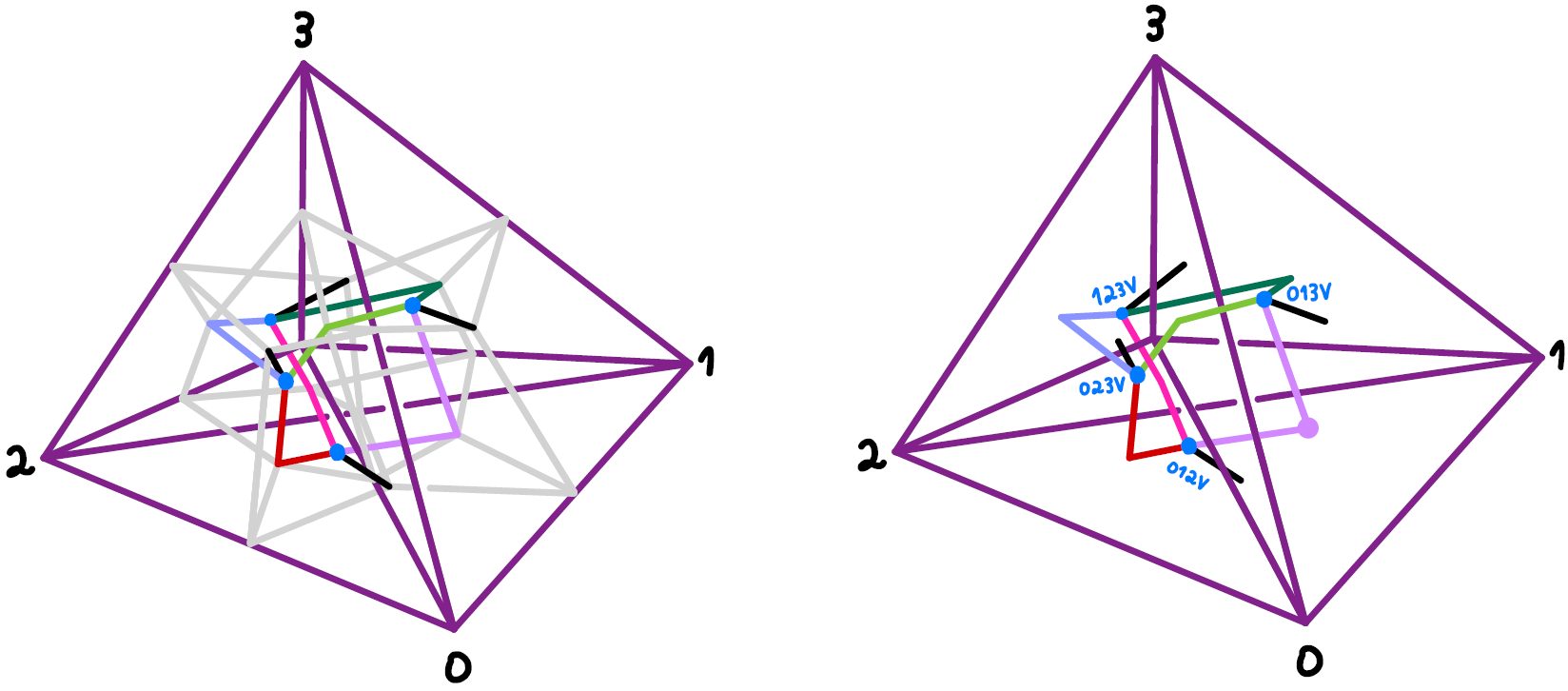}
        \centering{}\caption{A membrane-net in $H(D^3,K,\tilde{\Gamma})\cong\mathcal{H}^\text{mem}_{K^{0,1,2}}(D^3,\tilde{\Gamma})\cong\mathcal{H}^\text{mem}_{K^{0,1}}(D^3,\tilde{\Gamma})$, where $K$ is the combinatorial structure on the 3-disk $D^3$ depicted in the previous figure. Object sheets are colored in gray (and omitted in the right figure), external 1-morphism wires are colored in black, internal 1-morphism wires are colored in different colors (for later use), and 2-morphism vertices are colored in blue (labeled by the corresponding dual tetrahedra in the right figure). The 1-morphism and object labels on the external wires and sheets (respectively) are fixed by the partial state $\tilde{\Gamma}$ while the rest are arbitrary. There are 6 external and 4 internal sheets, 4 external and 6 internal wires, and 4 internal vertices. Three sheets end on each 1-morphism wire, and we only illustrate these trijunctions at the center of each of the 4 external triangles and 6 internal triangles (indicated in the right figure only by the sharp corners in the internal wires). Each triangle is Poincare dual to a unique wire (and vice versa) so each such trijunction specifies the source/target of a unique 1-morphism wire.}
        \label{fig:memstar}
\end{figure}

In terms of the membrane-nets themselves the origami decomposition results in a simple visual description of $\berkanan_v$'s action, which is essentially just the Poincare dual of the previous description. Given some specific vertex $v$, an element in $H(\text{St}(v),\Gamma|_{\text{St}(v)})$ is a membrane-net (simplest example illustrated in figure \ref{fig:memstar}) with some number of "external" 1-morphism wires ending on $\text{Link}(v)$, each such wire connects $\text{Link}(v)$ to a 2-morphism vertex and these 2-morphism vertices connect to three other vertices through internal wires. The external sheets and external 1-morphisms are fixed by $\tilde \Gamma$ and the rest are arbitrary (up to allowed compositions). With this canonical form for the input membrane-net, $\berkanan_v$ acts by replacing each 2-morphism vertex (which are always connected to four 1-morphism wires) by a small tetrahedron such that each 2-morphism vertex of this new tetrahedron is attached to a unique 1-morphism wire among the four that ended on the old vertex, and the other three 1-morphisms wires coming out of this new vertex are edges of the new small tetrahedron (in particular they are internal). The internal wires and internal sheets are local and global Kirby colored (respectively)\footnote{Up to a small but important subtlety that we describe later, since there are already a lot of details in this description.}, and the whole membrane-net is normalized by $\frac{1}{\dim(\mathcal{C})}$\footnote{Note that indeed the weights in the definition of the local and global Kirby colors are precisely the $\phi(e),\phi(s)$ labels from the state sum formula, and the $\frac{1}{\dim(\mathcal{C})}$ normalization is exactly the $\phi(\mathcal{C})$ from the state sum formula.}. The labels on new 2-morphism vertices on which external 1-morphism wires end are labeled by the same label as the old 2-morphism vertex that gave rise to the corresponding tetrahedron, and the rest of the vertices are contracted along local Kirby wires (as in gluing rule I).

In the next few figures we depict this description visually.

\begin{figure}[H]
        \includegraphics[width=0.5\textwidth]{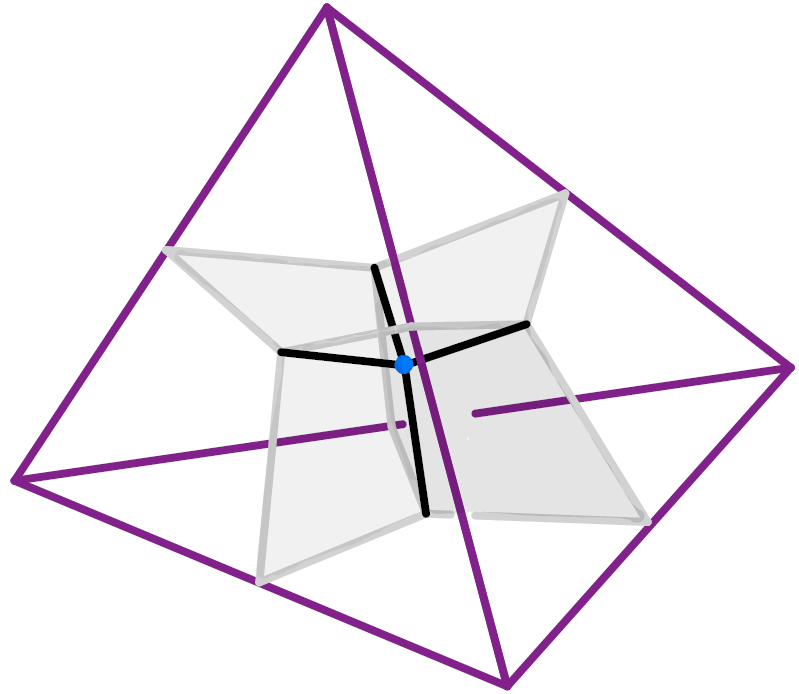}
        \centering{}\caption{A tetrahedron $\kappa$ (purple) and a membrane-net in $H(\kappa)$ (object sheets in gray, 1-morphism wires in black and 2-morphism vertices in blue). This canonical form for membrane-nets (which defined the $H$ spaces in the first place) can be seen as a labeling of the Poincare dual of $\kappa$ induced by a classical state on $\kappa$, with labels transferred from each simplex to the corresponding Poincare-dual cell.}
\end{figure}

\begin{figure}[H]
        \includegraphics[width=0.95\textwidth]{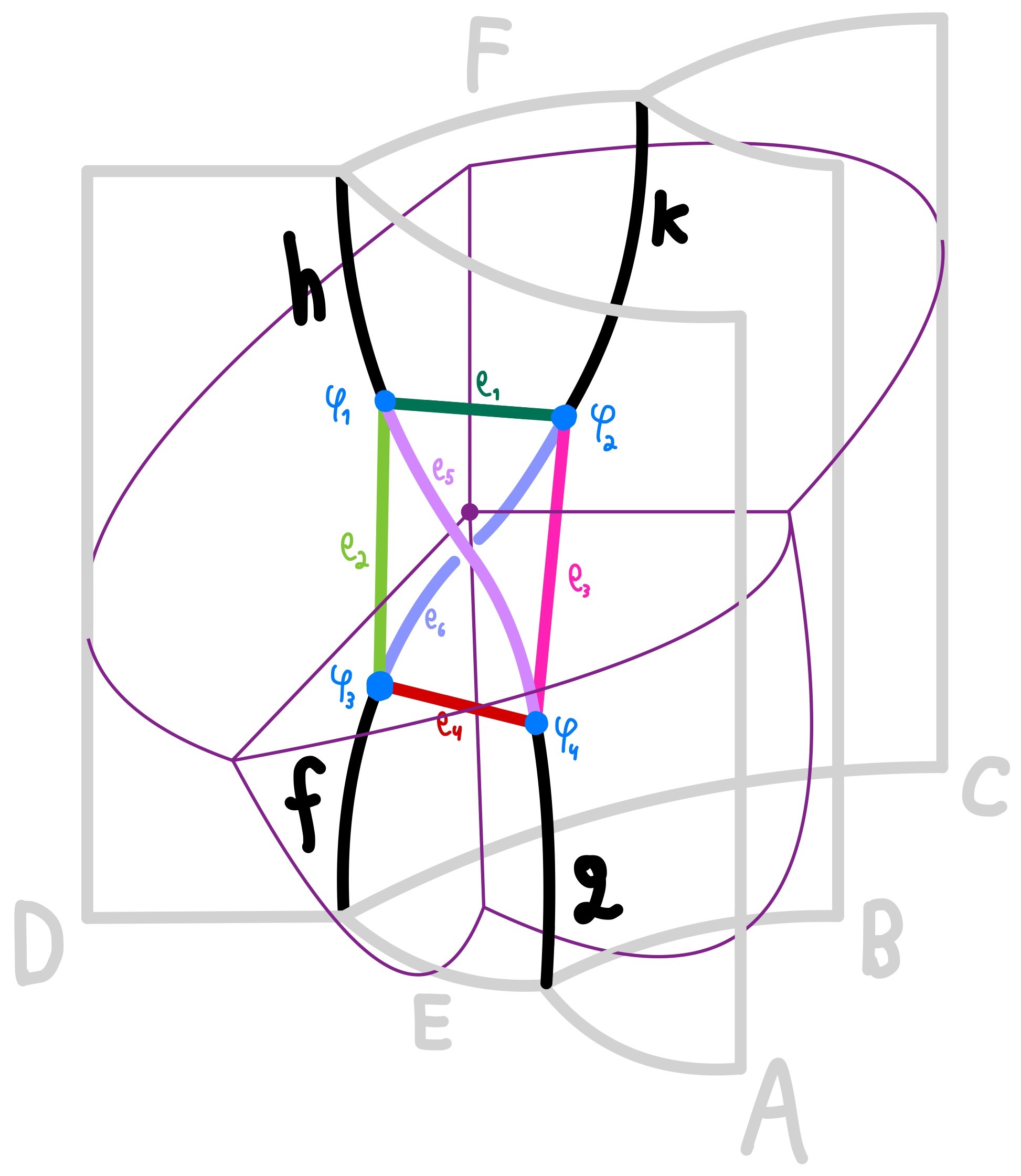}
        \centering{}\caption{The last figure is dense and complicated, and not in the ideal form for reading out the membrane-net sitting inside $K$. We thus deform the last figure into the shape shown in this figure. The objects $A,B,C,D,E,F$ are the external edge labels determined by $\tilde{\Gamma}$ and the 1-morphisms $f,g,h,k$ are the external triangle labels determined by $\tilde{\Gamma}$. We choose a specific classical state $|\psi\rangle$ (where classical means an element of $H$ instead of $\mathcal{H}$ as usual) by picking 1-morphism labels $e_1,\dots,e_6$ for the internal wires, 2-morphism labels $\varphi_1,\dots,\varphi_4$ for the internal vertices and simple object labels for internal sheets. The latter are not drawn in the figure to keep it from getting too cluttered, but there are four such sheets stretching between the triples of internal edges that form closed loops - the sheet ending on $e_1,e_3,e_5$ is labeled by some simple $t_{135}\in\mathcal{C}^0$, the $e_1,e_2,e_6$ sheet is labeled by some simple object $t_{256}$ and we similarly pick simple $t_{245},t_{346}\in\mathcal{C}^0$. We will work with this particular classical state from now on in the figures, aiming to describe $\berkanan_v|\psi\rangle$ in subsequent figures. As for the purple simplices, the purple vertex is $v$, the 4 purple edges coming out of it are the internal edges and so they each intersect a unique internal object sheet (by the Poincare duality), and the 6 purple edges between the other ends of the internal purple edges are the external purple edges, each of which pierces the boundary of a unique external sheet. One advantage of the deformation shown in this figure is that it presents the membrane-net in the form appropriate for the graphical calculus of fusion 2-categories, so we can use it to conveniently read off the source/target of the 1-,2-morphisms.}
        \label{fig:coloredexternal}
\end{figure}

From now on, let $v\in\Delta^0$ be a particular vertex, let $\tilde{\Gamma}$ be a partial state on $\text{Link}(v)$ and all the regions of $M$ outside $\text{St}(v)$ and let $|\psi\rangle\in H(\text{St}(v),\tilde{\Gamma})$ be a classical state obeying the boundary conditions determined by $\tilde{\Gamma}$. Figure \ref{fig:coloredexternal} depicts such a state in detail, and a version of it without the external edges is shown in the figure below:

\begin{figure}[H]
        \includegraphics[width=0.99\textwidth]{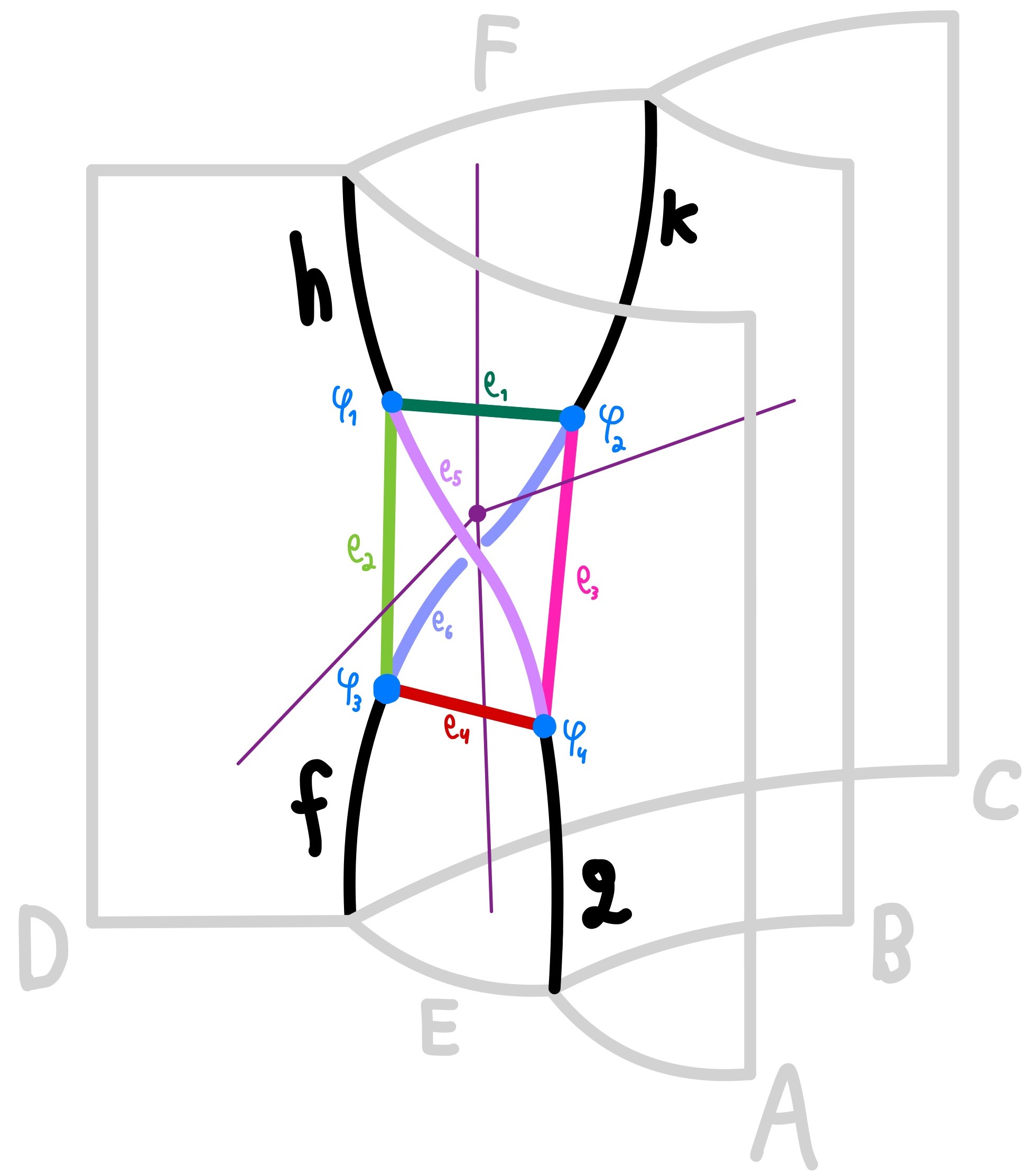}
        \centering{}\caption{The classical state $|\psi\rangle$ from figure \ref{fig:coloredexternal}, drawn without the external edges of $\text{St}(v)$.}
\end{figure}
According to the previous detailed description of $\berkanan_v$'s action on membrane-nets (inherited from the origami decomposition), we get the following form for $\berkanan_v|\psi\rangle$:

\begin{figure}[H]
        \includegraphics[width=0.77\textwidth]{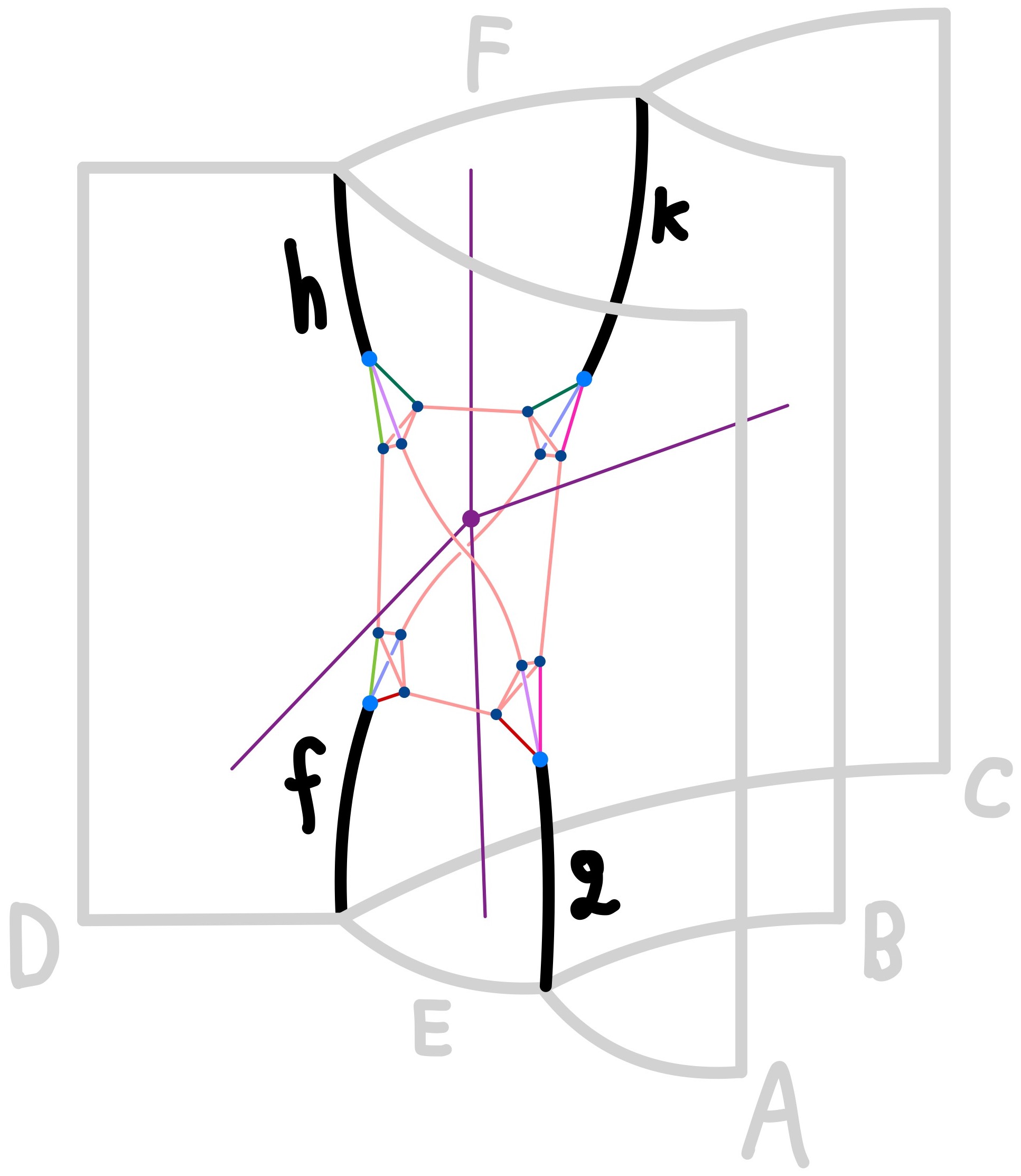}
        \centering{}\caption{The result of applying $\berkanan_v$ to the classical state $|\psi\rangle$ depicted in the previous figure. Each 2-morphism vertex is replaced by a small tetrahedron and all the internal labels change. The colored non-pink wires represent the same 1-morphism labels as on the corresponding colored wires in the previous figure (for instance the two green wires are labeled by $e_1$). Note that there are now two wires labeled by each of the $e_i$s instead of one. The pink wires connecting the small tetrahedra are local Kirby colored (i.e. $\omega$-colored). The bright blue vertices are labeled by the 2-morphisms on the corresponding vertex in the previous figure (the one that inflated into the corresponding tetrahedron in this figure). The internal object sheets are not shown in this figure (as in the previous two figures), but for every three 1-morphism wires forming a closed loop (a deformed triangle) there is a sheet ending on them, so all the 1-morphism loops bound object-labeled disks. The labels on sheets ending on two $e_i$s and one $\omega$ are fixed by the fact that the bright blue vertices correspond to the same 2-morphisms as before. For instance, the (unique) disk ending on $e_1,e_5$ and on a local Kirby wire (this disk is part of the small tetrahedron in the top left corner) is labeled by $t_{135}$, and so are the disks ending on $e_1,e_3,\omega$ (part of the top right tetrahedron) and $e_3,e_5,\omega$ (part of the bottom right tetrahedron). The sheets ending on one $e_i$ and two $\omega$s on the other hand are global Kirby colored. Finally, the dark blue vertices are contracted along $\omega$-wires in pairs, where each dark blue vertex is paired with the unique other dark blue vertex on a different tetrahedron that connects to it via an $\omega$ wire. The four sheets ending on three $\omega$s (bases of the tetrahedra) as well as the four large sheets ending on six $\omega$s are both $\Omega$-colored. Note that the four sheets ending on three $\omega$s are correlated Kirby sheets (see \ref{KirbyDef}) while the rest are uncorrelated, and similarly the three Kirby wires in each small tetrahedra are each correlated with a pair of Kirby wires in \emph{different} small tetrahedra, as required for the contraction of the 2-morphism vertices that they end on. An overall $\frac{1}{\dim(\mathcal{C})}$ normalizes the membrane-net.}
\end{figure}

\begin{figure}[H]
        \includegraphics[width=1\textwidth]{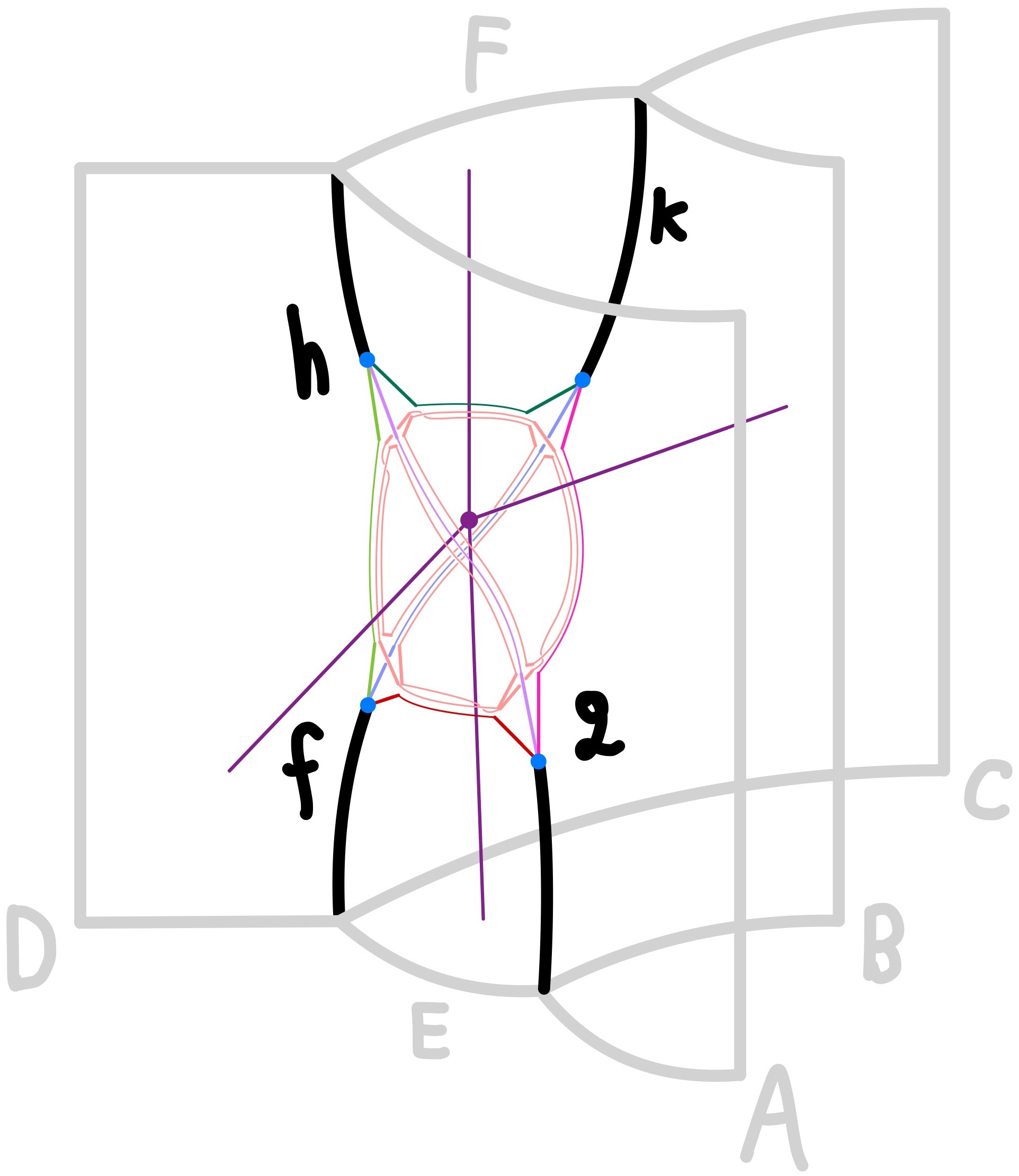}
        \centering{}\caption{Our next step is to use gluing rule I to perform a surgery on the dark blue vertex pairs, using its "untwining" version to remove those vertices and the 6 $\omega$-wires connecting different small tetrahedra. The outcome of this 6-fold untwining is shown in this figure. Note how the $\omega$-wires connecting the tetrahedra are now replaced by triples of wires forming long prism-shaped tubes. This surgery removes and reconnects wires, and as a result will stitch certain sheets to one another. Among the $\Omega$-sheets, only the four correlated ones are stitched together, so we are left with five uncorrelated Kirby sheets (note that they we indeed have the correct normalization for five $\Omega$s). Also, each of the $t$-sheet pairs is stitched to single $t$-sheets. Notice in this figure how every internal purple edge is surrounded by a unique $\omega$-loop - these loops exactly correspond to the triples of correlated $\omega$ wires, so all correlation disappears after the 6-fold surgery, hence the Kirby wires in this figure are genuinely local Kirby colored (instead of the correlated variant). The next figure will highlight some of the sheets in this figure.}
\end{figure}

\begin{figure}[H]
        \includegraphics[width=1\textwidth]{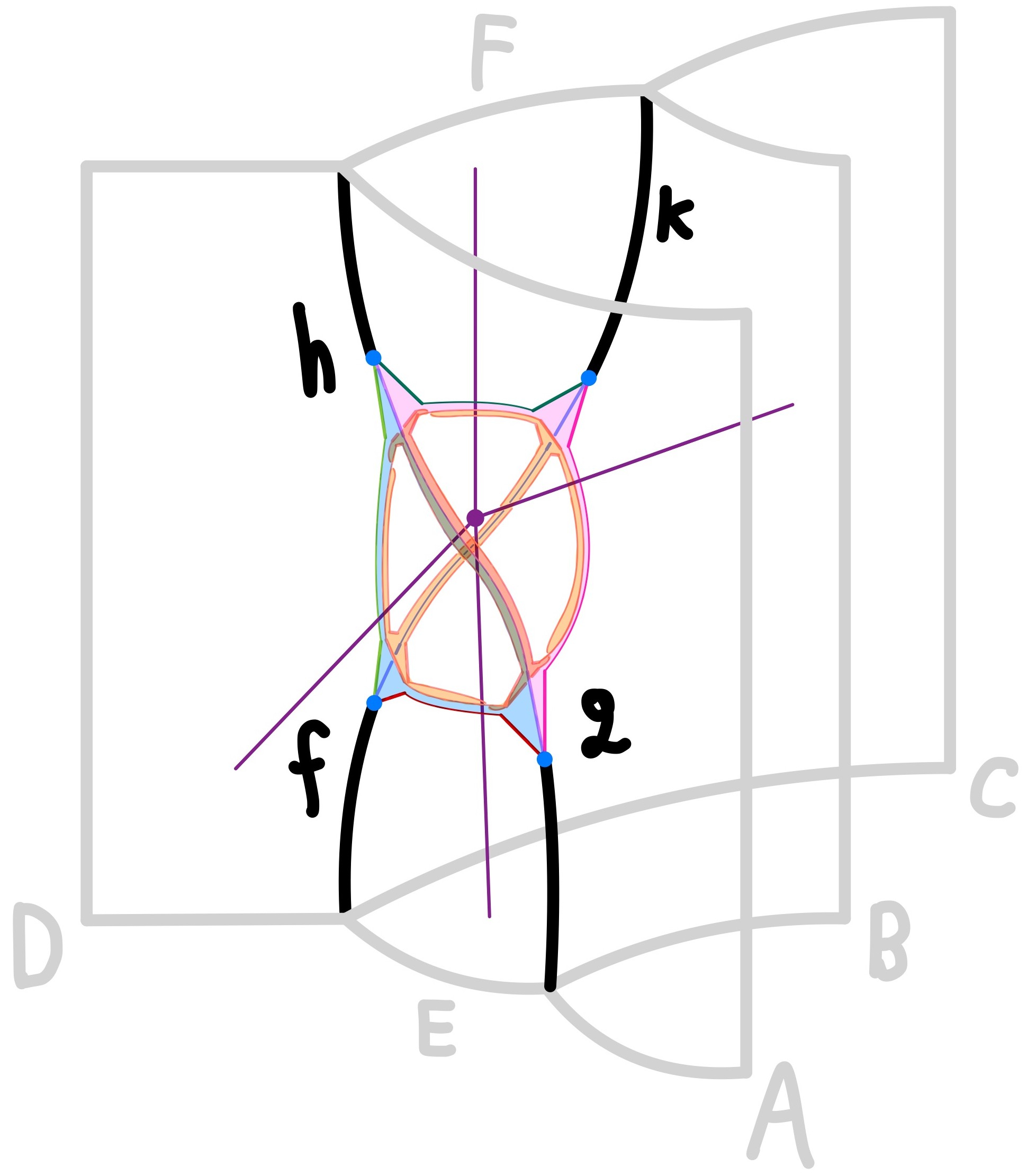}
        \centering{}\caption{This figure is identical to the previous one, except we mark three object-sheets in transparent marker. In pink we highlight a $t_{135}$-sheet, in blue we highlight a $t_{245}$-sheet, and the other two $t$-sheets are unhighlighted and sit in the back. Notice that the four $t$-sheets in this diagram of $\berkanan_v|\psi\rangle$ face the same direction as the $t$-sheets in the original diagram of $|\psi\rangle$. In yellowish-orange we color a global Kirby sheet, which formed from stitching together the four correlated Kirby sheets as a result of the first gluing operation. The boundary of that $\Omega$ sheet consists of four $\omega$ loops, each constricting a unique internal edge (drawn in purple as usual for the simplices of the triangulation). There are also four unhighlighted global Kirby disks whose boundary is exactly the constricting loops mentioned in the last sentence, filling the holes pierced by the internal edges.}
\end{figure}
Now comes the key point. The $\omega$ loops give us $\ehwaz_{E_j}$ projectors for each of the internal edges $E_1,E_2,\dots$ as in \ref{wormhole rod}, so we can identify this diagram with its $\prod_jq_{E_j}$ image (using transparency lemma II), which removes the internal edges and allows us contract those loops using property 4 of the Kirby colors, leaving us with the following membrane net in $\mathcal{H}^\text{mem}_{\Delta^{0,1}_{E_1,E_2,\dots}}$:

\begin{figure}[H]
        \includegraphics[width=0.97\textwidth]{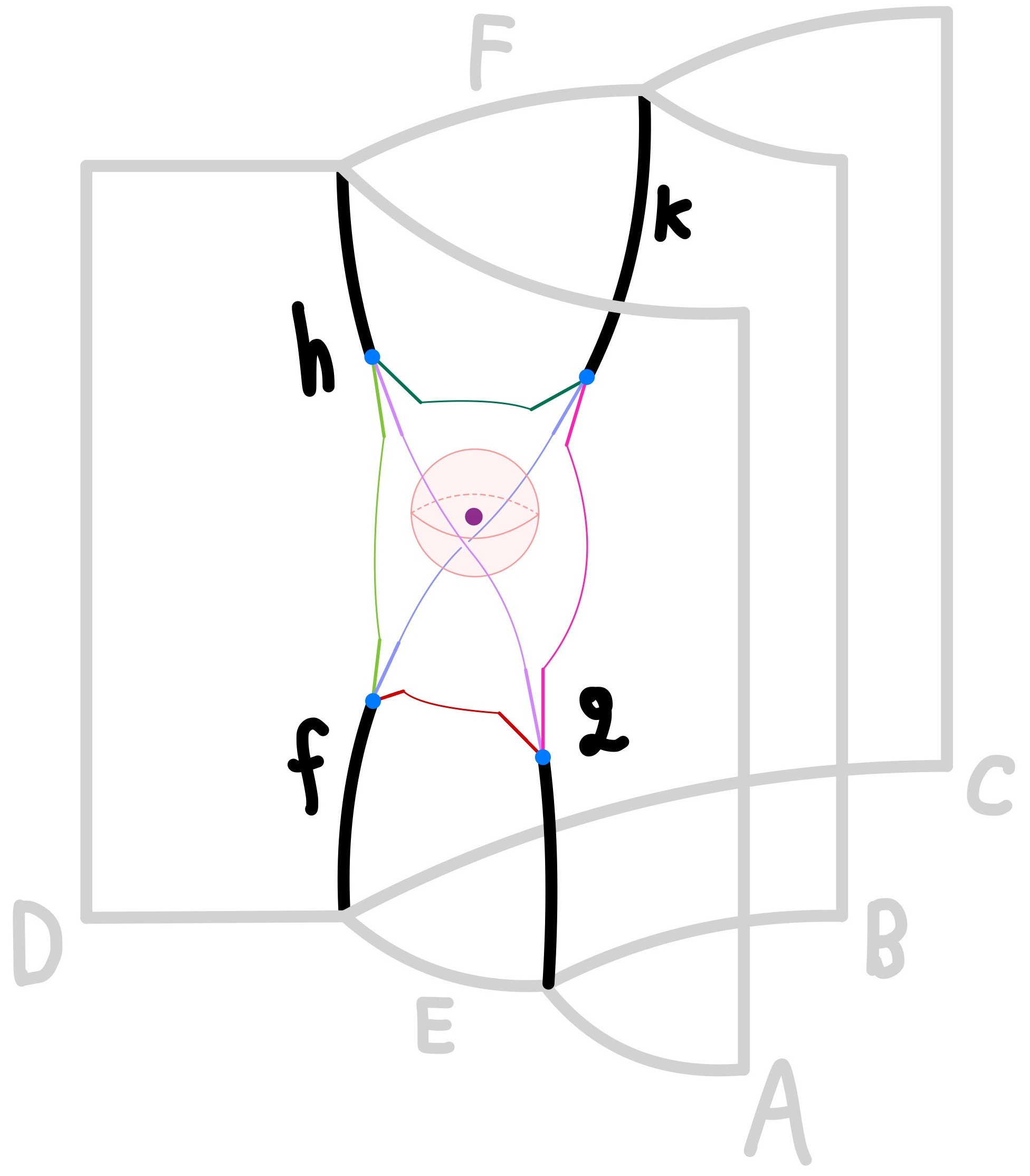}
        \centering{}\caption{After collapsing the constricting loops to $\ehwaz_{E_j}$ projectors that remove the internal edges, the yellowish-orange sheet with four circular holes becomes a global Kirby sphere disconnected from the rest of the diagram, engulfing $v$ and nothing else. The rest of the diagram is exactly the embedding of $|\psi\rangle$ into $\mathcal{H}^\text{mem}_{\Delta^{0,1}_{E_1,E_2,\dots}}$. Note that we still have an extra $\frac{1}{\dim(\mathcal{C})}$ normalization not shown in the figure.}
\end{figure}
This diagram is precisely the embedding of $|\psi\rangle$ into $\mathcal{H}^\text{mem}_{\Delta^{0,1}_{E_1,E_2,\dots}}$, together with a global Kirby sphere engulfing $v$ and nothing else, normalized by a factor of $\frac{1}{\dim(\mathcal{C})}$. In other words, it's an insertion of $\pertho_v$ into the embedding of $|\psi\rangle$, so we can remove the sphere, the vertex $v$ and the $\frac{1}{\dim(\mathcal{C})}$ factor, leaving us with the original diagram embedded in $\mathcal{H}^\text{mem}_{\Delta^{0,1}_{v,E_1,E_2,\dots}}$:

\begin{figure}[H]
        \includegraphics[width=1\textwidth]{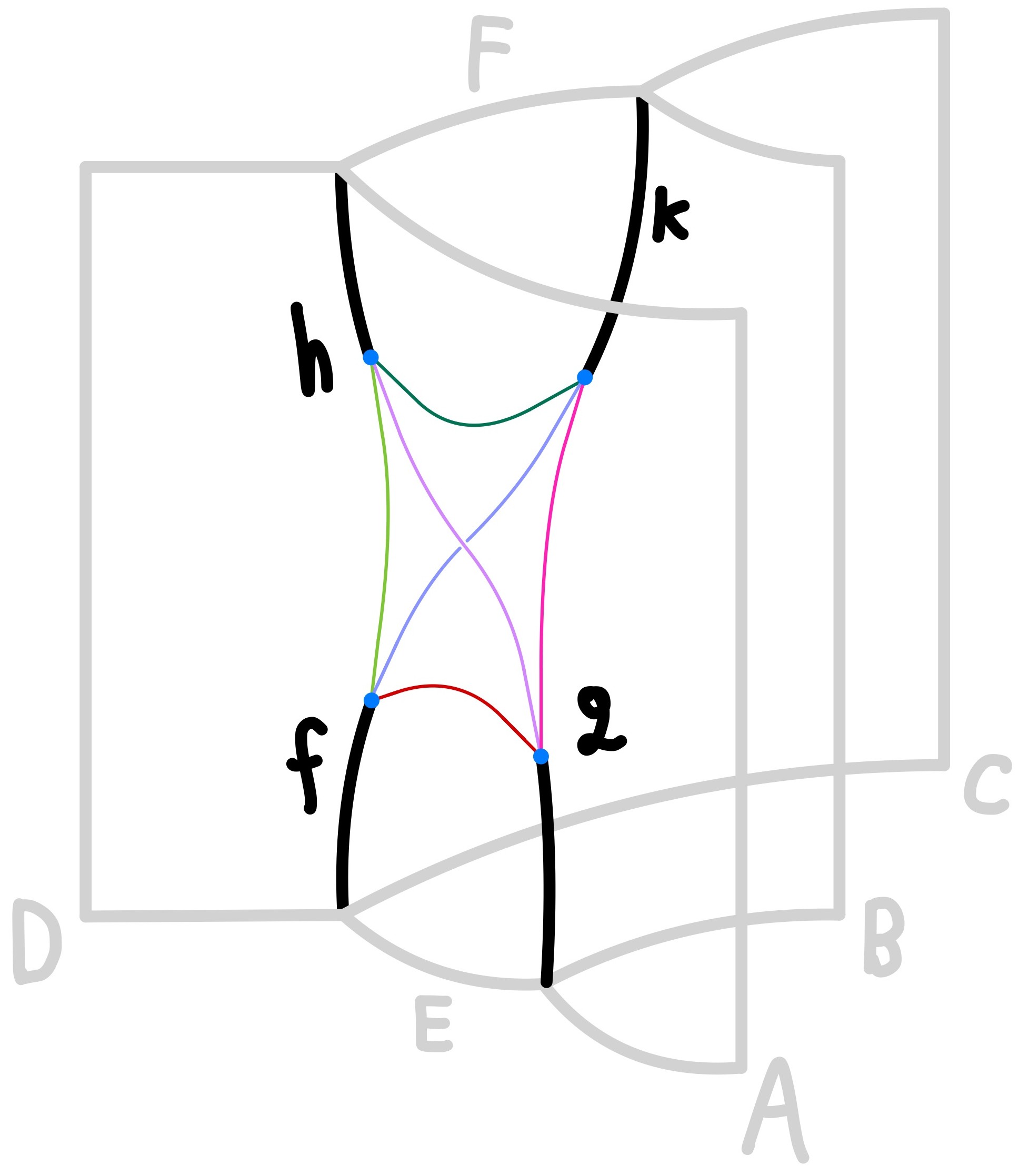}
        \centering{}\caption{After applying transparency lemma III, we are left with this diagram, which is isotopic to the diagram we started with.}
\end{figure}

In the general case\footnote{For which the exact same proof works, as we only used this simple case for visualization and to make the counting of components explicit.} with an arbitrarily complicated star, we will obtain a global Kirby sphere with many tetrahedral spikes, glued to the original diagram along a large number of local kirby loops with global Kirby interior that constrict the internal edges. Clearly, even in this more general case with many simplices defining the star, we can still collapse the constricting loops to apply $\ehwaz_e$ projectors and remove the edges, then we remain with a single $\pertho_v$ operator that removes $v$.

Finishing off, note that we have just shown that $\prod_v\berkanan_v$ is precisely the product of \emph{all} $\ehwaz_e$ and $\pertho_v$ projectors, up to a small subtlety. By the remark after transparency lemma III, the $\pertho_v$ projectors are well defined whenever no edges end on $v$, and only in that situation, so this product makes sense only for some orders of the $\ehwaz_e,\ \pertho_v$ projectors. However, we have shown that each $\berkanan_v$ insertion decomposes into an application of all $\ehwaz_{E_j}$s coming from edges $E_j$ ending on $v$, \emph{then} an application of $\pertho_v$, and this product of course make sense independently of the order of the $\ehwaz_{E_j}$s. Moreover, from the same remark (after transparency lemma III), \emph{any ordered product of $\ehwaz_e,\ \pertho_v$ operators that makes sense} formed from the same collection of $\pertho_v,\ehwaz_e$ operators yields the same overall operator, and we have shown that all $\ehwaz_e,\pertho_v$ projectors appear somewhere in the (well-typed) product $\prod_v\berkanan_v$, so we get that:

\begin{equation}
    \prod_{v\in\Delta^0}\berkanan_v = (\prod_{p\in\Delta^0}\pertho_p) \circ (\prod_{e\in\Delta^1}\ehwaz_e).
\end{equation}
As noted earlier, the l.h.s equals the $\tilde{U}_{\Delta\Delta}$ projector (which projects $H(M,\Delta)$ onto $\mathcal{H}(M)$) and the r.h.s is a projector onto $\mathcal{H}^\text{mem}(M)$, so this proves the desired isomorphism:

\begin{equation}
    \mathcal{H}(M) \cong \mathcal{H}^\text{mem}(M).
\end{equation}

\subsection{The \jrunes{ᛉ} operator, \jrunes{ᚠ}-symbol and a topological lattice model}\label{tensor network}

As mentioned before, the proof using the $\berkanan_v,\ \pertho_p,\ \ehwaz_e$ projectors hints at a tensor network construction, which we expect to be relevant for fault-tolerant quantum computation. From a different but equivalent perspective, the key role of these local commuting projectors suggests that we should be able to extract from our proof a concrete topological lattice model whose Hamiltonian is a sum of terms which are the 3+1D analogs of the corresponding projectors defining the Levin-Wen model. The purpose of this section is to provide the description of this lattice model. For the remainder of this section, fix a closed oriented 3-manifold $M$ and a combinatorial structure $\Delta$ on it.

\subsubsection{Input data}

As in previous parts of the text we fix a complete set of simple objects $I\coloneqq\{A_1,\dots, A_n\}$ (one representative from each equivalence class of simples) and for every pair of objects $A_i,\ A_j$ we fix a complete set of simple 1-morphisms $I_{ij}\coloneqq\{f^1_{ij},\dots,f^{m_{ij}}_{ij}\}$ for the $\text{Hom}$ 1-category of $A_i,\ A_j$ (one from each isomorphism class of simples). Recall that the Hom spaces between of pair of simple 1-morphisms $\text{Hom}(f^a_{ij},f^b_{ij})$ are zero for $a\neq b$ and one dimensional for $a=b$, with a canonical basis element given by the identity 2-morphism $1_{f^a_{ij}}$ which gives a canonical identification $\text{End}(f^a_{ij})\coloneqq\text{Hom}(f^a_{ij},f^a_{ij})\cong\mathbb{C}$. For each pair of opposite Hom spaces $\text{Hom}(f,g),\ \text{Hom}(g,f)$ (with $f,g$ not necessarily simple) we had a non-degenerate pairing $\langle\alpha,\beta\rangle\coloneqq\text{Tr}(\alpha\cdot\beta)$, which in particular defines the quantum dimensions $\dim\mathcal{C},\ \dim(f^a_{ij}),\ \dim(A_i),\ d(A_i)$ as in \ref{quantum dimensions}. Given a pair of objects $A_i,A_j$, we can form their monoidal product and decompose it into simple objects. We will denote the coefficients in this expansion as follows:

\begin{equation}
    A_i\Box A_j = A_1^{\boxplus N_{ij}^1}\boxplus\cdots\boxplus A_n^{\boxplus N^n_{ij}}.
\end{equation}
Given a triple of objects $A_i,A_j,A_k$ we then have a decomposition:

\begin{equation}
    \text{Hom}(A_i\Box A_j,A_k) \cong\bigoplus_{l=1}^n\text{Hom}(A_l,A_k)^{N^l_{ij}}
\end{equation}
which picks for us a complete set of simple 1-morphisms:

\begin{equation}
    I_{ijk}\coloneqq\left\{f^{laa'}_{ijk}\right\}_{l\leq n,\ a\leq m_{lk},\ a'\leq N^l_{ij}}.
\end{equation}
Note that for fixed $i,j,k,l$ the 1-morphisms $\left\{f^{laa'}_{ijk}\right\}_{a\leq m_{lk},\ a'\leq N^l_{ij}}$
all live in different copies of $\text{Hom}(A_l,A_k)$, and if we further fix $a$ then the simple 1-morphisms $\left\{f^{laa'}_{ijk}\right\}_{a'\leq N^l_{ij}}$ are the images of $f^a_{lk}$ under the $N^l_{ij}$ corresponding.

Given six simple objects $A_i,A_j,A_k,A_l,A_m,A_n$ and simple 1-morphisms:
\begin{equation}
    f^{Aaa'}_{ijl}:A_i\Box A_j \to A_l,\quad f^{Bbb'}_{lkm}:A_l\Box A_k\to A_m,\quad f^{Ccc'}_{jkn}:A_j\Box A_k \to A_n,\quad f^{Ddd'}_{inm}:A_i\Box A_n\to A_m
\end{equation}
we pick some basis of $\text{Hom}(f^{Bbb'}_{lkm}\circ (f^{Aaa'}_{ijl}\Box A_k),f^{Ddd'}_{inm}\circ(A_i\Box f^{Ccc'}_{jkn}))$:
\begin{equation}
    I_{ijklmn}\coloneqq\left\{\alpha^{Aaa',Bbb',Ccc',Ddd'}_{ijklmn}(1),\dots,\alpha^{Aaa',Bbb',Ccc',Ddd'}_{ijklmn}(\Phi^{Aaa',Bbb',Ccc',Ddd'}_{ijklmn})\right\}
\end{equation}
which induces a dual basis $\left\{\overline\alpha^{Aaa',Bbb',Ccc',Ddd'}_{ijklmn}(x)\right\}_{1\leq x\leq\Phi^{Aaa',Bbb',Ccc',Ddd'}_{ijklmn}}$ of the opposite Hom space w.r.t to the trace pairing $\langle\_,\_\rangle$.

We will also need the 10j symbols of $\mathcal{C}$ in the above canonical bases. Given ten simple objects $A_{j_{[01]}}A_{j_{[02]}},A_{j_{[12]}},\dots,A_{j_{[34]}}$ labeling the edges of an ordered 4-simplex $[01234]$, 10 simple morphisms:
\begin{equation}
    f^{l_{ijk}a_{ijk}a'_{ijk}}_{[ij][jk][ik]}:A_{j_{[ij]}}\Box A_{j_{[jk]}}\to A_{j_{[ik]}}
\end{equation}
and 5 basis elements (one for each 3-simplex $[ijkl]$):

\begin{equation}
    \alpha^{l_{012}a_{012}a'_{012},l_{013}a_{013}a'_{013},l_{123}a_{123}a'_{123},l_{023}a_{023}a'_{023}}_{j_{[01]},j_{[02]},j_{[12]},j_{[03]},j_{[13]},j_{[23]}}(\beta_{0123}),\cdots,\alpha^{l_{123}a_{123}a'_{123},l_{124}a_{124}a'_{124},l_{234}a_{234}a'_{234},l_{134}a_{134}a'_{134}}_{j_{[12]},j_{[13]},j_{[23]},j_{[14]},j_{[24]},j_{[34]}}(\beta_{1234})
\end{equation}
we get a canonical complex number:
\begin{equation}
    \fehu^{j_{[01]},\dots,j_{[34]}}_{l_{012}a_{012}a'_{012},\dots,l_{234}a_{234}a'_{234}}(\beta_{0123},\dots,\beta_{1234})\in\mathbb{C}
\end{equation}
by tracing the composition of the corresponding $\alpha$s which we call the \emph{$\fehu$-symbol} or the \emph{10j-symbol}. This is a (huge) tensor analogous to the $F$-symbol of fusion 1-categories. We will often enumerate its indices as $\fehu^{j_1,\dots,j_{10}}_{l_1a_1a'_1,\dots,l_{10}a_{10}a'_{10}}(\beta_1,\dots,\beta_5)$, though one should keep in mind that despite the similar symbols for the indices $\fehu$ doesn't have a simple permutation symmetry, but rather a more intricate list of symmetries coming from properties of the pentagonator. The four tensors:

\begin{equation}
    m_{ij},\quad N_{ij}^k,\quad \fehu^{j_1,\dots,j_{10}}_{l_1a_1a'_1,\dots,l_{10}a_{10}a'_{10}}(\beta_1,\dots,\beta_5),\quad \Phi^{Aaa',Bbb',Ccc',Ddd'}_{ijklmn}
\end{equation}
together with the quantum dimensions contain all of the relevant data of the (pre)fusion 2-category, and we will use them to define the tensor network/lattice realization of our TQFT.

\subsubsection{The lattice state space}
Recall that we have also fixed a triangulation $\Delta$ on $M$. We will define for each $e\in\Delta^1,\ t\in\Delta^2,\ \kappa\in\Delta^3$ spaces of states $\tilde{H}_1,\ \tilde{H}_2,\ \tilde{H}_3$, then tensor all of them to get the space of "lattice states" $\tilde{H}(M,\Delta)$.
\begin{itemize}
    \item To each edge $e\in\Delta^1$ we assign the space freely spanned by the vectors $|A_1\rangle,\dots,|A_n\rangle$, one for each simple object.
    \item To each face $t\in\Delta^2$ we assign the space freely spanned by vectors of the form $|f^{laa'}_{ijk}\rangle$ where $i,j,k\leq n$ are arbitrary, $l\leq n$ is an arbitrary index for which $m_{lk}\neq 0$ and $a\leq m_{lk},\ a'\leq N^l_{ij}$ are arbitrary.
    \item To each tetrahedron $\kappa\in\Delta^3$ we assign the space freely spanned by vectors of the form $|\alpha^{Aaa',Bbb',Ccc',Ddd'}_{ijklmn}(\beta)\rangle$ where the lower indices are six arbitrary indices $\leq n$, the four capital indices $A,B,C,D$ are arbitrary indices $\leq n$ such that $m_{Al},m_{Bm},m_{Cn},m_{Dm}$ are all nonzero, the rest of the upper indices are arbitrary indices satisfying:
    \begin{equation}
        a\leq m_{Al},\ b\leq m_{Bm},\ c\leq m_{Cn},\ d\leq m_{Dm},\ a'\leq N^A_{ij},\ b'\leq N^B_{lk},\ c'\leq N^C_{jk},\ d'\leq N^D_{in}
    \end{equation}
    and $\beta$ is an arbitrary index $\leq \Phi^{Aaa',Bbb',Ccc',Ddd'}_{ijklmn}$.
\end{itemize}

The space $\tilde{H}(M,\Delta)$ is then:
\begin{equation}\label{lattice space}
    \tilde{H}(M,\Delta)\coloneqq \tilde{H}_1^{\otimes|\Delta^1|}\otimes \tilde{H}_2^{\otimes|\Delta^2|}\otimes \tilde{H}_3^{\otimes|\Delta^3|}.
\end{equation}

\subsubsection{The family of Hamiltonians}

Finally, we define a Hamiltonian on \ref{lattice space} analogous to the Levin-Wen Hamiltonian by summing \emph{$\algiz_\kappa,\ \thurisaz_t$ projectors} enforcing local fusion rules, then adding lattice realizations of the $\berkanan_v$ operators to get a ground state isomorphic to $\mathcal{H}^\text{mem}(M)$.
\begin{itemize}
    \item For every face $t\coloneqq[IJK]\in\Delta^2$, we define the $\thurisaz_t$ operator to be:
    \begin{equation}
        \thurisaz_t\coloneqq\sum_{i,j,k=1}^n\sum_{l\leq n\ \text{s.t.}\ m_{lk}\neq0}\sum_{a\leq m_{lk},\ a'\leq N^l_{ij}}|A_iA_jA_kf^{laa'}_{ijk}\rangle\langle A_iA_jA_kf^{laa'}_{ijk}|
    \end{equation}
         where the state $|A_iA_jA_kf^{laa'}_{ijk}\rangle$ is a state with $[IJ]$ labeled by $A_i$, $[JK]$ labeled by $A_j$, $[IK]$ labeled by $A_k$, $t$ labeled by $f^{laa'}_{ijk}$ and the rest $\Delta$'s components labeled arbitrarily (there is an implied summation over their labels).
    \item For every volume $\kappa\coloneqq[IJKL]\in\Delta^3$, we define the $\algiz_\kappa$ operator to be:
    \begin{equation}
        \algiz_\kappa\coloneqq\sum_{}|f^{Aaa'}_{ijl}f^{Bbb'}_{lkm}f^{Ccc'}_{jkn}f^{Ddd'}_{inm}\alpha^{Aaa',Bbb',Ccc',Ddd'}_{ijklmn}(\beta)\rangle\langle f^{Aaa'}_{ijl}f^{Bbb'}_{lkm}f^{Ccc'}_{jkn}f^{Ddd'}_{inm}\alpha^{Aaa',Bbb',Ccc',Ddd'}_{ijklmn}(\beta)|
    \end{equation}
    where the sum is over all upper indices satisfying the conditions in the definition of $I_{ijklmn}$ and over all $\beta\leq\Phi^{Aaa',Bbb',Ccc',Ddd'}_{ijklmn}$. The state $|f^{Aaa'}_{ijl}f^{Bbb'}_{lkm}f^{Ccc'}_{jkn}f^{Ddd'}_{inm}\alpha^{Aaa',Bbb',Ccc',Ddd'}_{ijklmn}(\beta)\rangle$ is a state with the faces $[IJK],\dots,[JKL]$ labeled by the four $f$s and the volume $\kappa$ labeled by the $\alpha$ written inside the state, with an implied summation over all other labels.
\end{itemize}
The $\thurisaz_t,\algiz_\kappa$ operators are clearly projectors, they commute with each other and their image is $H(M,\Delta)$. We can then realize the $\berkanan_v$ operators on the lattice using the origami decomposition of \ref{origami}, taking a state in $H(M,\Delta)$ to a sum over states weighted by a product of many $\fehu$-symbols (one for each $\kappa\in\text{St}(v)^3$) whose indices are contracted as in the origami decomposition - one contraction with each label of a certain tetrahedron in the input state, one free index for each label of the output state and the rest contracted with the other $\fehu$ symbols. We also weigh by quantum dimensions as in the definition of $\tilde{U}$ to obtain the $\berkanan_v$ operator from before. We extend $\berkanan_v$ to states in the kernel of the $\thurisaz_t,\algiz_\kappa$ projectors by defining it to be zero operator, which by linearity extends $\berkanan_v$ to an operator on $\tilde{H}(M,\Delta)$. The $\berkanan_v$ operators are commuting projectors that commute with the $\thurisaz_t,\ \algiz_\kappa$ projectors, and together the product of all these projectors is a projector $\tilde{H}(M,\Delta)\to\mathcal{H}^\text{mem}(M)$.

The Hamiltonian is then defined to be:
\begin{equation}
    H\coloneqq-\sum_{t\in\Delta^2}\thurisaz_t-\sum_{\kappa\in\Delta^3}\algiz_\kappa-\sum_{v\in\Delta^0}\berkanan_v
\end{equation}
and its space of ground states (i.e. the vacuum sector) is:
\begin{equation}
    \text{Im}\left(\left(\prod_t\thurisaz_t\right)\left(\prod_\kappa\algiz_\kappa\right)\left(\prod_v\berkanan_v\right)\right) = \text{Im}\tilde{U}_{\Delta\Delta}=\mathcal{H}(M)\cong \mathcal{H}^\text{mem}(M).
\end{equation}
where the last isomorphism is the canonical one of \ref{Main theorem}. We can slightly generalize the above Hamiltonian to the family:
\begin{equation}
    H_{a_t,b_\kappa,c_v}\coloneqq-\sum_{t\in\Delta^2}a_t\thurisaz_t-\sum_{\kappa\in\Delta^3}b_\kappa\algiz_\kappa-\sum_{v\in\Delta^0}c_v\berkanan_v
\end{equation}
where $a_t,b_\kappa,c_v$ are arbitrary positive real numbers, obtaining a more general family of Hamiltonians $H_{a_t,b_\kappa,c_v}$ whose vacuum sector is independent of $a_t,b_\kappa,c_v$. Note that the vacuum sector is spanned by the images of basis states in $\tilde{H}(M,\Delta)$ under the tensor-network operator $\tilde{U}_{\Delta\Delta}$ hence are tensor-network states, the 3+1D analog of \cite{GLSW09}.

\subsubsection{Hermiticity/unitarity and a continuum picture for all general 3D topological orders with gapped boundaries}
The above definition of the state spaces did not provide an inner product, and in particular we don't have a notion of unitarity or hermiticity in general. To define physical lattice models, we need a \emph{unitary structure} on our spherical (pre)fusion 2-category $\mathcal{C}$, which should consist of a dagger structure:
\begin{equation}
    \text{Hom}(f,g)\ni\alpha\mapsto\alpha^\dagger\in\text{Hom}(g,f)
\end{equation}
for which $\text{Tr}(\alpha^\dagger\cdot\beta)$ is positive definite, s.t. the dagger structure respects the rest of the structure of the fusion 2-category (say the pivotal structure). The author of this text is unaware of a full definition in the literature, but under any reasonable definition $\berkanan_v$ should become hermitian, and if we choose the above bases to be orthogonal w.r.t the inner products $\text{Tr}(\_^\dagger\cdot\_)$, $H$ will become hermitian as well, hence time evolution will be unitary. We do not pursue this specialization in the present work, but note that given such a definition our lattice models should specialize in the unitary and $a_t=b_\kappa=c_v=1$ case to \cite{XLLC21}'s construction if we also specialize $\Delta$ to their specific lattice. \cite{XLLC21} argue that their construction is a lattice realization of all general 3-dimensional topological orders with gapped boundaries, thus our canonical $\mathcal{H}(M)\cong \mathcal{H}^\text{mem}(M)$ isomorphism should yield a continuum description for all such topological orders.

\section*{Acknowledgements}
The author of this paper would like to thank Yaron Oz for his advisorship, review of this paper, frequent and insightful discussions, and inspiring me to learn about and relate this paper to the Levin-Wen story. This work is supported by the Israeli Science Foundation Quantum Science and Technology Grant No.1265/26, the German
Israeli Foundation For Scientific Research \& Development (GIF) Grant No. G-3013-303.7/202, and
the US-Israel Binational Science Foundation (BSF) Grant No. 2022110.

\textbf{AI disclosure} Claude was used to review and look for mistakes in the final draft of this paper, but contributed none to the content, writing or figures of this paper.

\bibliographystyle{JHEP}
\bibliography{ref}

\end{document}